\documentclass[aps,pre,showpacs,noshowkeys,amsmath,amssymb,amsfonts,superscriptaddress,longbibliography,reprint]{revtex4-1}
\usepackage[english]{babel}

\usepackage{graphicx}
\usepackage{bm}
\usepackage{physics}
\usepackage{mathtools}
\usepackage{soul}

\setcitestyle{super}
\usepackage{caption}
\usepackage{subcaption}
\DeclareCaptionLabelSeparator{bar}{~\rule[-0.4ex]{0.2ex}{1em}~}
\DeclareCaptionLabelFormat{subfor}{\textbf{#2}}
\newcommand*\bfcaption[2]{\caption[#1]{\textbf{#1.}#2}}
\usepackage{xcolor}
\definecolor{UBcolor}{HTML}{007CC1}
\usepackage{placeins}

\newcommand{\Movie}[1]{\hyperref[movies]{Movie\ S{#1}}}
\newcommand{\Movies}[2]{\hyperref[movies]{Movies\ S#1}\ and\ \hyperref[movies]{S#2}}
\newcommand{\Animation}[1]{\hyperref[animations]{Animation\ S{#1}}}
\newcommand{\Animations}[2]{\hyperref[animations]{Animations\ S#1}\ and\ \hyperref[animations]{S#2}}
\newcommand{\Animationrange}[2]{\hyperref[animations]{Animations\ S#1}\ to\ \hyperref[animations]{S#2}}

\newcommand{\bq}{\mathbf{q}}

\newcommand{\R}{\mathbb{R}}

\newtheorem{theorem}{Theorem}[section]

\newtheorem{remark}[theorem]{Remark}

\usepackage[colorlinks=true,pdfnewwindow=true,linkcolor=UBcolor,citecolor=UBcolor,urlcolor=UBcolor,breaklinks=true,linktocpage]{hyperref}
\usepackage[all]{hypcap}
\usepackage[nameinlink,capitalise]{cleveref}
\crefname{SI section}{SI Section}{SI Sections}
\Crefname{SI section}{SI Section}{SI Sections}

\begin{document}

\title{Dynamical arrest in active nematic turbulence}

\author{Ido Lavi}
\email{ilavi@flatironinstitute.org}
\altaffiliation{Present address: Flatiron Institute, New York, NY, USA}
\affiliation{Center for Computational Biology, Flatiron Institute, 162 5th Ave, New York, NY 10010, USA}
\affiliation{Departament de F\'{i}sica de la Mat\`{e}ria Condensada, Universitat de Barcelona, Barcelona, Spain}

\author{Ricard Alert}
\affiliation{Max Planck Institute for the Physics of Complex Systems, N\"{o}thnitzerst. 38, 01187 Dresden, Germany}
\affiliation{Center for Systems Biology Dresden, Pfotenhauerst. 108, 01307 Dresden, Germany}
\affiliation{Cluster of Excellence Physics of Life, TU Dresden, 01062 Dresden, Germany}

\author{Jean-Fran\c{c}ois Joanny}
\affiliation{Coll\`{e}ge de France, Paris, France}
\affiliation{Laboratoire PhysicoChimie Curie, Institut Curie, Universit\'{e} Paris Sciences \& Lettres (PSL), Sorbonne Universit\'{e}s, Universit\'{e} Pierre et Marie Curie (UPMC), Paris, France}

\author{Jaume Casademunt}
\affiliation{Departament de F\'{i}sica de la Mat\`{e}ria Condensada, Universitat de Barcelona, Barcelona, Spain}
\affiliation{Universitat de Barcelona Institute of Complex Systems (UBICS), Barcelona, Spain}

\date{\today}

\begin{abstract}
Active fluids display spontaneous turbulent-like flows known as active turbulence. Recent work revealed that these flows have universal features, independent of the material properties and of the presence of topological defects. However, the differences between defect-laden and defect-free active turbulence remain largely unexplored. Here, by means of large-scale numerical simulations, we show that defect-free active nematic turbulence can undergo dynamical arrest. This state is characterized by an emergent network of nematic domain walls that channels coherent streams and suppresses chaotic flows. As the system evolves, the branched wall network produces a large-scale pattern with tree-like topological properties. We find that flow alignment---the tendency of nematics to reorient under shear---enhances large-scale chaotic jets in contractile rodlike systems while promoting dynamical arrest in extensile systems. We further show that dynamical arrest persists regardless of whether defects are prohibited by construction or simply fail to form due to a high energy cost of defect cores. Taken together, our findings reveal a striking pattern-formation mechanism, with labyrinths emerging from active turbulence, and illuminate the rich transitional regime between defect-free and defect-laden dynamics. These behaviors call for the experimental realization of active nematics at vanishing or low defect densities, and underscore that, in extensile rodlike nematics, topological defects enable turbulence by preventing dynamical arrest.
\end{abstract}

\maketitle

\section{Introduction}
Active fluids are driven internally by their own components---be they molecular motors, cells, or synthetic colloidal particles. This internal driving produces spontaneous flows, which often exhibit spatio-temporal chaos at high activity \cite{Alert2022b}. Chaotic flows have been observed in a variety of systems, such as bacterial suspensions \cite{Dombrowski2004,Cisneros2007,Ishikawa2011,Wensink2012,Dunkel2013,Patteson2018,Li2019,Peng2021,Liu2021c,Wei2024}, sperm \cite{Creppy2015}, mixtures of cytoskeletal components \cite{Sanchez2012,Henkin2014,Guillamat2017,Ellis2018,Kumar2018,Lemma2019,Martinez-Prat2019,Tan2019,Duclos2020,Martinez-Prat2021}, cell monolayers \cite{Doostmohammadi2015,Yang2016a,Blanch-Mercader2018,Lin2021,Li2022f}, and artificial self-propelled particles \cite{Nishiguchi2015,Kokot2017,Karani2019,Bourgoin2020}. Although all these systems operate at low Reynolds number, for which inertia is negligible, the flows are reminiscent of classic inertial turbulence. Hence, active chaotic flows are generically referred to as active turbulence \cite{Alert2022b}.

Active turbulence encompasses systems with different symmetries, including polar or nematic orientational order, each exhibiting unique statistical properties \cite{Alert2022b}. Recent work showed that the velocity power spectrum of active nematic turbulence features scaling laws with universal exponents independent of the material properties of the fluid such as its viscosity and activity \cite{Giomi2015,Alert2020a,Martinez-Prat2021}. In particular, the $q^{-1}$ scaling of the velocity power spectrum at small wavenumbers $q$ was shown to be highly robust in simulations, provided the Reynolds number remains low \cite{rorai2022coexistence}. Whereas the $-1$ exponent was first predicted for defect-laden dynamics \cite{Giomi2015}, turbulence with the same universal exponent was also found for strongly-ordered nematics in the absence of defects \cite{Alert2020a}. However, how the material properties and the presence or absence of defects affect other, non-universal features of the flow field remains unclear.

Here, we show that defect-free active nematic turbulence exhibits arrested patterns not seen in the presence of defects. We reveal these patterns by studying the effect of the flow-alignment parameter $\nu$, which characterizes the tendency of liquid crystals to reorient under shear. As expected, we find that the scaling exponent of the velocity power spectrum is unaffected when varying $\nu$, confirming its universal character \cite{Alert2020a,Alert2022b}. Despite keeping the same scaling, we find that flow-alignment strongly affects both the strength and the spatiotemporal structure of the flows.

\begin{figure*}[tbhp!]
\begin{center}
\includegraphics[scale=0.48]{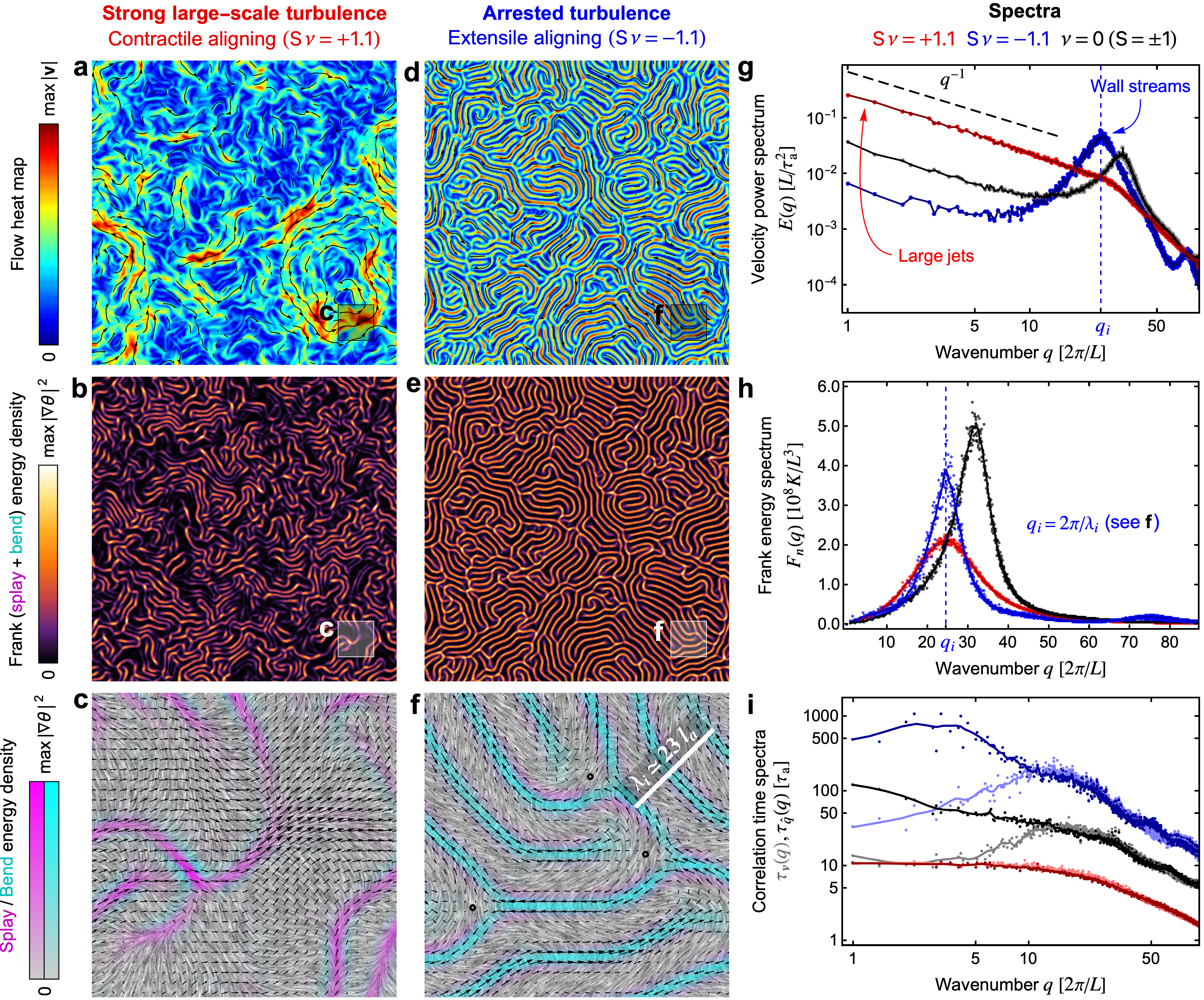}
\end{center}
  {\phantomsubcaption\label{Fig velocity-non-arrested}}
  {\phantomsubcaption\label{Fig Frank-non-arrested}}
  {\phantomsubcaption\label{Fig splay-bend-non-arrested}}
  {\phantomsubcaption\label{Fig velocity-arrested}}
  {\phantomsubcaption\label{Fig Frank-arrested}}
  {\phantomsubcaption\label{Fig splay-bend-arrested}}
  {\phantomsubcaption\label{Fig velocity-spectra}}
  {\phantomsubcaption\label{Fig Frank-spectra}}
  {\phantomsubcaption\label{Fig correlation-times}}
\bfcaption{Strong and arrested regimes of active nematic turbulence}{ \subref*{Fig velocity-non-arrested}--\subref*{Fig splay-bend-arrested}, Snapshots from simulations of defect-free active nematic turbulence in contractile (\subref*{Fig velocity-non-arrested}--\subref*{Fig splay-bend-non-arrested}) and extensile (\subref*{Fig velocity-arrested}--\subref*{Fig splay-bend-arrested}) flow-aligning systems. Parameter values were set to $R=1$, $\nu=-1.1$, and $A=3.2\times10^5$. Top panels (\subref*{Fig velocity-non-arrested},\subref*{Fig velocity-arrested}) show the flow field; black curves are streamlines, and the color indicates the speed (see \Movies{1}{2}). Middle panels (\subref*{Fig Frank-non-arrested},\subref*{Fig Frank-arrested}) show the Frank free energy density $\sim|\nabla\theta|^2$, with high-intensity lines corresponding to nematic domain walls (see \Movies{1}{2}). Bottom panels (\subref*{Fig splay-bend-non-arrested},\subref*{Fig splay-bend-arrested}) are zooms highlighting the type of nematic distortion as well as the interplay between nematic walls and flows. The gray-scale background is the line integral convolution representation of the director field $\bm{n}$. Magenta and cyan intensities respectively represent splay $(\nabla\cdot\bm{n})^2$ and bend $|\nabla\times\bm{n}|^2$ contributions to the Frank energy density. The black arrows represent the flow field $\bm{v}$, which localizes along the nematic walls in the arrested regime. Black circles indicate stagnation points of the flow. White scale bar represents the selected wavelength $\lambda_i$. \subref*{Fig velocity-spectra}--\subref*{Fig correlation-times}, Spectra characterizing fully-developed active nematic turbulence (see details in \cref{spectra}). The lines in (\subref*{Fig Frank-spectra},\subref*{Fig correlation-times}) represent a smoothed (Gaussian) interpolation of the computed data points. We compare the flow-aligning contractile (red, as in \subref*{Fig velocity-non-arrested}--\subref*{Fig splay-bend-non-arrested} and \Movie{1}) and extensile (blue, as in \subref*{Fig velocity-arrested}--\subref*{Fig splay-bend-arrested} and \Movie{2}) cases with the $\nu=0$ case (black, as in Fig.\ S1\cite{SM} and \Movie{3}), for which contractile and extensile stresses are equivalent up to a rotation\cite{Edwards2009,Giomi2014a,Lavi2024} . \subref*{Fig velocity-spectra}, Velocity power spectrum on a log-log scale, showing (i) the universal low-$q$ scaling law and (ii) the distinct organization of flows across scales in the different cases. The wider scaling regime in the contractile case captures the strong large-scale jets (see \subref*{Fig velocity-non-arrested}). The peak in the extensile case is representative of wall streams (see \subref*{Fig velocity-arrested}). \subref*{Fig Frank-spectra}, Frank energy spectrum, showing that (i) the selected wavelength (peak position) depends on $\nu$ but not on the sign of active stress, and (ii) the peak width depends on the sign of active stress when $\nu\neq0$. \subref*{Fig correlation-times}, Spectrum of correlation times associated with the flow $\bm{v}$ (light colored points and lines) and the nematic tensor $\hat{q}_{\alpha\beta}$ (darker points and lines). This log-log plot reveals strong differences in decay times between the regimes, as well as the differences between the flow and nematic tensor within a regime. Correlation times are extracted from exponential fits to the corresponding space-time autocorrelation functions in Fourier space (see \cref{spectra}). } 
\label{Fig regimes}
\end{figure*}

In the absence of defects, we identify two distinct regimes depending on the parameter combination $\text{sign}(\zeta) \nu$, where $\zeta$ is the active stress parameter, with $\zeta > 0$ for extensile and $\zeta < 0$ for contractile stresses, respectively. We discuss the case of rod-like nematic components, for which $\nu \leq 0$. In the flow-aligning regime, where $\nu< -1$, we find that contractile nematics (with $\zeta \nu>0$) exhibit strongly chaotic flows and nematic distortions, which largely decorrelate from one another. In this regime, nematic domain walls---lines of strong variation of the director---appear, split, and dissolve, forming a dynamically reorganizing pattern (\crefrange{Fig velocity-non-arrested}{Fig splay-bend-non-arrested}, \Movie{1}). In contrast, for extensile aligning nematics ($\zeta \nu<0$), we find that domain walls are strongly stabilized by the flow they entrain. The walls typically grow and branch but neither break nor merge. As walls avoid each other, they become arrested (or grid-locked) in a space-filling tree-like pattern (\crefrange{Fig velocity-arrested}{Fig splay-bend-arrested}, \Movie{2}). Once arrested, the pattern exhibits slow residual dynamics that shares some features with ageing phenomena in glassy systems. The special case $\nu=0$ has been the focus of previous work\cite{Alert2020a} and is considered here as a reference. The dynamics in this case lie between these two extremes while also manifesting comparatively shorter wavelengths (Fig.\ S1\cite{SM}, \Movie{3}).

We uncover that the dynamical arrest stems from the emergence of an effective topology, characterized by a set of connectivity rules for the domain walls imposed by the combined effects of activity and flow alignment. In particular, we demonstrate that, in the absence of topological defects, the arrested state is organized by local nematic structures, which we name pseudo-defects. We discuss their topological interpretation and demonstrate that they coexist with actual defects in experiments on microtubule-based active nematics. 

Finally, we show that dynamical arrest is not exclusive to our minimal defect-free model, but rather a robust phenomenon that also arises in the full Q-tensor formulation of active nematics. In this more general framework, an additional control parameter---the defect core size---governs whether topological defects are energetically suppressed or favored. By tuning this parameter, we uncover a transition in which defect nucleation ramps up sharply near a threshold. As defects proliferate, the flow crosses over from a dynamically arrested state, organized into streams that are channeled through the wall network, to the more familiar regime of chaotic, vortex-dominated flows. These simulations further highlight the role of defects in enabling active turbulence and reveal a rich transitional regime where defects and domain walls can coexist.

\section{Director-based defect-free model} 

We generalize our previous minimal model of active nematic turbulence \cite{Alert2020a} by adding flow alignment and the Ericksen stress \cite{DeGennes-Prost}. This yields the active Ericksen-Leslie liquid crystal model, widely regarded as a paradigm in this field\cite{Alert2022b}. 

We work at vanishing Reynolds number, for which momentum balance reads: 
\begin{equation} 
0 = -\partial_\alpha P + \partial_\beta \left( \sigma_{\alpha\beta}^{\text{a}} + \sigma_{\alpha\beta}^{\text{E}} + \sigma_{\alpha\beta} \right). \label{eq force-balance}
\end{equation}
The pressure $P$ enforces the incompressibility condition $\partial_\alpha v_\alpha =0$ of the flow field $\bm{v}$. The next two terms are the divergence of the antisymmetric part of the stress tensor, and of the symmetric Ericksen stress, respectively\cite{DeGennes-Prost,Julicher2018}:
\begin{equation} \label{eq anti-Ericksen}
\sigma_{\alpha\beta}^{\text{a}} = \frac{1}{2} \left( n_\alpha h_\beta - h_\alpha n_\beta \right), \quad \sigma_{\alpha\beta}^{\text{E}} = - \frac{\delta F_n}{\delta \left(\partial_\alpha n_\gamma \right)} \partial_\beta n_\gamma.
\end{equation}
These stresses arise from the elastic distortions of the director field $\bm{n}$. Distortions produce an orientational field $h_\alpha = - \delta F_n/\delta n_\alpha = K\nabla^2 n_\alpha + h_\parallel^0 n_\alpha$ computed from the Frank free energy which, in the one-constant $K$ approximation, reads
\begin{equation} \label{eq Frank}
F_n = \int \left[ \frac{K}{2} \left(\partial_\alpha n_\beta \right) \left(\partial_\alpha n_\beta \right) - \frac{1}{2}h_\parallel^0 n_\alpha n_\alpha \right] \dd^2\bm{r}.
\end{equation}
Here, we assumed that the fluid is deep in the nematic phase so that the director field has a fixed modulus $|\bm{n}| = 1$. This constraint is imposed by the second term through the Lagrange multiplier $h_\parallel^0$. We start with a continuous director field, which precludes the generation of topological defects \cite{Alert2020a,Lavi2024}. Note that this constraint is lifted in our analysis of the full Q-tensor model, presented in the final section of this paper.

The last term in \cref{eq force-balance} arises from the symmetric part of the deviatoric stress, which is given by the constitutive equation \cite{Kruse2005,Marchetti2013,Prost2015,Julicher2018}
\begin{equation} \label{eq symmetric-stress}
\sigma_{\alpha\beta} = 2 \eta v_{\alpha\beta} - \zeta \hat{q}_{\alpha\beta} + \frac{\nu}{2} \left( n_\alpha h_\beta + h_\alpha n_\beta - n_\gamma h_\gamma \delta_{\alpha\beta} \right).
\end{equation}
Here, $\eta$ is the shear viscosity, $v_{\alpha\beta} = 1/2 \left(\partial_\alpha v_\beta + \partial_\beta v_\alpha \right)$ is the symmetric part of the strain-rate tensor, $\zeta$ is the active stress parameter, and $\hat{q}_{\alpha\beta} = n_\alpha n_\beta - 1/2\, \delta_{\alpha\beta}$ is the nematic orientation tensor defined by the director $\bm{n}$. The last term describes stresses due to flow alignment, with parameter $\nu$.

As in passive nematics, the dynamics of the director field are given by
\begin{equation} 
\partial_t n_\alpha + v_\beta \partial_\beta n_\alpha + \omega_{\alpha\beta} n_\beta = \frac{1}{\gamma} h_\alpha - \nu v_{\alpha\beta} n_\beta, \label{eq director}
\end{equation}
where $\omega_{\alpha\beta} = 1/2 \left(\partial_\alpha v_\beta - \partial_\beta v_\alpha \right)$ is the vorticity tensor, and $\gamma$ is the rotational viscosity. The left-hand side is the co-rotational material derivative of the director field. On the right-hand side, the orientational field $h_\alpha$ specifies the elastic torque acting on the director, whereas the last term captures its reorientation due to extensional flow, i.e., the flow-alignment effect.

Nondimensionalizing the equations, we rescale length by the system size $L$, time by the active time $\tau_{\text{a}} = \eta/|\zeta|$, pressure by the active stress $|\zeta|$, and orientational field by $K/L^2$. To eliminate the pressure, we take the curl of \cref{eq force-balance} and obtain a Poisson equation for the vorticity $\omega$, which we write in terms of the stream function $\psi$ (defined by $v_x = \partial_y \psi$ and $v_y = - \partial_x \psi$) \cite{Lavi2024}, as $\omega = -\nabla^2 \psi$:
\begin{multline} \label{eq vorticity}
\nabla^2 \omega = - \nabla^4 \psi = - S \partial_\alpha \partial_\beta \bar q_{\alpha\beta} + R \nu \partial_\alpha \partial_\beta ( \bar q_{\alpha\beta} h_\parallel - \hat{q}_{\alpha\beta} h_\perp ) \\
+ \frac{R}{2} \nabla^2 h_\perp + R \epsilon_{\alpha\beta} (\partial_\alpha h_\perp ) \partial_\beta \theta.
\end{multline}
Here, $S$ is the sign of the active stresses, with $S=\pm1$ for extensile and contractile stresses, respectively, and $\theta$ is the nematic director angle: $\bm{n} = (\cos \theta,\,\sin\theta)$. The right-hand side contains four sources of vorticity, arising respectively from active, flow-alignment, antisymmetric, and Ericksen stresses. We used the totally antisymmetric tensor $\epsilon_{\alpha\beta}$ to define $\bar q_{\alpha\beta} \equiv - \epsilon_{\alpha\gamma} \hat{q}_{\gamma\beta}$. We also introduced the parallel and perpendicular components of the orientational field \cite{Lavi2024}:
\begin{equation} \label{eq h}
h_\perp = \frac{1}{A} \nabla^2 \theta,\qquad h_\parallel = \nu \bar q_{\alpha\beta} \partial_\alpha \partial_\beta \psi.
\end{equation}
\Cref{eq vorticity,eq h} contain three dimensionless parameters: the activity number $A = L^2/\ell_{\text{a}}^2$, the viscosity ratio $R = \gamma/\eta$, and the flow-alignment parameter $\nu$. The activity number compares the system size $L$ with the active length $\ell_{\text{a}} = \sqrt{ K/(|\zeta| R)}$ defined by the balance between active and elastic nematic stresses.

Finally, the director dynamics \cref{eq director} breaks down into a component parallel to $\bm{n}$, which specifies $h_\parallel$ as given in \cref{eq h}, and a component perpendicular to $\bm{n}$, which determines the dynamics of the nematic angle $\theta$ \cite{Lavi2024}:
\begin{equation} \label{eq theta}
\partial_t \theta - \epsilon_{\alpha\beta} \partial_\alpha \psi \, \partial_\beta \theta + \frac{1}{2}\nabla^2 \psi = h_\perp + \nu \hat{q}_{\alpha\beta} \partial_\alpha \partial_\beta \psi.
\end{equation}
\Crefrange{eq vorticity}{eq theta} give the hydrodynamics of our active fluid in terms two fields: the angle $\theta$ and the stream function $\psi$.

We numerically integrate \crefrange{eq vorticity}{eq theta} in a square with periodic boundary conditions using a hybrid numerical scheme that combines a pseudo-spectral method for \cref{eq vorticity} with the finite element method for \cref{eq theta} (\cref{numerical}). Unless otherwise specified, we set $R=1$ and $A=3.2\times 10^5$, for which the system was shown to develop large-scale active turbulence when $\nu=0$ \cite{Alert2020a}. The dynamics are invariant under the transformation ($\nu \rightarrow -\nu$, $S \rightarrow -S$, $\theta \rightarrow \theta \pm \pi/2$), meaning that changing the signs of $\nu$ and $S$ simultaneously leaves the system invariant up to a rotation \cite{Edwards2009,Giomi2014a,Lavi2024}. 

\vspace{10pt}
\section{Regimes of defect-free active turbulence}
\subsection{Strong large-scale turbulence}

We first study the case $S\nu >1$, realized with contractile flow-aligning rods ($S=-1$ and $\nu = -1.1$). We find a chaotic flow field with vortices at many scales and strong (albeit transient) large-scale jets (\cref{Fig velocity-non-arrested}). These flow patterns stem from long-range hydrodynamic interactions and appear to be locally uncorrelated with the distortion pattern of the director field (\cref{Fig splay-bend-non-arrested}), which also evolves chaotically. The Frank energy density concentrates along domain walls (\cref{Fig Frank-non-arrested,Fig splay-bend-non-arrested})---well-known regions of strong orientation gradients \cite{Thampi2014a,Thampi2014b,Thampi2016a}. Such walls feature a typical separation length, corresponding to the peak position in the Frank energy spectrum (\cref{Fig Frank-spectra}, red), coinciding with the peak of the nematic tensor power spectrum (\cref{Fig Q spectrum}, red). Compared with the $\nu=0$ case\cite{Alert2020a}, the nematic pattern in the $S\nu>1$ regime appears more broken, with fragmented walls coexisting with wall-free regions (dark patches in \cref{Fig Frank-non-arrested}). As a consequence, the total Frank free energy is smaller, and its peak is shifted to larger scales (\cref{Fig Frank-spectra}, red). A dynamically-rearranging pattern emerges as walls spontaneously appear, split, and dissolve (\Movies{1}{4}, left). Both the flows and the nematic tensor are very dynamic, as confirmed by their short correlation time at all length scales (\cref{Fig correlation-times}, red). 

In recent work we have shown that, for $S\nu>1$, both the quiescent state with uniform orientation and the flowing state with splaying domain walls are non-linearly unstable against one-dimensional modulations\cite{Lavi2024}. Thus, strong enough perturbations can trigger sudden transitions between these states. For this reason, we assert that domain wall patterns can be disrupted by the strong turbulent flows. In the two-dimensional simulations shown here, we observe the coexistence of uniform and wall-laden patches that continuously rearrange. We conclude that the metastibility associated with this regime strongly enhances the chaotic flows, particularly at large-scales, as indicated by the wide $q^{-1}$ scaling regime in \cref{Fig velocity-spectra}, red. 

\subsection{Arrested turbulence}

We now study the case $S\nu < -1$, realized with extensile flow-aligning rods ($S=+1$ and $\nu=-1.1$). While the system still evolves chaotically, large-scale flows are much weaker than in the other cases, with the $q^{-1}$ power law spanning a narrower range of scales (\cref{Fig velocity-spectra}, blue). Rather than large-scale jets, the dominant flows are at shorter length scales, as indicated by the enhanced peak of the velocity power spectrum (\cref{Fig velocity-spectra}, blue). These smaller-scale flows form streams that are strongly localized along the bending domain walls where the active stress gradients are larger (\crefrange{Fig velocity-arrested}{Fig splay-bend-arrested}).

In contrast to the non-arrested case, here a connected wall pattern extends across the entire system (\cref{Fig Frank-arrested}). The Frank free energy spectrum is more narrowly peaked at the characteristic length scale (\cref{Fig Frank-spectra}, blue), compared to the non-arrested case (\cref{Fig Frank-spectra}, red). The same peak is also prominent in the nematic tensor power spectrum (\cref{Fig Q spectrum}, blue). Strikingly, the nematic wall pattern becomes far less dynamic; the walls and their associated streams become locked in a tree-like pattern. We term this phenomenon dynamical arrest. The weak large-scale chaotic flows make the pattern fluctuate only slightly (\Movies{2}{4}, right). The dynamical arrest manifests in much longer correlation times, especially of the nematic tensor field at large scales (\cref{Fig correlation-times}, blue). The tree-like pattern of domain walls is characterized by unique connectivity properties, which we elaborate on later. We conclude that, in the absence of topological defects, the combined effects of extensile stress and rod-like flow-alignment give rise to the arrest of active nematic turbulence.

\begin{figure*}[tbhp!]
\begin{center}
\includegraphics[scale=0.48]{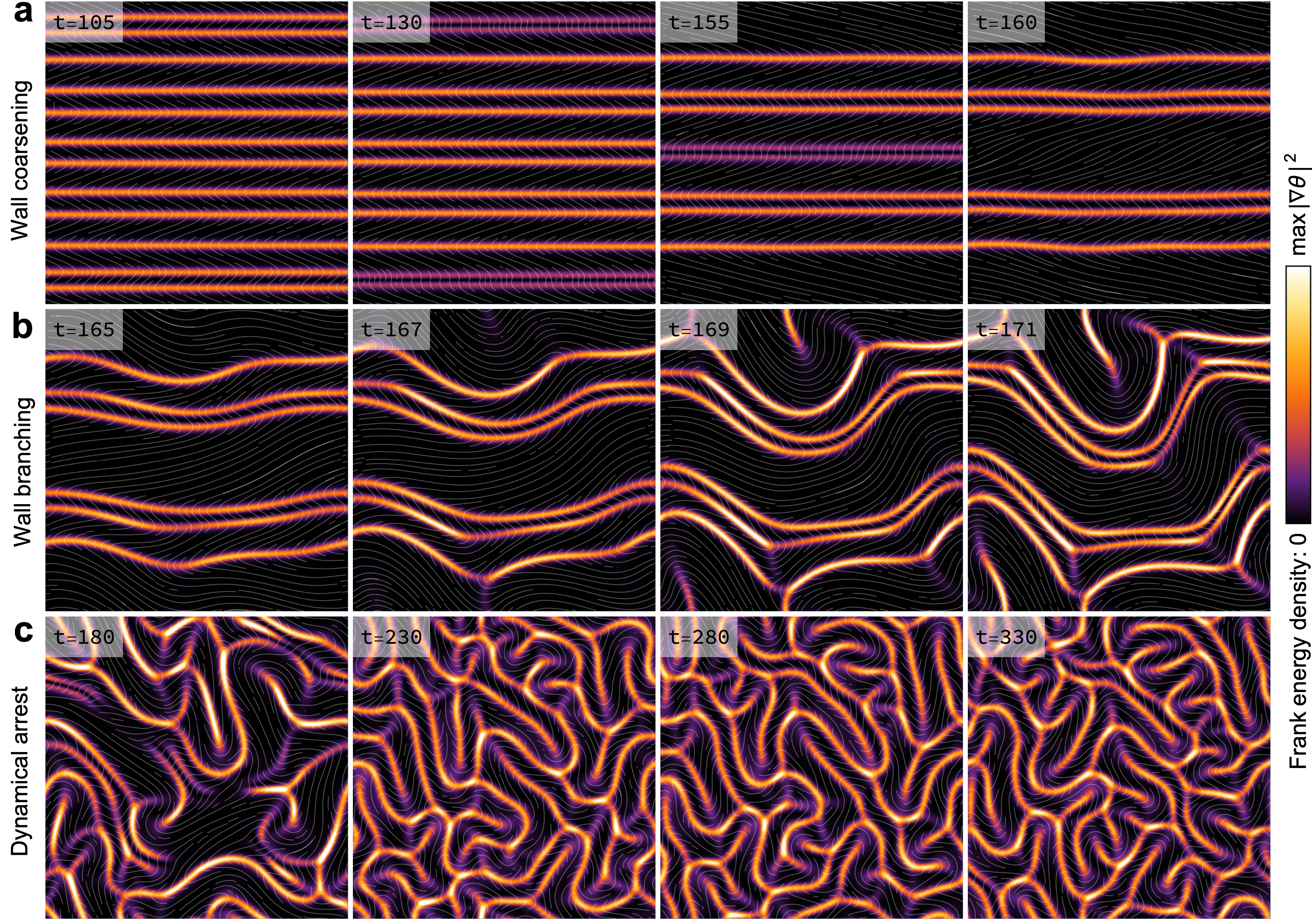}
\end{center}
  {\phantomsubcaption\label{Fig wall-coarsening}}
  {\phantomsubcaption\label{Fig wall-branching}}
  {\phantomsubcaption\label{Fig wall-arrest}}
\bfcaption{Wall coarsening, branching, and dynamical arrest}{ Sequential snapshots from a single simulation in the extensile flow-aligning regime, showing the coarsening of bending walls (\subref*{Fig wall-coarsening}, top row), the zigzag instability of the coarsened configuration, followed by wall branching and rapid tip growth (\subref*{Fig wall-branching}, second row), and the evolution towards the arrested tree pattern (\subref*{Fig wall-arrest}, bottom row). The initial condition was set to a striped pattern of wavelength that matches the typical length selected in the chaotic regime (see first and final panels). A perturbation along the $y$-axis triggers the coarsening of one-dimensional stripes, confirming that straight anti-parallel walls do not have a preferred separation. The length selected in the tree-like pattern is determined by two-dimensional interactions, with growing branches avoiding preexisting walls. Parameter values are: $R=1$, $S=1$, $\nu=-1.1$, and $A=19692$ (chosen so that the system size roughly equals six times the selected wavelength).}
\label{Fig walls}
\end{figure*}

To understand the mechanism of arrest, we note that the spontaneous-flow instability, which ultimately drives turbulence, generates shear flow \cite{Voituriez2005}. This shear produces antiparallel flows that concentrate along the nematic walls\cite{Voituriez2005,Edwards2009,Giomi2014a,Ramaswamy2016,Alert2020a,Lavi2024}. The walls are directed lines, with their direction defined by the vector $(\partial_y\theta,-\partial_x\theta)$, which is perpendicular to the gradient of the angle and coincides with the flow direction in the arrested regime. 

Quasi one-dimensional wall patterns tend to coarsen, leading to larger domains (\cref{Fig wall-coarsening} and \Movie{4}, right). Since the director orientation is roughly uniform within such domains, the same instability prompts distortions in a perpendicular direction\cite{Alert2020a,Ramaswamy2016}. This process induces sequential `zig-zag' instabilities, leading to wall branching and progressively smaller domains, until a characteristic wavelength is selected (\cref{Fig wall-branching} and \Movie{4}, right) \cite{Martinez-Prat2019,Alert2020a}. When an open end of a branch approaches another wall perpendicularly, it is deflected and becomes antiparallel to that wall. The growth of branches typically stops when they are blocked by a preexisting branch point. Ultimately, these dynamics lead to a grid-locked pattern (\cref{Fig wall-arrest} and \Movie{4}, right). 

When a growing open-ended wall encounters a branch point ahead of it, the outflowing stream at the end point of that wall is diverted symmetrically in perpendicular directions so that material can join the streams of the two walls that split away from the branch point (see \cref{Fig splay-bend-arrested}, around the black circles). In effect, this diversion creates a local stagnation point: flows arrive from the direction of the open-ended wall and depart in the perpendicular direction, which is also the direction to which rods tend to align by flow-alignment. For bend walls---generated by an extensile stress---the director along the incident wall is perpendicular to the wall itself, and thus naturally aligned with the extensional axis of the flow at the stagnation point. We therefore propose that the structure around the stagnation point acts as a trap, and its stability is central to understanding how flow alignment promotes the dynamical arrest of active turbulence.

\begin{figure}[tbhp!]
\begin{center}
\includegraphics[scale=0.48]{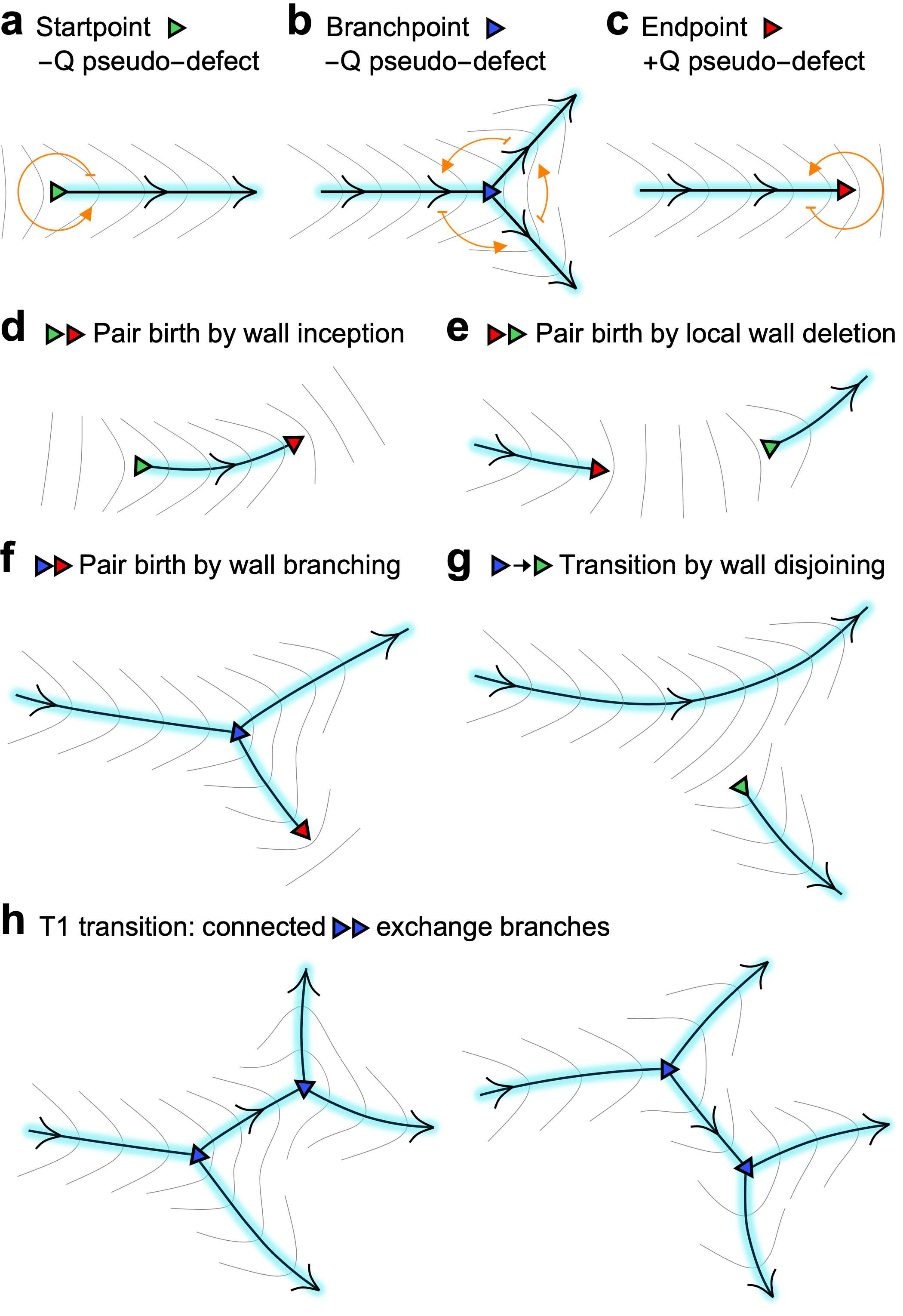}
\end{center}
  {\phantomsubcaption\label{Fig startpoint}}
  {\phantomsubcaption\label{Fig branch point}}
  {\phantomsubcaption\label{Fig endpoint}}
  {\phantomsubcaption\label{Fig inception}}
  {\phantomsubcaption\label{Fig deletion}}
  {\phantomsubcaption\label{Fig branching}}
  {\phantomsubcaption\label{Fig dislocation}}
  {\phantomsubcaption\label{Fig T1}}
\bfcaption{Domain wall nodes and their pseudo-topology}{ In all diagrams, black directed lines represent nematic bending walls, with gray lines tracing the director and cyan indicating the bending energy. \subref*{Fig startpoint}--\subref*{Fig endpoint}, Nodes are either startpoints (green, negative pseudo-defects), branchpoints (blue, negative pseudo-defects), or endpoints (red, positive pseudo-defects). The orange curves illustrate the paths excluding walls used to define the pseudo-charge (\cref{pseudo-charge}). \subref*{Fig inception}--\subref*{Fig deletion}, Two ways in which a startpoint-endpoint pair of opposite charge can be born. The inverse processes of (\subref*{Fig inception}) (complete wall dissolution) and (\subref*{Fig deletion}) (local wall completion) result in the annihilation of such pairs. \subref*{Fig branching}, Wall branching gives birth to a branchpoint-endpoint pair, also of opposite charge. A connected pair may annihilate via branch retraction. \subref*{Fig dislocation}, A branchpoint (\subref*{Fig branch point}) transitions into a startpoint when one outgoing wall disjoins. The inverse process corresponds to a startpoint (\subref*{Fig startpoint}) joining with a bare wall. \subref*{Fig T1}, Connected branchpoints can shrink their connecting branch, transiently creating a $-2Q$ pseudo-charged structure, exchange outgoing walls and extend again in the perpendicular direction. The processes (\subref*{Fig inception}--\subref*{Fig T1}) are further illustrated in \Animationrange{1}{5}.}
\label{Fig topology}
\end{figure}

\section{Pseudo-topology of domain wall networks}

\subsection{Wall nodes define pseudo-defects}

The domain walls emerging in our system are typically of a single type (either splaying or bending, see \cref{Fig splay-bend-non-arrested,Fig splay-bend-arrested}). Such walls generically feature three types of nodes:
\begin{enumerate}
    \item Startpoints, with one outgoing wall and no incoming wall (\cref{Fig startpoint}),
    \item Branchpoints, with one incoming wall and two outgoing walls (\cref{Fig branch point}),
    \item Endpoints, with one incoming wall and no outgoing wall (\cref{Fig endpoint}).
\end{enumerate}
These recurring node structures arise naturally in our simulations and serve as an effective framework for characterizing defect-free dynamics. We note that a node with two incoming walls and one outgoing wall (a mergepoint) is only possible by combining different wall types (Fig.\ S2\cite{SM}). Additionally, we observethat startpoints are very common in the non-arrested (contractile aligning) regime but become increasingly rare over time in the arrested (extensile aligning) regime.

We define these three types of nodes as pseudo-defects. Unlike proper topological defects, they are not phase singularities of the director field and do not carry a topological charge. Nevertheless, the director varies around them in a manner that resembles true nematic defects, except for its sharp counter variation across the walls. We thus assign the nodes a pseudo-charge, defined as the number of turns of the director along a path that surrounds the pseudo-defect, excluding the sharp variation across all intersecting walls (\cref{Fig startpoint,Fig branch point,Fig endpoint}). The excluded walls in this definition are analogous to the so-called Dirac strings, which are constructs that enable magnetic monopoles in Maxwell's equations\cite{Dirac1931} (see discussion in \cref{pseudo-charge}). With this definition, endpoints have a positive pseudo-charge $+Q$ (\cref{Fig endpoint}) and both startpoints and branchpoints have pseudo-charge $-Q$ (\cref{Fig startpoint,Fig branch point}). The pseudo-charge depends on $\nu$. For $S\nu> -1$, we have $Q\simeq 1/2$, while for $S\nu\leq -1$, we have $Q\simeq\frac{1}{2}\left[1-\frac{1}{\pi}\arccos{(|\nu|^{-1}}) \right]$ (see details in \cref{pseudo-charge}). The total pseudo-charge contained in a closed domain is $mQ$ where $m$ is the difference between the number of walls crossing the path inwards and outwards (\cref{pseudo-charge}). All pseudo defects have polarity, defined by the direction of the outgoing wall for startpoints (\cref{Fig startpoint}) and the direction of the incoming wall for branchpoints and endpoints (\cref{Fig branch point,Fig endpoint}).

The topological pseudo-charge is preserved by the dynamics of the director. Pseudo-defects with opposite charges are created in pairs and annihilated in pairs (\cref{Fig inception,Fig deletion,Fig branching}, \Animationrange{1}{3}), similarly to actual nematic defects. A wall endpoint cannot merge with a preexisting wall. Instead, it can connect with a startpoint ahead of it (the inverse process of \cref{Fig deletion}, \Animation{2}) or, through wall shrinkage, it can annihilate with the nearest negative pseudo-defect it is \emph{connected to} (the inverse processes of \cref{Fig inception,Fig branching}, \Animations{1}{3}). Notably, this nearest connected point does not need to be its birth pair and may be quite a long distance away, as we often observe in the arrested regime. In addition to pair birth and annihilation, negative pseudo-defects can transition from one to the other while conserving the pseudo-charge (\cref{Fig dislocation} and its inverse process, \Animation{4}). In another pseudo-topological transition, connected branchpoints can exchange branches (\cref{Fig T1}, \Animation{5}). This process is akin to a T1 transition in foams and tissues, except that our lines are directed, and branching nodes are restricted to having one incoming wall and two outgoing walls. As a result, T2 transitions are not possible.

One can imagine other transitions between wall nodes that are permitted by the pseudo-topological rules. To the best of our understanding, such processes can be viewed as either analogues or sequences of the basic transitions shown in \cref{Fig topology}. Moreover, all processes involving startpoints are rare in arrested wall networks. 

\begin{figure*}[tbhp!]
\begin{center}
\includegraphics[scale=0.48]{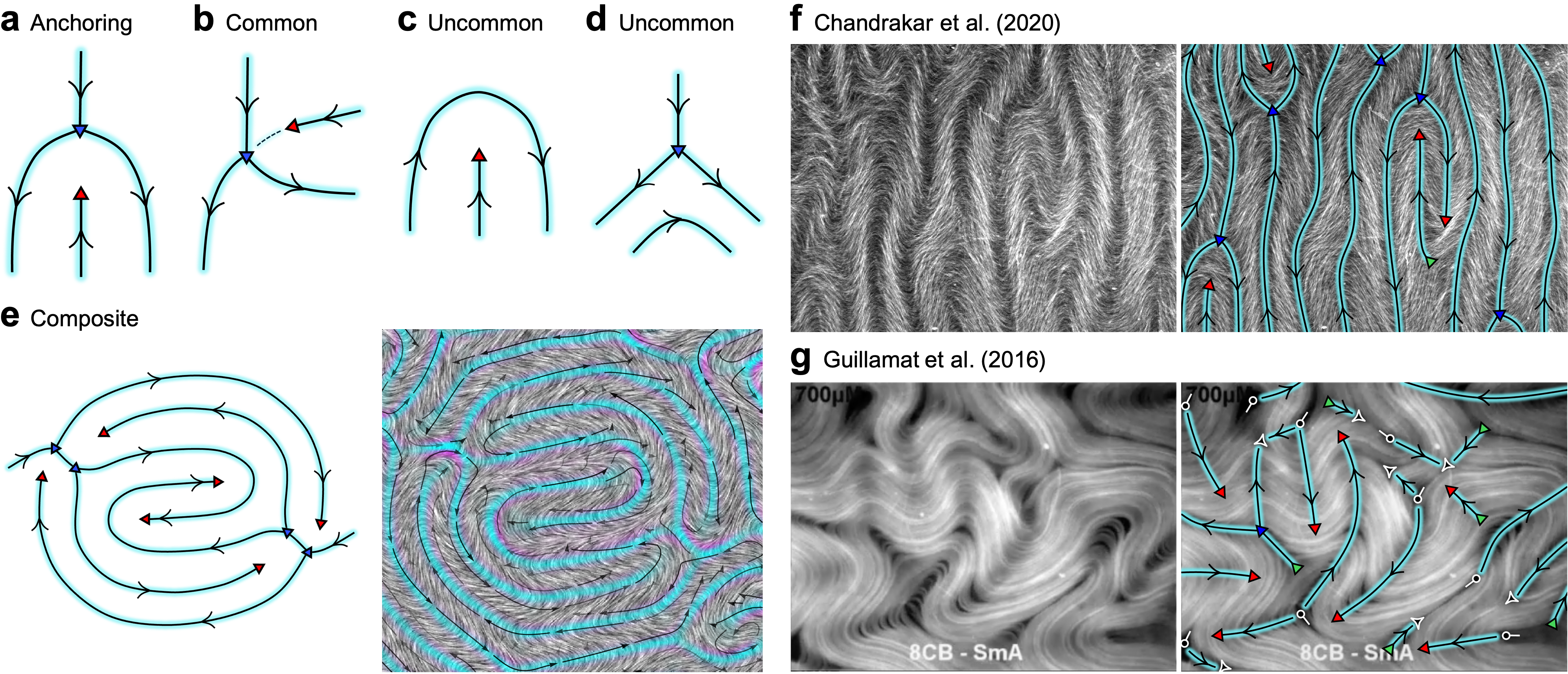}
\end{center}
  {\phantomsubcaption\label{Fig anchoring}}
  {\phantomsubcaption\label{Fig common}}
  {\phantomsubcaption\label{Fig uncommon1}}
  {\phantomsubcaption\label{Fig uncommon2}}
  {\phantomsubcaption\label{Fig Composite}}
  {\phantomsubcaption\label{Fig Chandrakar}}
  {\phantomsubcaption\label{Fig Guillamat}}
\bfcaption{Motifs of arrested wall networks and experimental snapshots of microtubule-kinesin active nematics}{ In all diagrams, lines, nodes and colors are as defined in \cref{Fig topology}. \subref*{Fig anchoring}--\subref*{Fig uncommon2}, Basic network motifs. The anchoring motif (\subref*{Fig anchoring}) is made of an endpoint and branchpoint that meet head-on, with the endpoint trapped between the two outgoing walls of the branchpoint. In motif (\subref*{Fig common}) the endpoint meets the branchpoint from one of its sides, i.e. between the incoming wall and an outgoing wall. The dashed line traces a weak distortion, indicating that the wall associated with the endpoint tends to align its direction with the outgoing wall on the opposite side of the branchpoint. The motifs (\subref*{Fig uncommon1},\subref*{Fig uncommon2}) involve a single pseudo-defect interacting with a bare wall. These, along with (\subref*{Fig common}), do not follow the tendency to have strictly antiparallel walls. \subref*{Fig Composite}, Composite motif schematic (left) and one formed spontaneously in a simulation (right). The stream plot on the right represents the flow, with black indicating maximal $|\bm{v}|$ and full transparency indicating $|\bm{v}|=0$. The gray background is the line-integral-convolution representation of the nematic director $\bm{n}$. Parameter values and color legend for splay and bend distortions are as in \cref{Fig splay-bend-arrested}. \subref*{Fig Chandrakar}--\subref*{Fig Guillamat}, Raw fluorescence images from experiments (left panels) and overlaid schematic drawings (right panels) depicting domain walls, pseudo-defects and actual $\pm1/2$ defects in white. \subref*{Fig Chandrakar}, Taken from a movie in \cite{Chandrakar2020} (courtesy of Guillaume Duclos), which shows the evolution of the microtubule-based nematic following the bending instability of the aligned state. \subref*{Fig Guillamat}, Taken from a movie in \cite{Guillamat2016} (courtesy of Pau Guillamat), which shows a turbulent transient with all types of pseudo-defects and actual nematic defects. Note how walls may also originate from true $+1/2$ defects and be absorbed by true $-1/2$ defects.}
\label{Fig motifs}
\end{figure*}

\subsection{Network motifs} 

In the arrested turbulence regime (extensile rods with $S\nu<0$), the processes of wall folding, branching, extension, and arrest produce a directed tree-like network of self-avoiding walls (\cref{Fig walls}). This network does not form closed loops in a simply-connected domain because such a loop would enclose a topological charge of 1, implying actual nematic defects, which our system forbids.

We find that endpoints propagate rapidly in the direction of their polarity, until they are blocked by other walls. Startpoints, on the other hand, move quickly in the opposite direction of their polarity until their wall dissolves or they merge with another wall, transitioning into a branchpoint. Unlike endpoints and startpoints, branchpoints do not move in the direction parallel or antiparallel to their polarity. Rather, when they are created (\cref{Fig branching}), they tend to deflect only slightly in the direction of the newly formed branch (\cref{Fig wall-branching}). Hence, in the sense of their mobility, endpoints are akin to actual $+1/2$ nematic defects and branchpoints are similar to $-1/2$ defects, which are not mobile. Despite these similarities, we emphasize that actual defects generate fundamentally different dynamics, likely due to fewer constraints on their mobility and the transitions they can undergo. This contrast becomes evident in the final section, where we perform full Q-tensor simulations that permit the nucleation of true topological defects.

Ultimately, the pseudo-defects---primarily endpoints and branchpoints (in arrested states)---maintain a steady distance from neighbouring walls and organize into motifs like those shown in \crefrange{Fig anchoring}{Fig uncommon2}. A frequent motif, which we call the anchoring motif (\cref{Fig anchoring}), is the one discussed previously in relation to the stagnation points in \cref{Fig splay-bend-arrested}. We postulate that this structure might be particularly stable because neighbouring walls are anti-parallel, as promoted by the spontaneous-flow instability. In contrast, the other basic motifs (\crefrange{Fig common}{Fig uncommon2}) feature some neighbouring walls that are parallel. For large system sizes, a variety of composite structures combining all types of motifs can emerge and persist for long times (see \Movie{2}). One such structure is shown in \cref{Fig Composite}.

We identified bending domain walls, pseudo-defects and our motifs in images and videos of microtubule-based active nematics \cite{DeCamp2015,Guillamat2016,Guillamat2017,Chandrakar2020,Lemma2021,Ardavseva2024} (\crefrange{Fig Chandrakar}{Fig Guillamat}). In \cref{Fig Chandrakar}, the spontaneous-flow instability leads to the growth of bending domain walls and the emergence of pseudo-defects. These tend to organize in interlaced anchoring motifs, which are also frequently observed in our simulations.  In \cref{Fig Guillamat}, pseudo-defects are seen to coexist with actual nematic defects (white). Since the latter may also act as origin points or termination points of walls, there may be an interesting interplay between both types of structures in organizing the dynamics of the system. We take initial steps toward exploring this interaction in the final section, where we analyze the full Q-tensor model. We further note that a very recent study \cite{Ardavseva2024} observed that a series of folding events generates an extensive network of bending domain walls, where both types of nodes (pseudo-defects and true defects) are clearly present. Importantly, in these experimental examples, the system does not maintain an arrested state. We expand on this aspect further in the final discussion.

\begin{figure}[tbp!]
\begin{center}
\includegraphics[scale=0.48]{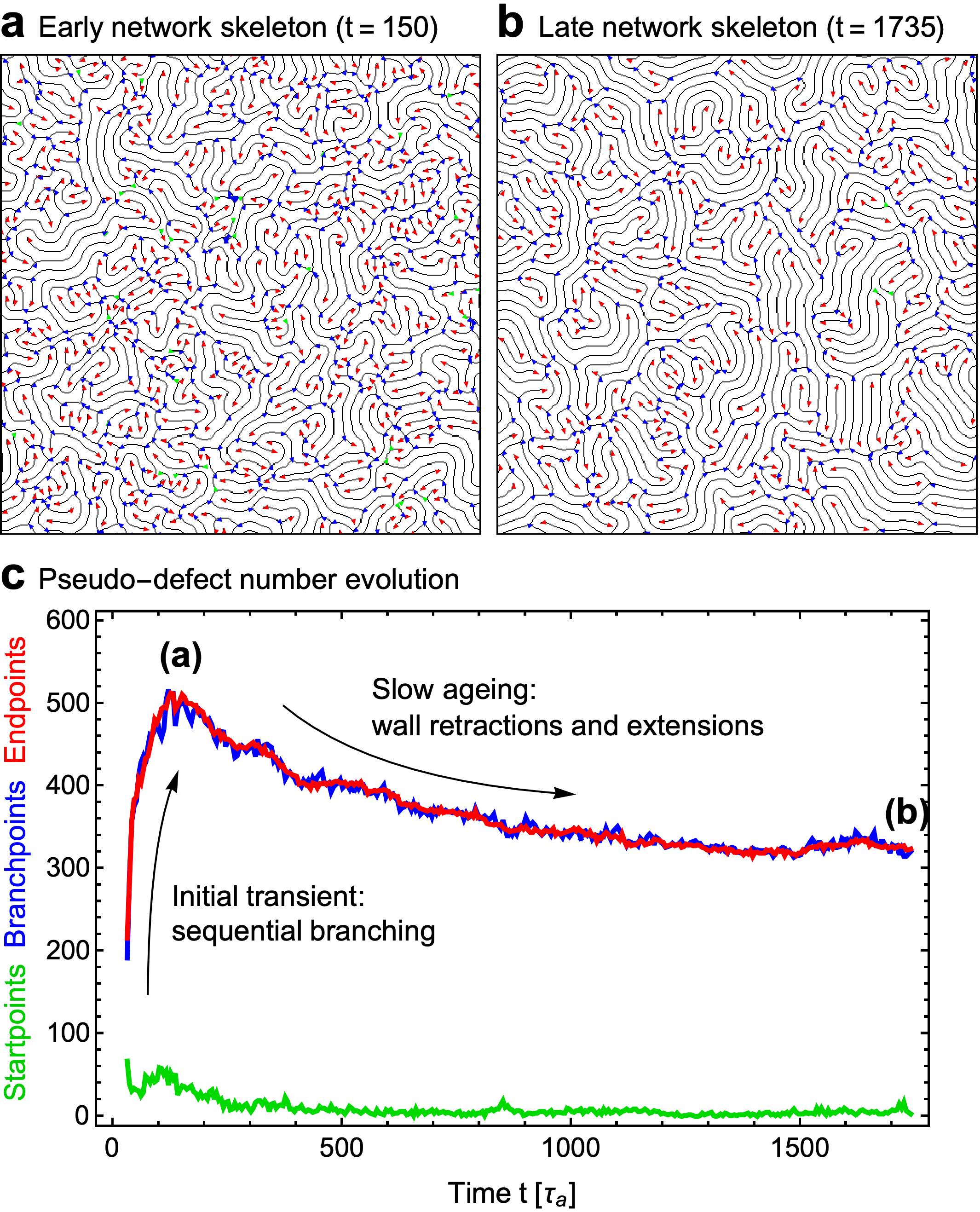}
\end{center}
  {\phantomsubcaption\label{Fig initial}}
  {\phantomsubcaption\label{Fig final}}
  {\phantomsubcaption\label{Fig branchpoint-evolution}}
\bfcaption{Ageing of the arrested wall network}{ \subref*{Fig initial},\subref*{Fig final}, Skeleton of the domain walls (black) with startpoints, branchpoints and endpoints (green, blue and red triangular nodes) at an early time (\subref*{Fig initial}) and a late time (\subref*{Fig final}). The detection of the network skeleton and its nodes is described in \cref{image processing}, \cref{Fig node detection}. \subref*{Fig branchpoint-evolution}, Evolution of the number of startpoints (green), branchpoints (blue) and endpoints (red). In the initial transient, sequential `zig-zag' instabilities result in the proliferation of both branchpoints and endpoints. Once the wall pattern establishes a wavelength, the system ages slowly as some endpoints retract and annihilate with their connected branchpoint, while others extend (\Movie{5}). Throughout the simulation, there are frequent transitions between branchpoints and startpoints, though the number of startpoints remains low. Additionally, the detection algorithm is not perfect, occasionally misidentifying endpoints or branchpoints as startpoints and vice versa. Parameter values were set to $R=1$, $S=1$, $\nu=-0.9$, and $A=3.2\times 10^5$.}
\label{Fig aging}
\end{figure}

\subsection{Ageing-like dynamics}

The dynamics of active nematics are not relaxational, and the arrested patterns emerging in the extensile case co-exist with a background of weak large-scale chaotic flows. Nevertheless, these dynamics share interesting similarities with relaxational systems that exhibit frustration. In particular, trapping motifs (\crefrange{Fig anchoring}{Fig common}) might promote arrest similarly to how cages and locally-favoured structures promote arrest in gels and glasses \cite{Li2020c,Royall2008,Royall2015a}. After an initial transient in which sequential branching establishes the network, we find that the wall network undergoes a slow relaxation that is reminiscent of the ageing of gels and glasses.
The background turbulent-like flows produce small low-$q$ fluctuations that slowly evolve the pattern. This includes jittering the walls and inducing some pseudo-topological transitions in the network; primarily those depicted in \cref{Fig branching,Fig dislocation,Fig T1} and their inverse processes. Crucially, these represent a highly restrictive subset of transitions compared to the full range of processes enabled when startpoints are more prominently present. Through this fluctuating process, we find that the number of pseudo-defects gradually decreases (\cref{Fig aging}, \Movie{5}), mostly via annihilation of connected branchpoint-endpoint pairs. As some walls shrink and disappear from the network, others extend, thereby maintaining the selected wavelength. At long times ($\sim 10^3\,\tau_\text{a}$), the evolution slows down and the number of pseudo-defects seems to approach a steady state (\cref{Fig branchpoint-evolution}). This relaxation occurs on timescales of hundreds of $\tau_\text{a}$ and appears to be a specific feature of arrested wall networks. It is absent in the contractile aligning regime, and in the presence of true topological defects---as shown in the final section.

\begin{figure*}[tbhp!]
\begin{center}
\includegraphics[scale=0.48]{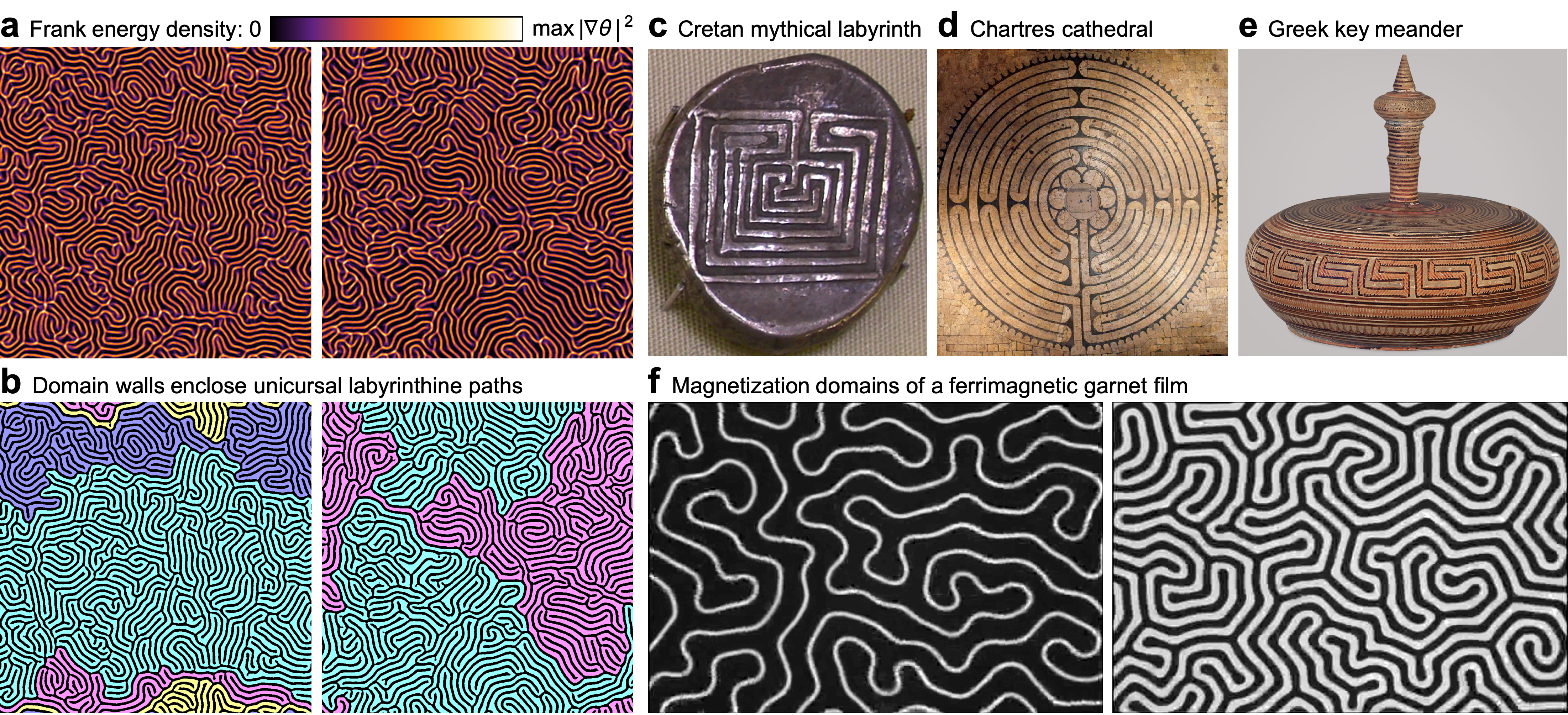}
\end{center}
  {\phantomsubcaption\label{Fig wall-labyrinths}}
  {\phantomsubcaption\label{Fig unicursal}}
  {\phantomsubcaption\label{Fig Crete}}
  {\phantomsubcaption\label{Fig Chartres}}
  {\phantomsubcaption\label{Fig Meander}}
  {\phantomsubcaption\label{Fig Garnet}}
\bfcaption{Unicursal labyrinths}{ \subref*{Fig wall-labyrinths}, Two examples of arrested tree-like patterns formed by domain walls, with $S\nu=-0.9$ (left) and $S\nu=-1.1$ (right). \subref*{Fig unicursal}, Processed negatives of (\subref*{Fig wall-labyrinths}) showing how domain walls (black) enclose a small and even number of unicursal labyrinths (4 connected regions on the left, 2 on the right) that span the entire system (\cref{image processing}, \cref{Fig labyrinth detection}). The walls separating different labyrinths are self-connected across the periodic boundaries, unlike all other branches which feature endpoints. \subref*{Fig Crete}, The mythical labyrinth of Crete, shown on a silver coin from around 400 BC (by AlMare, reproduced from \href{https://commons.wikimedia.org/w/index.php?curid=5111302}{Wikimedia Commons}, creative commons license \href{https://creativecommons.org/licenses/by-sa/3.0/deed.en}{CC BY-SA 3.0}). \subref*{Fig Chartres}, The labyrinth in the Chartres cathedral (courtesy of Père Emmanuel Blondeau and the secretariat de la Cathédrale Notre-Dame de Chartres). \subref*{Fig Meander}, Meander pattern known as the Greek key encircling a box (pyxis) from around 850 BC (reproduced from the \href{https://www.metmuseum.org/art/collection/search/254598}{Metropolitan Museum of Art}). \subref*{Fig Garnet}, Out-of-plain magnetization domains of a ferrimagnetic garnet film, with black and white regions indicating opposite magnetizations. The amplitude of an external magnetic field favouring the black domain is gradually reduced from left to right, resulting in growth by folding of the white unicursal domain. Frames are desaturated snapshots from a video courtesy of Pietro Tierno.}
\label{Fig labyrinths}
\end{figure*}

\subsection{Unicursal labyrinths}

When there are no startpoints, which is typical in the extensile aligning regime, the \emph{gaps between the walls} form unicursal labyrinths of a type known as meanders, defined by having a single path with neither bifurcations nor dead ends. A typical wall network encloses a small and even number of such labyrinths, with each one closing on itself through the periodic boundaries, i.e., through the hole of the torus, which is not a simply-connected domain (\cref{Fig wall-labyrinths,Fig unicursal}). On the other hand, the wall network bounding these unicursal paths forms a maze, defined as having multiple bifurcations and dead ends, corresponding to our branchpoints and endpoints. The specific walls separating the \emph{different} unicursal labyrinths are initially generated by the spontaneous-flow instability. There are as many of these \emph{primary walls} as there are unicursal labyrinths, and they also close on themselves through the periodic boundaries. Since primary walls are the only paths on the network that avoid dead ends, they can be thought of as the solutions to the maze. 

Paintings, rock-carvings, ornaments and monuments depicting unicursal labyrinths predate recorded history and have remained common in artistic and spiritual designs \cite{Billock2016}. Renowned examples include the Cretan labyrinth of Greek mythology (\cref{Fig Crete}), and the labyrinth decorating the floor of Chartres cathedral (\cref{Fig Chartres}). However, unlike ours, these and most other Labyrinth designs feature a dead end in the center. On the other hand, meanders (\cref{Fig Meander}) may close on themselves through the hole of the torus. In pattern-forming systems, unicursal labyrinths are rare. They were obtained under special conditions in simulations of a modified Swift-Hohenberg equation \cite{Bordeu2015}, and in experiments with ferrimagnetic garnet films in specific experimental protocols \cite{Seul1991,Seul1992,Reimann2002} (\cref{Fig Garnet}). In our case, they result from an emerging topology of domain wall networks and they occur spontaneously and generically in the arrested regime.

\begin{figure*}[tbhp!]
\begin{center}
\includegraphics[scale=0.47]{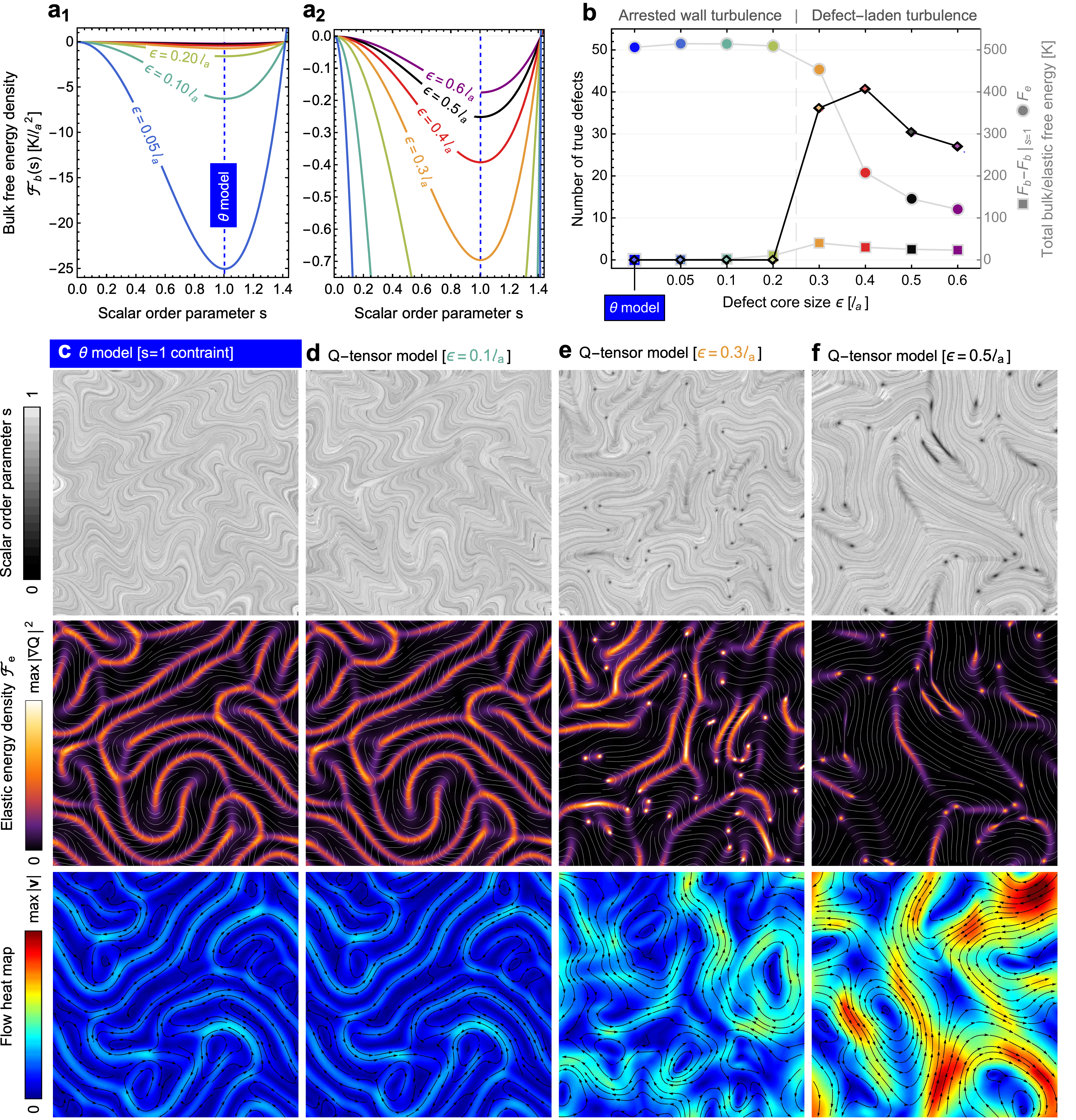}
\end{center}
  {\phantomsubcaption\label{Fig Fb(s)}}
  {\phantomsubcaption\label{Fig no defects}}
  {\phantomsubcaption\label{Fig Theta model}}
  {\phantomsubcaption\label{Fig eps 0.1}}
  {\phantomsubcaption\label{Fig eps 0.3}}
  {\phantomsubcaption\label{Fig eps 0.5}}
\bfcaption{Persistence of dynamical arrest and the transition to defect-laden turbulence}{ {\subref*{Fig Fb(s)}, Plots of the bulk free energy as a function of the scalar order parameter $s$ for the values of $\epsilon$ explored in our simulations. Dashed blue line represents the $\theta$ model, for which $s=1$ is enforced. The energy is shown in units of $K/\ell_\text{a}^2$, characterizing elastic distortions. Note the difference in magnitude between panel (\subref*{Fig Fb(s)}$_\textbf{1}$), highlighting cases for which defects do no not form, and panel (\subref*{Fig Fb(s)}$_\textbf{2}$), for which they do. \subref*{Fig no defects}, Statistical means of the total defect number (diamond markers, black line), the total bulk free energy (square markers, gray line), and total elastic free energy (circle markers, gray line), as a function of $\epsilon$. These  temporal averages characterize the fully developed turbulent state in our simulations (discarding initial transients). For timeseries and more detailed statistical breakdown see \cref{Fig timeseries}. \subref*{Fig Theta model}, Typical snapshot from a simulation of the $\theta$ model, showing the line integral convolution representation of the director field $\bm{n}$ (top), the corresponding elastic free energy (center), and the flow heatmap with black streamlines (bottom). \subref*{Fig eps 0.1}--\subref*{Fig eps 0.5}, Same as (\subref*{Fig Theta model}) for the Q-tensor model, at increasing values of $\epsilon$. Note that the top panels in (\subref*{Fig eps 0.1}--\subref*{Fig eps 0.5}) show both $s$ and $\bm{n}$, computed as the principal eigenpair of $\mathbf{Q}$. True topological defects are seen as black spots where $s\to0$. In all simulations depicted here, we fixed $R=1$, $S=1$, $A=10000$, $\nu=-1$ (the extensile rod-aligning regime), and set the same initial condition: a quiescent nematic state ($s=1$) with a small angular perturbation. See also \Movies{6}{7}.}}
\label{Fig Q-tensor}
\end{figure*}

\section{Unconstrained Q-tensor model}
To place our results in a broader context, we now turn to the unconstrained active Q-tensor model, based on the Landau-de Gennes framework, which allows for the nucleation of true $\pm 1/2$ defect pairs. This enables us to (i) delineate the regime in which our defect-free findings remain applicable, and (ii) explore the transition to the more familiar defect-laden behavior that arises when those conditions are no longer met. As in many previous studies, e.g. \cite{Marenduzzo2007,Thampi2013,Doostmohammadi2016,Hemingway2016a,rorai2021active,rorai2022coexistence,head2024spontaneous}, we use the governing equations based on the Beris-Edwards model \cite{beris1994thermodynamics} for dimension $d=2$ and $\text{Re}=0$ with the addition of the active nematic stress. 
These equations, expressed in our notation, together with details of the numerical scheme used to evolve $\mathbf{Q}$, are provided in \cref{Qmodel}. For our discussion here, we define the nematic tensor in terms of the scalar order parameter $s$, quantifying local nematic alignment, and a unit director $\bm{n}$, as 
\[
Q_{\alpha\beta} = s \left(n_\alpha n_\beta - \frac{1}{2} \delta_{\alpha\beta}\right).
\]

With respect to our constrained director-based formulation presented earlier, the key difference lies in the system’s free energy:
\begin{equation}
    F=\int \left(\mathcal{F}_\text{e}+\mathcal{F}_\text{b}  \right) \dd^2\bm{r},
\end{equation}
where the elastic and bulk free energy densities are
\begin{align}
\mathcal{F}_\text{e}
&=\frac{K}{4} (\partial_\gamma Q_{\alpha\beta}) (\partial_\gamma Q_{\alpha\beta}), \nonumber \\
 \mathcal{F}_\text{b}
&=\frac{K}{4}\frac{1}{\epsilon^2}\left( -s_0^2 Q_{\alpha\beta}Q_{\alpha\beta}+(Q_{\alpha\beta}Q_{\alpha\beta})^2\right)
\nonumber \\
& =\frac{K}{16}\frac{1}{\epsilon^2}s^2\left(-2s_0^2+s^2\right). \nonumber
\end{align}

We note that the one-constant elastic free energy density $\mathcal{F}_\text{e}$ extends the Frank energy considered in \cref{eq Frank}. Now, rather than enforcing perfect nematic alignment, the bulk free energy density $\mathcal{F}_\text{b}$ depends explicitly on the scalar order parameter $s$, which is allowed to vary in space and time. For convenience, we parameterize $\mathcal{F}_\text{b}$ in terms of $K$, the defect core size $\epsilon$, and the equilibrium scalar order $s_0$. Deep in the nematic phase, one has $s_0\simeq1$, favoring an equilibrium state with $s \simeq 1$. We thus fixed $s_0=1$ in our simulations, but allow to $\epsilon$ vary as it depends on the constituents of the system. 

Because the with of domain walls scales as $\ell_\text{a}$ (the active length scale), the cost of elastic distortions per unit area at the walls scales as $K/\ell_\text{a}^2$. On the other hand, the depth of the bulk potential is controlled by the factor $K/\epsilon^2$ (see \cref{Fig Fb(s)}). Hence, a key control parameter that arises is the ratio $\epsilon^2/\ell_\text{a}^2$. While the activity number $A = L^2/\ell_\text{a}^2$ measures the strength of active forcing relative to system size, it is the ratio $\epsilon^2/\ell_\text{a}^2$ that governs whether domain-walls remain stable or give way to defect unbinding. We note that the regime $\epsilon\geq\ell_\text{a}$ leads to very different behavior beyond the scope of this work.

To quantitatively compare Q-tensor simulations with the constrained case, we re-derived the $\theta$ model directly from the Q-tensor framework under the constraint $s^2 = 1$ (\cref{Q theta model}). The resulting equations are identical to those derived from the director-based formulation under $\bm{n}^2=1$, except that the Lagrange multiplier enforcing the constraint is independent of the flow field---a consequence of the orthogonality of flow alignment in the standard Q-tensor model. We found that this difference does not alter any of the qualitative defect-free behaviors reported thus far.

\subsection{Arrested turbulence at finite defect core size}

In the regime $\epsilon/\ell_\text{a} \ll 1$, defect cores are narrow compared to the active forcing scale, so defects (wherein $s \to 0$) carry a relatively high free energy cost (\cref{Fig Fb(s)}$_1$). As a result, domain walls persist without provoking the nucleation of defects (\cref{Fig no defects}), and the nematic forms arrested wall networks---just as in our constrained $\theta$ model (\cref{Fig Theta model,Fig eps 0.1}, \Movie{6}). The persistence of this behavior at finite $\epsilon$ supports the view that dynamical arrest is a generic feature of active nematics, rather than a behavior limited to the constrained model. Beyond qualitative agreement, we find expected quantitative convergence to the $\theta$ model as $\epsilon \to 0$ (\cref{Q agreement}, \cref{Fig agreement}, \Movie{6}). Note that Q-tensor simulations become increasingly stiff and computationally expensive in this limit, making the $\theta$ model a more efficient and practical alternative.

\subsection{Transition to defect-laden dynamics}

When the ratio $\epsilon/\ell_\text{a}$ exceeds a critical threshold (approximately 0.25 for fixed $S\nu = -1$), the potential well favoring nematic alignment becomes shallow compared to the elastic energy cost of orientational distortions, whose peak scales as $K/\ell_\text{a}^2$ (\cref{Fig Fb(s)}$_2$). As a result, $\pm 1/2$ defects begin to nucleate (\cref{Fig no defects}), disrupting the formation of grid-locked wall patterns and giving rise to a more disordered, vortex-dominated turbulent state (\cref{Fig eps 0.3,Fig eps 0.5}, \Movie{7}). In the transitional regime, we find that a modest increase in bulk free energy due to defect cores enables a dramatic reduction in the stored elastic free energy (\cref{Fig no defects,Fig timeseries}). This effect arises because defects, despite their intrinsic energetic cost, efficiently dissolve the high-energy walls that make up the network. As a consequence of the walls being less robust, the system no longer exhibits slow relaxation dynamics: \cref{Fig defects}$_1$ shows that the number of true defects quickly reaches a statistically stationary state (albeit with strong fluctuations), in contrast to the slow relaxation of pseudo-defects in the defect-free case (\cref{Fig branchpoint-evolution}).

\section{Unraveling Arrest: Interplay of Domain Wall Networks and True Defects}
The branched domain-wall networks that emerge in the extensile, rod-aligning regime (with defects suppressed) exhibit pronounced local peaks of elastic free energy. These distortion hotspots consistently arise near branchpoint pseudo-defects, particularly along the incoming wall that splits into two outgoing branches (see, e.g., the final panels of \cref{Fig wall-arrest}). Viewing defect unbinding as a mechanism for efficiently lowering the system’s free energy, we hypothesized that these hotspots would be predisposed to serve as defect nucleation sites upon quenching the defect core size.

The simulation shown in \Movie{8} supports this hypothesis. Beginning from an arrested wall network generated at low $\epsilon/\ell_\text{a}$ (specifically, $\epsilon=0.2\,\ell_\text{a}$), we quenched the defect core size to above the critical threshold ($\epsilon=0.35\,\ell_\text{a}$). Subsequent defect unbinding events were consistently initiated near branchpoint pseudo-defects—though not all at once. Some branchpoints remain intact for some time, but these too are ultimately dissolved. Once unbound, the $+1/2$ defects are seen to travel along the bend walls on which they formed, effectively “unzipping” them as they move, as observed in previous computational studies \cite{Thampi2014a,Giomi2014a,Thampi2014b}. Upon reaching an end-point pseudo-defect (at the terminus of a branch), the motility of $+1/2$ defects can appear somewhat suppressed by the surrounding network geometry. Eventually, each $+1/2$ defect annihilates with a nearby $-1/2$ defect, creating aligned regions into which existing walls can expand or new walls can form. 
Near the critical threshold of $\epsilon/\ell_\text{a}$, this interchange continues in a highly dynamic fashion. Wall segments and pseudo-defects coexist with a fluctuating population of unbound topological defects, resulting in a hybrid state that combines features of wall networks and defect-mediated turbulence (\Movies{7}{8}). 

\section{Discussion}

In this work, we showed that flow alignment---the tendency of liquid crystals to reorient under shear---dramatically influences the spatiotemporal structure of defect-free active nematic turbulence (\cref{Fig regimes}). For contractile flow-aligning active nematics, the chaos is characterized by fragmented nematic patterns, reduced total Frank free energy, and strong self-similar flows across a broad range of scales. We associate this strong large-scale turbulence to the underlying metastability of uniform and wall-laden states \cite{Lavi2024}. On the other hand, despite their chaotic dynamics, we find that extensile aligning active nematics self-organize into an arrested tree-like pattern of nematic domain walls, with the flow strongly channeled along them (\cref{Fig velocity-arrested,Fig Frank-arrested}). This finding reveals a mechanism of pattern formation mediated by active turbulence, complementing those found in other studies \cite{James2018,DeWit2024,Xu2024}. In particular, our patterns enclose unicursal labyrinths (\cref{Fig labyrinths}), which do not appear spontaneously in other pattern-forming systems. 

The labyrinths in \cref{Fig unicursal} span the entire system, but how far can they possibly extend? Further increasing the system size widens the range of chaotic self-similar flows, identified by the $q^{-1}$ scaling of the velocity power spectrum (\cref{Fig velocity-spectra}, blue). These flows become stronger at larger scales, and hence they could potentially disrupt the arrested wall pattern for sufficiently large systems. At the scale where chaotic large-scale flows become as strong as the wall streams, which sustain the domain walls, the tree-like network could lose its connectivity and fluctuate more strongly. 

To test this idea, we performed defect-free simulations at even larger system sizes (Figs.~S3 and S4\cite{SM}, \Movies{9}{10}). Since these simulations required substantially greater computational effort, we used the numerically advantageous $\theta$ model derived from the Q-tensor formulation (\cref{Q theta model}). In the extensile, rod-aligning regime ($S\nu = -1.1$), we reached a system size at which the magnitude of chaotic, system-scale flows equaled that of the flows associated with the peak at the characteristic $q$ (Fig.~S3\cite{SM}). As anticipated, a widespread wall network develops with a selected wavelength (Fig.~S4\cite{SM}), but frequent fracturing events prevent it from persistently spanning the entire system (\Movie{9}). By contrast, in the contractile aligning case ($S\nu = +1.1$), dynamical arrest was not at play to begin with, so increasing the system size did not lead to qualitative changes (\Movie{10}).

The arrested state does not show long-range correlations in the nematic order parameter (\cref{Fig Q spectrum}, blue). This is reminiscent of some passive systems, such as spin ice, that feature frustration in the absence of long-range geometrical correlations\cite{wang2006artificial}. Similarly, our pattern features short-range disorder and is yet connected over long distances. Therefore, we propose that the unicursal labyrinths exhibit a form of long-range topological order stemming from emergent connectivity rules associated with the folding, branching, extension and self-avoidance of the domain walls. 

In the fully-developed turbulent state, both the arrested and non-arrested regimes manifest a selected wavelength. This wavelength is significantly greater than the active length $\ell_\text{a}$ (\cref{Fig splay-bend-arrested}), and it also exceeds the critical wavelength of the spontaneous-flow instability. The wavelength selection mechanism is inherently nonlinear and two-dimensional; it is based on a balance between the coarsening of local striped patterns, which tends to enlarge the characteristic wavelength (\cref{Fig wall-coarsening}), and the bending and folding of walls that tends to reduce it (\cref{Fig wall-branching}). A deeper understanding of the selection mechanism and its dependence on parameters remains an open question.

Whereas defect-free flows highlight new and surprising phenomena, most experiments and simulations in active nematics have been conducted in defect-laden regimes. Yet, the physics of the two regimes may intersect---as shown in our Q-tensor simulations (\cref{Fig eps 0.3}, \Movie{7}). This crossover somewhat resembles the one found in polar fluids \cite{Chatterjee2021,Andersen2023}. Actual nematic defects tend to nucleate along nematic domain walls, thereby dissolving them \cite{Thampi2014a,Thampi2014b,Thampi2016a,Doostmohammadi2018} (\Movies{7}{8}). This can be theoretically traced to the explicit form of the molecular field, $\mathbf{H}=-\delta F/\delta\mathbf{Q}+\tfrac{1}{2}\text{tr}\left[\delta F/\delta\mathbf{Q}\right]\mathbf{I}$. As shown in \cref{Q nemato} (Remark \ref{remark H}), $\mathbf{H}$ has a component orthogonal to $\mathbf{Q}$, which drives rotations, and a component parallel to $\mathbf{Q}$ that drives variations in the scalar order parameter $s$. The latter is given by
\[
\mathbf{H}_\parallel=\frac{K}{2}\left(\frac{1}{\epsilon^2} (1-s^2)+\frac{\nabla^2 s}{s}-4(\nabla\theta)^2 \right)\mathbf{Q},
\]
where $\mathbf{Q}:=s\left(\bm{n}\bm{n}-\tfrac{1}{2}\mathbf{I}\right)$ and $\bm{n}:=\left(\cos\theta,\,\sin\theta\right)$. The first term in parenthesis, stemming from the bulk free energy, restores $s$ toward unity (the nematic phase). The second term, from the elastic energy, relaxes gradients in $s$, giving defects a finite core size. Crucially, the last term, also elastic, tends to \emph{reduce} $s$ in regions of strong orientational gradients (e.g., sharp domain walls)—highlighting a clear link between elastic distortions and the potential production of defects.

Additionally, an aspect not considered in our current modeling is that nematic constituents such as microtubules are observed to be depleted at defects (\cref{Fig Guillamat}), which likely lowers their energy cost and favors their nucleation\cite{assante2023active}. The frequent dissolution of domain walls by defects explains why, in the instances where our wall network motifs can be seen in experiments (\crefrange{Fig Chandrakar}{Fig Guillamat}), they only have a short life time. We therefore argue that, in addition to their well-known role as \emph{drivers} of turbulence by generating vorticity \cite{Giomi2015}, defects also play an important role as \emph{enablers} of stronger turbulent flows by breaking the potential grid-lock, i.e., preventing dynamical arrest (\cref{Fig Q-tensor}, \Movies{7}{8}).

Validating our predictions for defect-free active nematics presents an experimental challenge. By examining existing experimental images, we have already observed some of the phenomena, including the emergence of the three distinct pseudo-defects and the formation of large labyrinth-like patterns by domain walls\cite{Ardavseva2024}. However, dynamical arrest---our primary prediction---has not yet been realized. This would require searching for active materials for which the nucleation of defect pairs were significantly more costly than in the experimental realizations studied thus far. As demonstrated in our Q-tensor simulations, defect-free active turbulence should be achievable for sufficiently large system sizes $L$, ensuring large dimensionless activity $A=(L/\ell_\text{a})^2 \gg 1$, while keeping the defect-core size $\epsilon$ small enough ($\ell_\text{a} \gg \epsilon$) to preclude defect nucleation. 

\section{Conclusion}

Our study of active nematic turbulence highlights the crucial role of topology beyond the existence of topological defects. We argue that the connectivity of domain walls, the existence of pseudo-defects, and their interactions are key to understanding the dynamics of active nematics, from strong large-scale turbulence to dynamical arrest. Furthermore, in the presence of true topological defects, their interplay with pseudo-defects adds another intriguing layer of complexity, as our Q-tensor simulations illustrate. Our results open new avenues for experimental analysis and design, particularly in tracing the intricate web of domain walls, their orientation, and their nodes. They also raise a series of challenging theoretical questions, such as understanding the properties of pseudo-defects, the nonlinear wavelength selection mechanism, and the nature of the long-range topological order of arrested states. 

\section*{Acknowledgments}

Funding from the Spanish Ministry for Science and Innovation MICCINN/FEDER (PID2022-137713NB-C22 to IL and JC) and the Generalitat de Catalunya (AGAUR SGR-2021-00450 to IL and JC and ICREA Academia award to JC) is acknowledged. IL acknowledges support from the Flatiron Instutute, part of the Simons Foundation. RA thanks the Isaac Newton Institute for Mathematical Sciences for support and hospitality during the Programme \textit{New Statistical Physics in Living Matter}, supported by EPSRC Grant Number EP/R014604/1, when work on this paper was undertaken. 

\appendix
\renewcommand{\appendixname}{APPENDIX}

\section{NUMERICAL SCHEME} 
\label{numerical}
Here we describe our method for numerically integrating \crefrange{eq vorticity}{eq theta}. In what follows, the index $n$ is the time iteration,  $dt$ is the timestep, and $h=N^{-1}$ is the vertex spacing on our $N\times N$ square grid.

\subsection{Solving momentum balance (pseudo-spectral method)}
In each time iteration, we first solve for the streamfunction based on $\theta^n$, the array representing the current director angle on the vertices. After rearranging terms and substituting $\hat{q}_{\alpha\beta}(\theta)$ and $\bar{q}_{\alpha\beta}(\theta)$ in \cref{eq vorticity}, one has
\begin{align}
    \nabla^4\psi^n &+ G(\theta^n,\psi^n)= S \big(d_1 \sin 2\theta^n +d_2\cos2\theta^n\big)\nonumber \\
    &-\frac{R}{A}\Big(\frac{1}{2}\nabla^4\theta^n+(\partial_x\nabla^2\theta^n)\partial_y\theta^n -(\partial_y\nabla^2\theta^n)\partial_x\theta^n \nonumber \\
    & +\nu\left(d_1 \cos 2\theta^n \nabla^2 \theta^n - d_2 \sin 2\theta^n \nabla^2\theta^n\right)\Big), \label{eq vorticity num} 
\end{align}
where we defined $d_1:=\frac{1}{2}(\partial_y^2-\partial_x^2)$, $d_2:=\partial_y\partial_x$ and
\begin{equation}
    G(\theta,\psi):=R\nu\Big(d_1 \sin 2\theta h_\parallel(\theta,\psi) + d_2\cos 2\theta h_\parallel(\theta,\psi)\Big), \nonumber
\end{equation}
with
\begin{equation}
    h_\parallel(\theta,\psi)= \nu\left(\sin2\theta d_1\psi+ \cos 2\theta d_2\psi\right). \nonumber
\end{equation}

To find the solution $\psi^n$ we employ a pseudo-spectral method, with all derivatives calculated in the space of the two-dimensional discrete Fourier transform (DFT) and all nonlinear operations performed in real-space. Before each real-space multiplication, we apply the 2/3 de-aliasing rule, truncating the Fourier modes with $|q_x|$ or $|q_y|$ greater than $\frac{2}{3} \pi N$. 

We begin by computing the right-hand side (RHS) of \cref{eq vorticity num}, which is independent of $\psi^n$. Next, we note that $G(\theta^{n},\psi^n)$, the second term on the left-hand side (LHS) of \cref{eq vorticity num}, consists of non-constant coefficients combined with high-order derivatives. This defines a dense matrix in Fourier space that also varies in time. The high computation cost associated with building and inverting such a matrix at every time step renders direct inversion impractical. We therefore treat this term explicitly in a fixed-point iteration scheme, where we aim to compute a converging sequence $\left\{\psi^{n_k} \right\}_k$ (with $k=0,1,2,...$). Given an explicit approximation $\psi^{n_k}$, $G(\theta^n,\psi^{n_k})$ is computed pseudo-spectrally and transferred to the RHS. The subsequent iteration is then obtained by keeping the biharmonic term on the LHS implicit. By inverting the modified problem in Fourier space, the $k^\text{th}$ iteration is given by
\begin{equation}
    \mathcal{F}\left[{\psi}^{n_{k+1}}\right]=\frac{1}{q^4}\left(\mathcal{F}\left[\text{RHS}(\theta^n)\right]-\mathcal{F}\left[G(\theta^n,\psi^{n_k})\right]\right)\quad \forall q\neq 0,
\end{equation}
where $\mathcal{F}[\cdot]$ indicates the DFT and  $q=\|\bm{q}\|$. The mode $\bm{q}=0$ in the streamfunction is always truncated as it has no bearing on the flow. To aid convergence, we initiate the sequence with the solution obtained in the previous time step, i.e. $\psi^{n_0}= \psi^{n-1}$. Iterations on $k$ are performed until a convergence criterion for the residual error $\eta$ is met. Formally, we compute 
$$\eta_{k+1}=\frac{\int_\Omega\left(\psi^{n_{k+1}}-\psi^{n_k}\right)^2} {\int_\Omega (\psi^{n_k})^2}$$
and we check whether $\eta_{k+1}<\epsilon$, with $\epsilon$ a small computational tolerance (typically $10^{-8}$). While convergence is empirically tested, it is not guaranteed in the fixed-point method. With our chosen physical and numerical parameters, we find robust convergence so long as $|\nu| \lesssim 1.5$.

\subsection{Computing the flow, vorticity and flow-alignment rotations}

After obtaining a convergent streamfunction $\psi^n$, and before attending to \cref{eq theta}, we compute the following fields pseudo-spectrally:
\begin{align}
   &v_x^n = \partial_y \psi^n,\quad v_y^n=-\partial_x\psi^n,\quad \omega^n = -\nabla^2 \psi^n, \label{eq v omega c} \\
   &C^n=-\hat{q}_{\alpha\beta}(\theta^n)\partial_\alpha\partial_\beta\psi^n =\cos 2\theta^n d_1 \psi^n - \sin2\theta^n d_2 \psi^n. \label{eq v omega c} \nonumber
\end{align}
Respectively, these represent arrays of the flow components, vorticity, and a term proportional to flow alignment rotations ($C^n$), defined on the vertices of the $N\times N$ grid. 

\subsection{Evolving the angle field (finite element method)}
We chose to evolve \cref{eq theta} with the finite-element method (FEM), using the open source package FreeFem++ \cite{MR3043640}. The nodes of our triangular mesh $\mathcal{T}_h$, spanning the square domain $\Omega$, coincide with the regular grid points. With this arrangement, it is straightforward to map the data $v_\alpha^n$, $\omega^n$, and $C^n$ (\cref{eq v omega c}) to the amplitudes of linear finite-element `hat functions'. 

To ensure stability of the scheme, we first tackle the convective term in \cref{eq theta} independently via the Characteristics-Galerkin method\cite{lee1987characteristic}. Specifically, we solve the hyperbolic problem $\partial_t\theta+\bm{v}\cdot \nabla\theta =0$ with the built-in \textbf{convect} function\cite{MR3043640}: $\theta^n_\text{adv}=\text{convect}(\bm{v}^n,\,-dt,\, \theta^n)$.

Next, we formulate the time-discretized version of the rotations prescribed by \cref{eq theta},
\begin{equation}
        (\theta ^{n+1}-\theta^n_\text{adv})/dt-\frac{1}{2}\omega^n= \frac{1}{A}\nabla^2\theta^{n+1} -\nu C^n, \label{eq theta num 2}
\end{equation}
where we opted to handle the Laplacian term implicitly for enhanced stability.

To proceed with FEM, we write the weak form of \cref{eq theta num 2},
\begin{align}
    \int_\Omega \theta^{n+1}& \phi \,\dd^2\bm{r}+\frac{dt}{A} \int_\Omega \nabla \theta^{n+1}\cdot\nabla \phi\, \dd^2\bm{r}= \label{eq weak} \\
    &\int_\Omega \theta^{n}_\text{adv} \phi \,\dd^2\bm{r}+\frac{dt}{2} \int_\Omega \omega^{n} \phi\, \dd^2\bm{r}-dt\,  \nu \int_\Omega C^n \phi\,\dd^2\bm{r}. \nonumber
\end{align}
The problem consists in finding a solution $\theta^{n+1}\in H^1(\Omega)$ for any arbitrary smooth test function $\phi:\Omega\to \R$. 

The terms on the LHS of \cref{eq weak} are bilinear (implicit) while those on the RHS are linear (explicit). In our code, $\theta$ and $\phi$ are represented by the basis functions spanning the continuous \textbf{P2} finite-element space, i.e., polynomials of degree 2 defined piecewise on each element $K \in \mathcal{T}_h$. By imposing doubly-periodic boundary conditions, the boundary term coming from integration by parts vanishes. The matrix defined by the LHS is inverted using the default $\textbf{sparsesolver}$ in FreeFem++ \cite{MR3043640}.

\subsection{Adding noise}

To speed up the evolution towards fully-developed active turbulence, we introduce small-amplitude stochastic fluctuations to $\partial_t\theta$ in the spirit of previous work\cite{Alert2020a}. We use a Gaussian white noise field $\xi(\bm{r},t)$ with zero mean and uncorrelated in both space and time, i.e. $\left<\xi(\bm{r},t)\right>=0$ and $\left<\xi(\bm{r},t)\xi(\bm{r}',t')\right>=2D \delta(\bm{r}-\bm{r}')\delta(t-t')$ . 

In more detail, we define $\xi^n$ (the noise array at time $dt\, n$) in Fourier space by assigning to each mode a Gaussian random amplitude $a_{\bm{q}}$ and a random phase $\phi_{\bm{q}}$, expressed as $a_{\bm{q}} e^{i\phi_{\bm{q}}}$. The zero mode is truncated to impose a zero mean. To further counter aliasing, we also truncate the highest wave-numbers for which $|q_x|$ or $|q_y|$ are greater than $\frac{2}{3}\pi N$. After taking the real part of the inverse DFT, the resulting noise is scaled by $\frac{\sqrt{3D}}{\sqrt{dt h^2}}$. This factor ensures that the fluctuations are properly normalized, taking into account the space and time discretizations, as well as the $2/3$ filtering. The normalized noise $\xi^n$ is added to the RHS of \cref{eq theta num 2}, resulting in an explicit integral term in the weak formulation, \cref{eq weak}. 

We verified that spatio-temporal chaos persists also in the absence of noise. Additionally, we found that increasing the noise amplitude enhances the velocity fluctuations, especially at large scales, thereby expanding the $q^{-1}$ scaling region. 

\subsection{Additional notes}

\begin{itemize}
    \item The triangular FEM mesh $\mathcal{T}_h$ spans the entire periodic domain, resulting in $(N+1)\times(N+1)$ nodes. Thus, before proceeding to stage 3, we extend the real-space $N\times N$ arrays (\cref{eq v omega c} and the noise $\xi^n$) so that the added data along the $x=1$ and $y=1$ edges matches the data along the $x=0$ and $y=0$ edges, respectively.
    \item To allow the pseudo-spectral computations involved in stage 1, the data stored in the \textbf{P2} finite-element representation of $\theta^n$ is first downgraded to a \textbf{P1} representation to obtain the field values on the $N\times N$ vertices.
\end{itemize}

\subsubsection{Numerical parameters}

In the large-scale simulations (\cref{Fig regimes,Fig aging}, Fig.\ S1\cite{SM} and \hyperref[movies]{Movies 1--3, 5}) we fixed $N=256$, $dt=0.01$, and $D=0.0025^2$. In the simulations depicted in \cref{Fig walls}, \Movie{4}, we fixed $N=128$, $dt=0.02$, and $D=0$ (zero noise).

\begin{figure*}[tbhp!]
\begin{center}
\includegraphics[scale=0.48]{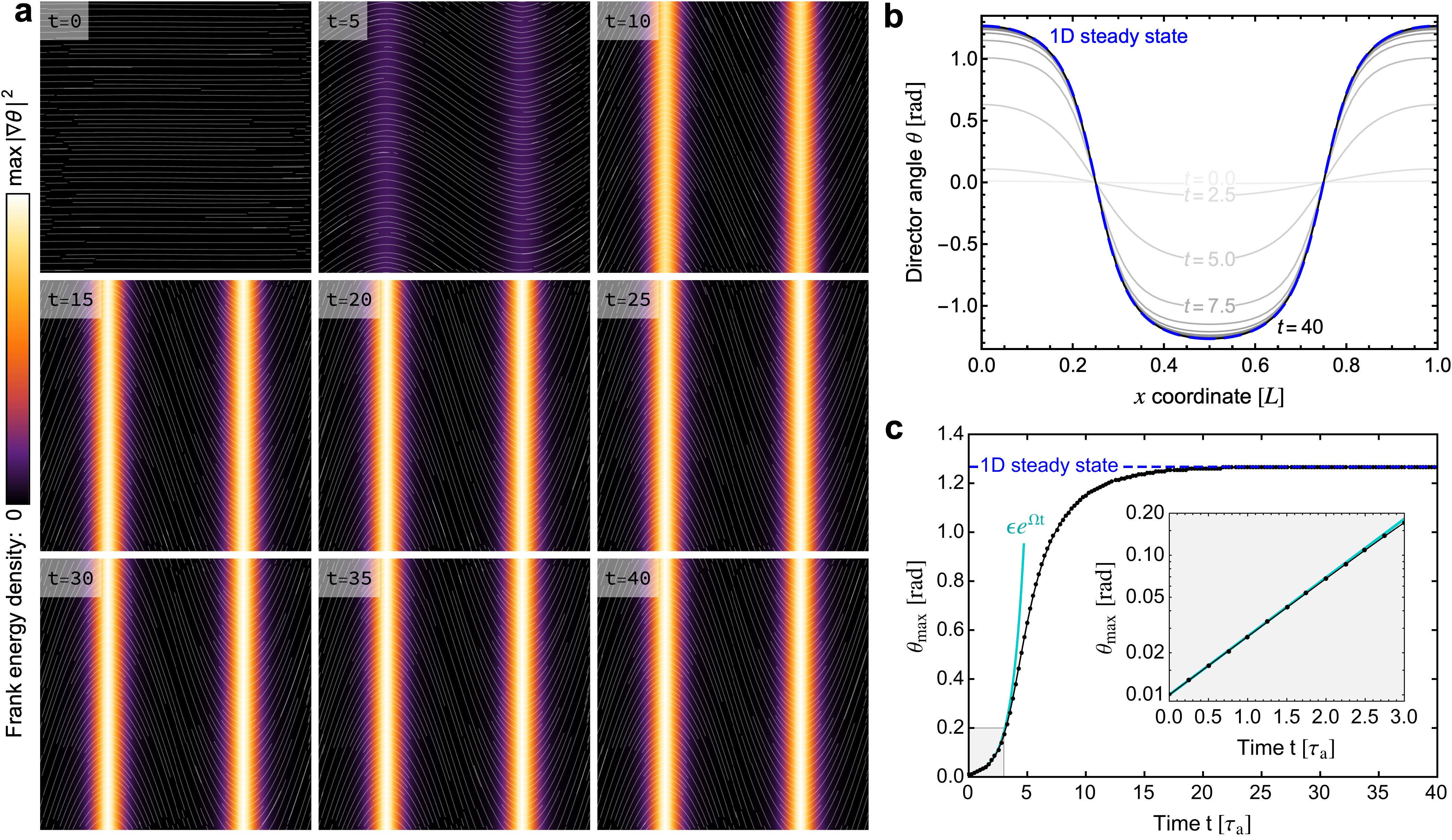}
{\phantomsubcaption\label{Fig snapshots}}
{\phantomsubcaption\label{Fig theta(x)}}
{\phantomsubcaption\label{Fig max theta}}
\end{center}
\bfcaption{Verification of the numerical scheme}{ \subref*{Fig snapshots}, Consecutive snapshots from a simulation initiated with a small bending modulation about the uniformly aligned state: $\theta(\bm{r},t=0)=\epsilon \cos(2\pi x)$, with $\epsilon=0.01$. Parameter values were set to $R=1$, $\nu=-1.1$, $A=1000$, and $S=1$ (extensile stress). Numerical parameters were set to $N=64$, $dt=0.01$, and $D=0$ (zero noise). Color indicates the Frank energy, $|\nabla\theta|^2$, and white lines trace the director $\bm{n}$. \subref*{Fig theta(x)}, We plot the numerical data corresponding to a horizontal slice, $\theta(x,y=0)$, for increasing times, starting from $t=0$ (lightest gray) up to $t=40\tau_\text{a}$ (black) in intervals of $2.5\tau_\text{a}$. It is shown that the angle profile relaxes on the predicted 1D steady state (dashed blue), obtained for the same parameter values as explained in our recent work\cite{Lavi2024}. \subref*{Fig max theta}, The maximum angle is plotted as function of time in our simulation (black). First, it is shown that the modulation grows exponentially in time with the growth rate $\Omega$ matching the linear stability dispersion relation \cite{Lavi2024} (cyan). The gray inset is a log-scale plot comparing the simulation results with this explicit prediction for small perturbations. Additionally, it is demonstrated that the maximal angle in the simulation relaxes at long times to the value predicted by the 1D steady state \cite{Lavi2024} (dashed blue). } 
\label{Fig verification}
\end{figure*}

\subsection{Verification}

We verified our computational solver for different parameter values. As demonstrated in \cref{Fig verification}, we compare the numerical output against two criteria: (i) long-time relaxation to predicted one-dimensional steady states (nonlinear striped patterns) that are stable at moderate activity levels\cite{Alert2020a,Lavi2024} (\cref{Fig theta(x)}), and (ii) the linear growth rate of small Fourier perturbations around the uniformly aligned quiescent state\cite{Alert2020a,Lavi2024} (\cref{Fig max theta}).

\section{SPECTRA} 
\label{spectra}
Here we detail the computation of the spectra presented in \crefrange{Fig velocity-spectra}{Fig correlation-times} and \cref{Fig Q spectrum}. In the equations below we use the following notations: 
\begin{itemize}
    \item $\tilde{f}(\bm{q})$ is the two-dimensional DFT of $f(\bm{r})$, i.e. $\tilde{f}=\mathcal{F}[f]$.
    \item $\left<f(\bm{r})\right>_{\bm{r}}$ is the spatial average of $f(\bm{r})$ over $\Omega$.
    \item $\left<\tilde{f}(\bm{q})\right>_\phi$ is the mean of all $\tilde{f}(\bm{q})$ for which $\bm{q}=q\,(\cos\phi,\sin\phi)$, i.e. an azimuthal average in Fourier space.
    \item $\left<\cdot \right>$ is an ensemble average within fully developed active turbulence. In our long simulations, this is practically computed as the temporal mean for $t > t_T$, with $t_\text{T}$ the onset time of a statistical steady state. To preclude initial transients, we empirically set $t_\text{T}=200 \tau_\text{a}$ ($2\times10^4$ computational time iterations).
\end{itemize}

\subsection{Velocity power spectrum}

The velocity power spectrum in the 2D Fourier space is defined by\cite{Alert2020a}
\begin{equation}
    E(\bm{q})=\frac{1}{2}\left<\tilde{v}_\alpha^*(\bm{q})\tilde{v}_\alpha(\bm{q})\right>=\frac{1}{2}q^2\left<\left|\tilde{\psi}(\bm{q})\right|^2\right>. \label{eq: E(bq)}
\end{equation}

Under the assumption of statistical isotropy, one  has
\begin{equation}
    E(q)=2\pi q \left<E(\bm{q})\right>_\phi=\pi q^3 \left<\left<\left|\tilde{\psi}(\bm{q})\right|^2\right>\right>_\phi. \label{eq: E(q)}
\end{equation}

Since $\bm{v}$ is in units of $[L/\tau_\text{a}]$ and $q$ is in units of $[L^{-1}]$, $E(q)$ is given in units of $[L/\tau_\text{a}^2]$ (\cref{Fig velocity-spectra}).

\subsection{Frank energy spectrum}

The Frank energy spectrum (or $\nabla\theta$ power spectrum) is defined by\cite{Alert2020a}
\begin{equation}
    F_n(\bm{q})=\frac{K}{2}\left<\mathcal{F}[\partial_\alpha\theta]^*(\bm{q})\mathcal{F}[\partial_\alpha\theta](\bm{q})\right>=\frac{K}{2}q^2\left<\left|\tilde{\theta}(\bm{q})\right|^2\right>.
\end{equation}

Under the assumption of statistical isotropy,
\begin{equation}
    F_n(q)=2\pi q \left< F_n(\bm{q})\right>_\phi=\pi K q^3 \left<\left<\left|\tilde{\theta}(\bm{q})\right|^2\right>\right>_\phi. \label{eq: Fn(q)}
\end{equation}

Because $q$ is in units of $[L^{-1}]$, $F_n(\bm{q})$ is given in units of $[K/L^3]$ (\cref{Fig Frank-spectra}).

\subsection{Correlation time spectra}

The nematic order parameter in our case, with the scalar amplitude fixed to $1$, is the tensor $\hat{q}_{\alpha\beta}=n_\alpha n_\beta-1/2\,\delta_{\alpha\beta}$ (not be confused with the wavevector $\bm{q}$). In terms of $\theta$,
\begin{equation}
    \hat{q}=\frac{1}{2}\begin{pmatrix}
    \cos 2\theta & \sin 2\theta \\
    \sin 2\theta & -\cos 2\theta
    \end{pmatrix}.
\end{equation}

We define $\mathcal{C}_{\hat{q}\hat{q}}(\bm{x},\tau)$ as the space-time autocorrelation function of $\hat{q}$,
\begin{equation}
    \mathcal{C}_{\hat{q}\hat{q}}(\bm{x},\tau)=\left<\left<\hat{q}_{\alpha\beta}(\bm{r},t)\hat{q}_{\alpha\beta}(\bm{r}+\bm{x},t+\tau)\right>_{\bm{r}}\right>,
\end{equation}
where $\bm{x}$ is the two point vector and $\tau$ is the lag time. For a given $\tau$, the frames considered in the time average correspond to $t_\text{T}\leq t\leq T-\tau$, with $T$ the total runtime of the simulation. 

The spatial Fourier transform of $\mathcal{C}_{\hat{q}\hat{q}}$ is then
\begin{align}
    \tilde{\mathcal{C}}_{\hat{q}\hat{q}}(\bm{q},\tau)=& 
    \left<\mathcal{F}\left[\hat{q}_{\alpha\beta}\right]^*(\bm{q},t)\mathcal{F}\left[\hat{q}_{\alpha\beta}\right](\bm{q},t+\tau)\right> \label{eq: Cqq}\\
    =&\,\frac{1}{2}\left<\mathcal{F}\left[\cos 2\theta\right]^*(\bm{q},t)\mathcal{F}\left[\cos 2\theta\right](\bm{q},t+\tau)\right> \nonumber \\
    &+\frac{1}{2}\left<\mathcal{F}\left[\sin 2\theta\right]^*(\bm{q},t)\mathcal{F}\left[\sin 2\theta\right](\bm{q},t+\tau)\right>. \nonumber
\end{align}

For each mode $\bm{q}$, we fit the time series $\tilde{\mathcal{C}}_{\hat{q}\hat{q}}(\bm{q},\tau)$ to an exponential $\tilde{\mathcal{C}}_{\hat{q}\hat{q}}(\bm{q},0)e^{-\tau/\tau_{\hat{q}}}$, with $\tau_{\hat{q}}$ the fitting parameter, using the \textbf{NonlinearModelFit} function in Mathematica (TM). This fitting gives $\tau_{\hat{q}}(\bm{q})$. The azimuthal average $\tau_{\hat{q}}(q)=\left<\tau_{\hat{q}}(\bm{q})\right>_\phi$ is shown in \cref{Fig correlation-times} (dark data points).

Next, we define the space-time autocorrelation function of the flow $\bm{v}$:
\begin{equation}
    \mathcal{C}_{\bm{v}\bm{v}}(\bm{x},\tau)=\left<\left<v_\alpha(\bm{r},\bm{t})v_\alpha(\bm{r}+\bm{x},t+\tau)\right>_{\bm{r}}\right>
\end{equation}

The spatial Fourier transform of $\mathcal{C}_{\bm{v}\bm{v}}$ is given by
\begin{align}
    \tilde{\mathcal{C}}_{\bm{v}\bm{v}}(\bm{q},\tau)=& \left<\tilde{v}_\alpha^*(\bm{q},t)\tilde{v}_\alpha(\bm{q},t+\tau)\right> \\
    =& \,q^2 \left<\tilde{\psi}^*(\bm{q},t) \tilde{\psi} (\bm{q},t+\tau)\right>. \nonumber
\end{align}
Note that $E(\bm{q})=\frac{1}{2}\tilde{\mathcal{C}}_{\bm{v}\bm{v}}(\bm{q},\tau=0)$ (see \cref{eq: E(bq)}). 

For each mode $\bm{q}$, we fit the time series $\tilde{\mathcal{C}}_{\bm{v}\bm{v}}(\bm{q},\tau)$ to an exponential $\tilde{\mathcal{C}}_{\bm{v}\bm{v}}(\bm{q},0)e^{-\tau/\tau_{\bm{v}}}$, with $\tau_{\bm{v}}$ the fitting parameter. This fitting gives $\tau_{\bm{v}}(\bm{q})$. The azimuthal average $\tau_{\bm{v}}(q)=\left<\tau_{\bm{v}}(\bm{q})\right>_\phi$ is shown in \cref{Fig correlation-times} (light data points).

\begin{figure}[tbhp!]
\begin{center}
\includegraphics[scale=0.48]{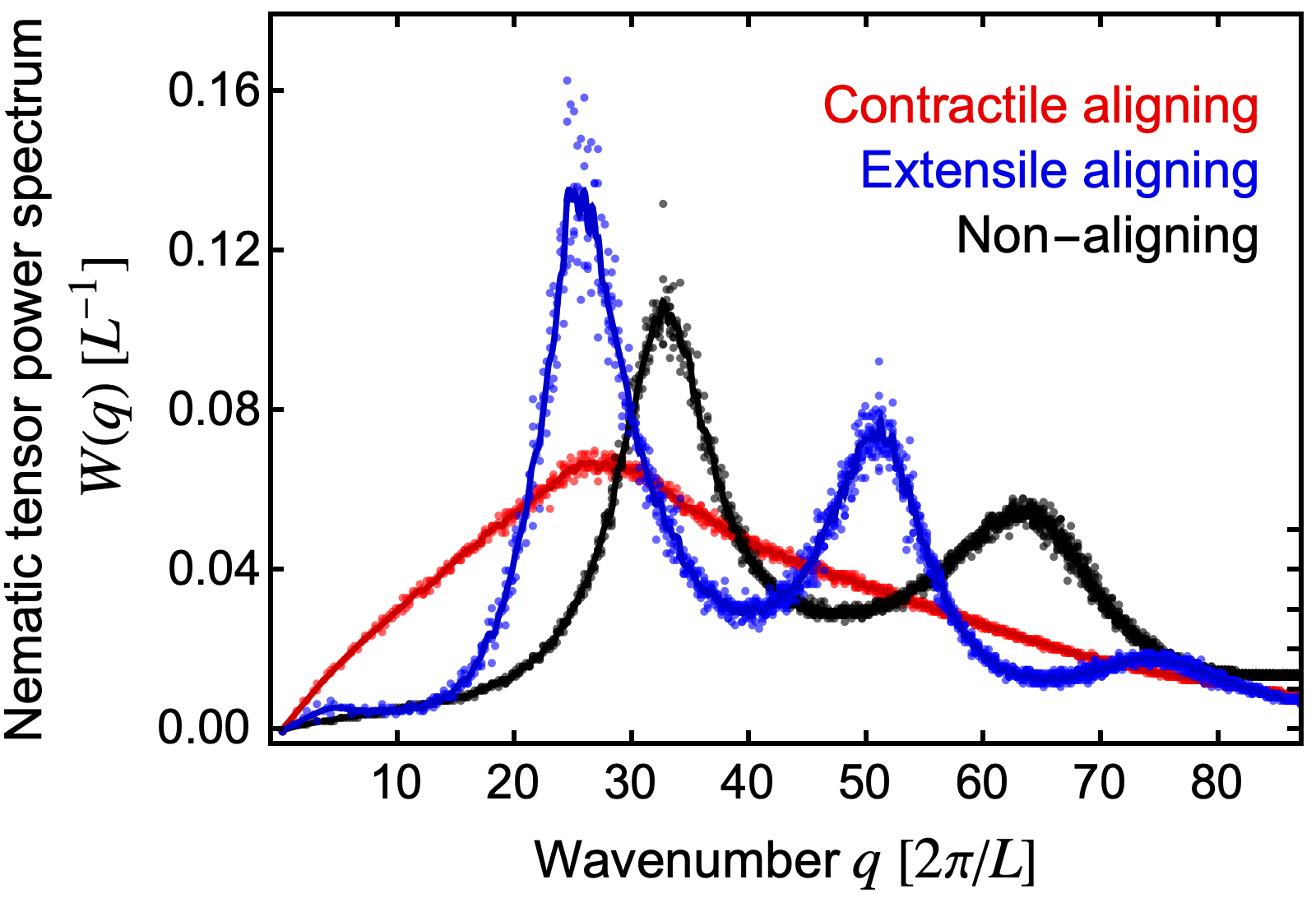}
\end{center}
\bfcaption{Power spectrum of the order parameter in active nematic turbulence}{ Shown is the power spectrum of the nematic tensor, $\hat{q}_{\alpha\beta}=n_\alpha n_\beta-1/2\,\delta_{\alpha\beta}$, in the contractile flow-aligning ($S\nu=+1.1$, red), extensile flow-aligning ($S\nu=-1.1$, blue), and non-aligning ($\nu=0$, black) cases. For the definition of $W(q)$ see \cref{eq:W(q)}. Lines represent a smoothed (Gaussian) interpolation of the computed data points. Parameter values are as in \cref{Fig regimes} and Fig.\ S1\cite{SM}. For arrested turbulence (blue), this spectrum features a narrow peak with its associated harmonics. For strong large-scale turbulence (red), such peaks are washed out. In all regimes, low-$q$ correlations are essentially absent. } 
\label{Fig Q spectrum}
\end{figure}

\subsection{Nematic tensor power spectrum}

The nematic tensor power spectrum is defined by \begin{equation}
    W(\bq)=\left<\mathcal{F}\left[\hat{q}_{\alpha\beta}\right]^*(\bm{q})\mathcal{F}\left[\hat{q}_{\alpha\beta}\right](\bm{q})\right>.
\end{equation}
By this definition, one has $W(\bm{q})=\tilde{\mathcal{C}}_{\hat{q}\hat{q}}(\bm{q},\tau=0)$ (see \cref{eq: Cqq}).

Under the assumption of statistical isotropy, 
\begin{align}
W(q)=&\,2\pi q \left<W(\bm{q})\right>_\phi \label{eq:W(q)}\\
=&\,\pi q\big<\left<\mathcal{F}\left[\cos 2\theta\right]^*(\bm{q})\mathcal{F}\left[\cos 2\theta\right](\bm{q})\right> \nonumber \\
&\quad + \left<\mathcal{F}\left[\sin 2\theta\right]^*(\bm{q})\mathcal{F}\left[\sin 2\theta\right](\bm{q})\right>\big>_\phi .\nonumber
\end{align}

Since $\hat{q}$ is dimensionless and $q$ is in units of $[L^{-1}]$, $W(\bm{q})$ is given in units $[L^{-1}]$ (\cref{Fig Q spectrum}).

\section{TOPOLOGICAL PSEUDO-CHARGE OF DOMAIN WALL NODES}
\label{pseudo-charge}
In arrested and nearly arrested states, arising in the extensile aligning regime, the domain walls are organized in a tree-like network. One can associate a planar, directed graph to this network, with three types of nodes: startpoints, endpoints, and branchpoints, as described in \crefrange{Fig startpoint}{Fig endpoint}. In the long-time asymptotics, the startpoints are only residual, and the ageing tree is mostly formed by endpoints and branchpoints (\cref{Fig aging}, \Movie{5}). While these nodes are not topological defects of the nematic field, one can assign to them a pseudo-charge which is conserved and contains useful information of topological nature, as discussed in this section. We emphasize that this pseudo-topology is an effective emergent property of the dynamics, that holds only for the appropriate parameter regimes and time asymptotics. Therefore, pseudo-defects are dynamically stabilized but are not topologically stable in the sense that they can be removed by a continuous perturbation of the nematic field.  

To define the topological pseudo-charge, we first consider the change of angle across a nematic wall. To this end, we look for steady-state solutions of the director field in 1D, that is, with translational invariance in the direction along the wall. In this case, the angle varies only along the transverse coordinate $s$. In the steady state, the director profile across the wall obeys the 1D stationary equation \cite{Lavi2024} 
\begin{equation}
  \ell_{\text{a}}^2  \partial_s^2\theta = \frac{S(1+\nu\cos 2\theta)(\sin 2\theta-\sin 2\bar{\theta})}{4+R+R\nu^2+2R\nu \cos 2\theta}, \label{eq:stationary}
\end{equation}
where $\bar\theta$ is the orientation of the nematic director at the center of the domain wall. This equation supports sigmoidal angle profiles with an interface of thickness of the order of the active length $\ell_\text{a}$ that continuously connects two plateaux of uniform angle. As the interfaces separate regions of uniform angle, they are referred to as domain walls, also called kinks and antikinks in other contexts. 

\begin{figure}[tbp!]
\begin{center}
\includegraphics[width=\columnwidth]{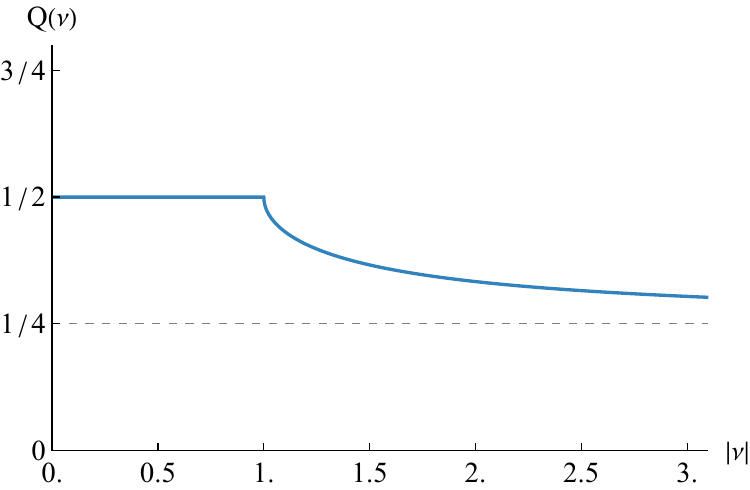}
\end{center}
\bfcaption{$Q$ as a function of $|\nu|$}{ Functional form of the elementary charge $Q (\nu)$. Positive pseudo-defects have charge $+Q$, and negative ones $-Q$. For the tumbling regime $Q=1/2$ and for the aligning regime $Q$ decreases monotonically from 1/2 to 1/4. }
\label{Fig pseudo-charge}
\end{figure}

For $S\nu>-1$, the two plateaux values of the nematic angle are $\theta=\bar{\theta} \pm \frac{\pi}{2}$, so the total rotation of the nematic across the wall is $\Delta \theta = \pm \pi$. For $S\nu<-1$, the plateaux angles correspond to the so-called Leslie angle defined by $1+\nu\cos 2\theta_L=0$, which amounts to an angle rotation of $\Delta \theta = \pm (\pi - \arccos{|\nu|^{-1}})$ across the wall. We remark that the total angle variation across domain walls is independent of both the active length $\ell_{\text{a}}$ and the viscosity ratio $R$. For a domain wall centered at $s=s_i$, we define the smooth step functions $\Theta^{\pm} (s-s_i)$ that connect the plateaux values via a total angle increment $\Delta \theta$, which is positive for $\Theta^{+}$ and negative for $\Theta^{-}$. The angle varies at a scale $\sim \ell_{\text{a}}$ about $s_i$. Similarly, the derivative with respect to the spatial coordinate $s$ defines two peaked functions $\dot\Theta^{\pm} (s-s_i) $ such that, in the sharp-wall limit, 
\begin{equation}
\lim_{\ell_{\text{a}} \rightarrow 0} \dot\Theta^{\pm}(s-s_i) = \Delta\theta\, \delta(s-s_i) \equiv \pm\frac{Q(\nu)}{2 \pi}\delta(s-s_i).
\end{equation}
Here, $Q(\nu)$ is the elementary pseudo-charge function, defined as
\begin{equation}
Q (\nu) \equiv \frac{1}{2\pi} \int_{-\infty}^{+\infty} \dot{\Theta}^{+}_0 \dd s > 0, \label{Qlimit}
\end{equation}
which gives, for $S\nu<0$, 
\begin{equation}
Q (\nu) = \frac{1}{2} \left( 1 - \frac{1}{\pi}{\cal H}(|\nu|-1) \arccos{|\nu|^{-1}} \right) > 0, 
\end{equation}
where ${\cal H}$ is the Heaviside step function. The elementary pseudo-charge function $Q(\nu)$ takes the value $1/2$ in the tumbling regime, and it spans the interval from $1/2$ to $1/4$ as $|\nu|$ is increased beyond $1$. The functional form of $Q(\nu)$ is plotted in \cref{Fig pseudo-charge}. 

To define a topological pseudo-charge of the pseudo-defects, and of any closed domain in general, we recall that the total angle variation along a closed loop must be
\begin{equation}
\oint \dd\theta = \oint \dot{\theta} \,\dd s = 0 , \label{eq defect-free}
\end{equation}
with $\dot{\theta} \dd s = \vec{\nabla} \theta \cdot \dd\vec{s} $ where $\dd\vec{s}$ describes the contour, with $s$ the arclength coordinate. We are interested in the limit of small $\ell_{\text{a}}$, or equivalently, to situations where the separation between consecutive walls is much larger than the wall thickness, so that there is no significant overlap between $\dot\Theta^{\pm}(s-s_i)$ and $\dot\Theta^{\pm}(s-s_j)$, i.e. $|s_i-s_j| \gg \ell_{\text{a} }$. Then, we can distinguish two scales of variation of the nematic orientation: the inner region of the domain wall where the nematic variation is fast ($\dot\theta \sim \ell_{\text{a}}^{-1} $), and the outer regions where it is slow ($\dot\theta \sim \ell_{\text{a}}^{0} $). Under these circumstances, we may define the pseudo-charge as the part of the director rotation along a closed loop that takes place at the slow-variation (outer regions of the domain walls). In other words, we consider the total director rotation excluding the fast variation across the walls that are intersected by the loop.

Consider, for instance, an isolated endpoint pseudo-defect (\cref{Fig endpoint}), formed by a wall that points towards it, and consider a counterclockwise closed path $\Gamma$ that contains it and intersects the center of the wall at $s=s_0$. We define the pseudo-charge of this pseudo-defect as
\begin{eqnarray}
Q_\Gamma &=& \frac{1}{2\pi} \oint_\Gamma \left(\dot{\theta}(s) - \dot{\Theta}^{-}(s-s_0) \right) \dd s \\ 
&=& \frac{1}{2\pi} \oint_\Gamma \dot{\Theta}^{+}(s-s_0) \,\dd s > 0, \label{eq def}
\end{eqnarray}
which coincides strictly with $Q(\nu)$ in the sharp-wall limit. To subtract the localized fast variation of $\theta$ across the wall we need to write $\Theta^- (s-s_0)$ because the nematic is rotating clockwise across the wall (decreasing angle), if the wall is oriented towards the interior of the path that encloses the endpoint. In general, the sense of rotation of the nematic across a wall is opposite (equal) to the sense of rotation of the closed loop when the wall points towards the interior (exterior) of the loop.

Similarly, for an isolated branchpoint we define the pseudo-charge as the rotation of the director in regions excluding the walls along a closed loop enclosing the branchpoint (\cref{Fig branch point}). In this case, we must subtract the fast variation across the three walls intersected by the integration path. Since there is always one wall pointing towards the interior and two pointing towards the exterior, the pseudo-charge for a branchpoint, in the sharp-wall limit, will be $-Q(\nu)$. 

For a general configuration with an arbitrary number of pseudo-defects and associated walls, the total pseudo-charge in the interior of a closed loop will given by the sum of the pseudo-charge of all the pseudo-defects in it. We remark that it is not necessary to know the detailed structure of the nematic field in the interior, but only in the contour, as in the definition of the actual topological charge. In fact, the total pseudo-charge in the interior of an arbitrary closed loop $\Gamma$ is also given by 
\begin{equation}
Q_\Gamma = (m_\text{in} - m_\text{out} ) Q(\nu),
\end{equation}
where $m_\text{in}$ and $m_\text{out}$ are the number of inward and outward domain walls crossed by the contour line, respectively. 

In the case of domain walls with a small but finite thickness compared to typical wall separation, the elementary pseudo-charge will slightly differ from the limiting value defined in \cref{Qlimit}. In that case, consecutive walls partially overlap so that the angle profile does not reach the asymptotic value of the plateau, and the magnitude of the angle variation across the walls, $\Delta \theta$, is slightly smaller. This is the case in the wall patterns studied in this work, so in the main manuscript we refer to the values taken by the pseudo-topological charge only as $\simeq Q(\nu)$. We remark, however, that the precise value of the elementary charge $Q(\nu)$ is not really informative of the relevant topological structure, that is, how walls are connected, but accounts for the detailed structure of the walls themselves. This may be important to determine their dynamical properties, such as their interactions or their rigidity to deformation. However, it is only the integer counts of positive and negative pseudo-defects (or the counts of inward and outward wall intersections), i.e., the multiples of the elementary charge, that are informative of the overall qualitative structure inside a given domain.

In the sharp-wall limit $\ell_{\text{a}} \rightarrow 0$, the idea of a pseudo-defect is reminiscent of Dirac's construction of magnetic monopoles, based on attaching to them the so-called Dirac string, which carries a quantized magnetic flux\cite{Dirac1931}. In that way, the magnetic monopole can be made consistent with Maxwell's equations. For instance, the total magnetic flux across a closed surface enclosing the magnetic monopole would be zero, as required by Gauss' theorem for the absence of net magnetic charge, because the flux generated by the monopole is canceled by that carried out by the string, which also crosses the surface. Similarly, in our problem, the wall attached to the pseudo-defect provides a localized (fast) rotation of the director that cancels the (slow) rotation outside the wall to ensure the lack of true topological charge. In this way, the problem satisfies the local constraint of zero topological charge while mimicking with pseudo-defects the director field around actual topological defects in a large region of space around them.

\section{Q-TENSOR MODEL AND SIMULATIONS} 

\label{Qmodel}

We complement our analysis of defect-free active nematics with simulations of the full Q-tensor model to demonstrate the broader applicability of our findings. Specifically, we show that arrested domain-wall turbulence is not merely a feature of the constrained director-based model, but rather a robust phenomenon that arises when defect nucleation is energetically suppressed. In the generic model, the ratio of the defect core size $\epsilon$ to the active length $\ell_\text{a}$ emerges as a key control parameter. When $\epsilon \ll \ell_\text{a}$, the free-energy cost of nucleating topological defects exceeds that of forming networks of domain walls threaded by pseudo-defects. Hence, large-scale labyrinthine patterns consistently arise in the extensile, rod-aligning regime when $L \gg \ell_\text{a} \gg \epsilon$.

We further show that as the core size exceeds a critical threshold, defect nucleation destabilizes the domain walls, ultimately destroying the grid-locked network and leading to qualitatively different dynamics. Near this threshold, we find that $\pm 1/2$ defects preferentially unbind at distortion hotspots associated with the branchpoint pseudo-defects. This finding highlights an interesting connection between the zigzagging and branching of domain walls and the subsequent formation of topological disclinations. 

\subsection{Definition and energetics}
\label{Q definition}

The nematic order parameter in 2D is a symmetric, traceless tensor defined by  
\[
Q_{\alpha\beta} = s \left(n_\alpha n_\beta - \frac{1}{2} \delta_{\alpha\beta}\right),
\]
where $s$ is the scalar order parameter and $\bm{n}$ is a unit director. The tensor $\mathbf{Q}$ then has eigenvalues $\pm s/2$, with $\bm{n}$ the eigenvector associated with the positive eigenvalue.

We adopt the Landau–de Gennes free energy,
\begin{equation}
    F=\int \left(\mathcal{F}_\text{e}+\mathcal{F}_\text{b}  \right) \dd^2\bm{r},
\end{equation}
where 
\begin{align}
& \mathcal{F}_\text{e}
=\frac{K}{4} (\partial_\gamma Q_{\alpha\beta}) (\partial_\gamma Q_{\alpha\beta}), \nonumber \\
& \mathcal{F}_\text{b}
=\frac{K}{4}\frac{1}{\epsilon^2}\left( -s_0^2 Q_{\alpha\beta}Q_{\alpha\beta}+(Q_{\alpha\beta}Q_{\alpha\beta})^2\right).
\nonumber 
\end{align}

Here we defined the elastic distortion energy $\mathcal{F}_\text{e}$ using a single constant $K$. The prefactor $1/4$ ensures that, when $s=1$, the expression reduces exactly to the one-constant Frank energy density of the director-based model: $\frac{K}{2}(\partial_\alpha n_\beta)(\partial_\alpha n_\beta)$, when $|\bm{n}|=1$ (see \cref{eq Frank}). The bulk free energy $\mathcal{F}_\text{b}$ includes quadratic and quartic thermotropic terms. The sign and ratio of the coefficients are chosen to correspond to a temperature below the isotropic–nematic transition. We parametrize these in terms of $K$ and a characteristic length scale $\epsilon$, known as the defect core size. The parameter $s_0$ sets the equilibrium value of the nematic scalar order. Hereafter we fix $s_0=1$, corresponding to a system that would, at equilibrium, lie deep in the nematic phase.

The orientational field $\mathbf{H}$ is a symmetric traceless tensor given by
\begin{align}
     H_{\alpha\beta}&= -\frac{\delta F}{\delta Q_{\alpha\beta}} + \frac{\delta_{\alpha\beta}}{2}\text{tr}\frac{\delta F}{\delta Q_{\gamma\delta}}  \label{eq H tensor}\\
     & = \frac{K}{2}\left( \nabla^2 Q_{\alpha\beta} + \epsilon^{-2}(1-s^2)Q_{\alpha\beta}\right),\nonumber
\end{align}
where we substituted $s^2=2\,\text{tr}\,\mathbf{Q}^2=2 Q_{\gamma\delta}Q_{\gamma\delta}$, clarifying how the bulk term in $\mathbf{H}$ acts to restore the equilibrium scalar order. We emphasize that the key difference from our constrained director-based model is that nematic alignment (here encoded in $s$) is no longer enforced by a Lagrange multiplier, but may vary dynamically.

\subsection{Hydrodynamics}
\label{Q hydro}

As in \cref{eq force-balance}, momentum balance reads 
\begin{equation}
    0 = -\partial_\alpha P + \partial_\beta \left( \sigma_{\alpha\beta}^{\text{a}} + \sigma_{\alpha\beta}^{\text{E}} + \sigma_{\alpha\beta} \right). \label{eq force-balance 2}
\end{equation}

In terms of $\mathbf{Q}$ and $\mathbf{H}$, and defining $\tilde{Q}_{\alpha\beta}\equiv Q_{\alpha\beta}+\delta_{\alpha\beta}/2$, the stress contributions are given by
\begin{align}
& \sigma^{\text{a}}_{\alpha\beta}=Q_{\alpha\gamma}H_{\gamma\beta}-H_{\alpha\gamma}Q_{\gamma\beta}, \\
& \sigma^{\text{E}}_{\alpha\beta} = -  \frac{\delta F}{\delta \left(\partial_\alpha Q_{\gamma\delta} \right)} \partial_\beta Q_{\gamma\delta}=-\frac{K}{2}(\partial_{\alpha}Q_{\gamma\delta})(\partial_\beta Q_{\gamma\delta}), \\
& \sigma_{\alpha\beta} = 2 \eta v_{\alpha\beta} - \zeta Q_{\alpha\beta} \\
&\qquad\,\,\,+ \nu \left( \tilde{Q} _{\alpha\gamma} H_{\gamma\beta} + H_{\alpha\gamma} \tilde{Q} _{\gamma\beta} - 2 (Q_{\gamma\delta} H_{\gamma\delta})\tilde{Q}_{\alpha\beta} \right), \nonumber
\end{align}
where $\eta$ is the shear viscosity, $v_{\alpha\beta}=(\partial_\alpha v_\beta+\partial_\beta v_\alpha)/2$, and $\zeta$ is the active stress parameter. 

\subsection{Nematodynamics}
\label{Q nemato}

As in the passive theory (Beris-Edwards model\cite{beris1994thermodynamics}), the Q-tensor evolves as
\begin{align}
    \partial_t Q_{\alpha\beta} &+ v_\gamma \partial_\gamma Q_{\alpha\beta} + \omega_{\alpha\gamma} Q_{\gamma\beta}- Q_{\alpha\gamma} \omega_{\gamma\beta}= \label{eq dt Q}\\
    &\Gamma H_{\alpha\beta} - \nu \left( v_{\alpha\gamma} \tilde{Q}_{\gamma\beta} + \tilde{Q}_{\alpha\gamma} v_{\gamma\beta} -2(Q_{\gamma\delta}\partial_{\gamma} v_{\delta}) \tilde{Q}_{\alpha\beta}\right),\nonumber
\end{align}
where $\omega_{\alpha\beta}=(\partial_\alpha v_\beta-\partial_\beta v_\alpha)/2$ and $\Gamma$ scales the response of $\mathbf{Q}$ to the molecular field $\mathbf{H}$. When $s=1$, we wish to recover the same expression as in the $|\bm{n}|=1$ limit of the director-based model. To that end, we set $\Gamma=2/\gamma$, with $\gamma$ the rotational viscosity.

{\remark[Orhogonal and parallel components of $\mathbf{H}$]
\label{remark H}
By substituting 
\[
Q_{\alpha\gamma}=s\left(n_\alpha n_\beta-\tfrac{1}{2}\delta_{\alpha\beta}\right)=\frac{s}{2}\begin{pmatrix}
    \cos2\theta & \sin 2\theta\\
    \sin 2\theta & -\cos 2\theta
\end{pmatrix}
\]
in \cref{eq H tensor}, we find that the contribution to $\partial_t\mathbf{Q}$ arising from the molecular field is given by
\begin{align}
\frac{2}{\gamma}\mathbf{H}&=\frac{K}{\gamma}\left(\nabla^2\mathbf{Q} + \frac{1}{\epsilon^2}(1-s^2)\mathbf{Q}\right) \nonumber \\
&=\frac{K}{\gamma}\left(\frac{4\nabla{s}\cdot\nabla{\theta}}{s}+2\nabla^2\theta\right)\mathbf{Q}^\perp  \nonumber \\ 
&\quad +\frac{K}{\gamma}\left(\frac{\nabla^2s}{s}-4(\nabla\theta)^2+\frac{1}{\epsilon^2}(1-s^2)\right)\mathbf{Q},\nonumber
\end{align}
where
\[
Q_{\alpha\beta}^\perp=Q_{\alpha\gamma}\epsilon_{\gamma\beta}=\frac{s}{2}\begin{pmatrix}
    -\sin2\theta & \cos2\theta\\
    \cos 2\theta & \sin2\theta
\end{pmatrix}.
\]
The terms proportional to $\mathbf{Q}^\perp$ drive rotations, while those along $\mathbf{Q}$ regulate the scalar order parameter $s$. In particular, the term $-4(\nabla \theta)^2\mathbf{Q}$ acts to reduce $s$, pushing the system away from perfect alignment. Assuming $\theta$ varies on the active length scale $\ell_\text{a}$, this contribution scales as $-1/\ell_\text{a}^2$. By contrast, the term proportional to $1/\epsilon^2$ tends to restore $s$ toward unity. This competition explains why the ratio $\epsilon/\ell_\text{a}$ plays a central role in determining whether the system remains uniformly nematic or gives way to defect nucleation.
}

\subsection{Dimensionless $(Q_{11}, Q_{12})$-$\psi$ formulation}
\label{Q dimensionless}

For consistency, we adopt the same nondimensionalization as in the director-based model: lengths are scaled by the system size $L$, time by the active timescale $\tau_\text{a} = \eta / |\zeta|$, stress by the activity scale $|\zeta|$, and the orientational field by $K / L^2$. We then define the problem in terms of the sign of active stress $S=\pm 1$ (positive for extensile), the viscosity ratio $R=\gamma/\eta$, the activity number $A=R L^2|\zeta|/K=L^2/\ell_\text{a}^2$, 
and the dimensionless defect core size $\varepsilon=\epsilon/L$. In the following, we denote adimensionalized fields and derivatives with top bar (e.g, $\bar{\nabla}=L\nabla$, $\bar{\partial}_t=\tau_\text{a}\partial_t$, $\bar{v}=(\tau_\text{a}/L)v$).

We now define $\mathbf{Q}$ and  the adimensional orientational field $\bar{\mathbf{H}}$ in terms of $Q_{11}$ and $Q_{12}$, for which we seek a closed-form model.
\begin{equation}
    \mathbf{Q}=
    \begin{pmatrix}
        Q_{11} & Q_{12} \\
        Q_{12} & -Q_{11}
    \end{pmatrix},\, 
    \bar{\mathbf{H}}=
    \begin{pmatrix}
        \bar{H}_{11} & \bar{H}_{12} \\
        \bar{H}_{12} & -\bar{H}_{11}
    \end{pmatrix},
\end{equation}    
where 
\begin{align}
        & \bar{H}_{11}=\frac{L^2}{K}H_{11}=\frac{1}{2}\left(\bar{\nabla}^2Q_{11}+\varepsilon^{-2}(1-s^2)Q_{11}\right), \label{eq H_11 adim}\\
        & \bar{H}_{12}=\frac{L^2}{K}H_{12}=\frac{1}{2}\left(\bar{\nabla}^2Q_{12}+\varepsilon^{-2}(1-s^2)Q_{12}\right), \label{eq H_12 adim}
\end{align}
and
\begin{equation}
    s^2=4(Q_{11}^2+Q_{12}^2).
\end{equation}

The adimensional stress terms read
\begin{align}
 \bar{\sigma}^{\text{a}}_{\alpha\beta}&=
 \frac{\sigma^{\text{a}}}{|\zeta|}=\frac{R}{A}\left(Q_{\alpha\gamma}\bar{H}_{\gamma\beta}-\bar{H}_{\alpha\gamma}Q_{\gamma\beta}\right), \label{eq antisymmetric adim} \\
 \bar{\sigma}^{\text{E}}_{\alpha\beta} &= \frac{\sigma^\text{E}}{|\zeta|}=-  \frac{R}{2 A}(\bar{\partial}_{\alpha}Q_{\gamma\delta})(\bar{\partial}_\beta Q_{\gamma\delta}), \label{eq ericksen adim} \\
 \bar{\sigma}_{\alpha\beta} &= \frac{\sigma}{|\zeta|}= 2 \bar{v}_{\alpha\beta} - S Q_{\alpha\beta}\label{eq sigma adim}\\
&\quad+ \frac{R\, \nu}{A} \left( \tilde{Q} _{\alpha\gamma} \bar{H}_{\gamma\beta} + \bar{H}_{\alpha\gamma} \tilde{Q} _{\gamma\beta} - 2 (Q_{\gamma\delta} \bar{H}_{\gamma\delta})\tilde{Q}_{\alpha\beta} \right). \nonumber
\end{align}

The adimensional version of \cref{eq dt Q} reads
\begin{align}
    &\bar{\partial}_t Q_{\alpha\beta} + \bar{v}_\gamma \bar{\partial}_\gamma Q_{\alpha\beta} + \bar{\omega}_{\alpha\gamma} Q_{\gamma\beta}- Q_{\alpha\gamma} \bar{\omega}_{\gamma\beta}=\\
    &\quad \frac{2}{A} \bar{H}_{\alpha \beta} - \nu \left( \bar{v}_{\alpha\gamma} \tilde{Q}_{\gamma\beta} + \tilde{Q}_{\alpha\gamma} \bar{v}_{\gamma\beta} -2(Q_{\gamma\delta}\bar{\partial}_{\gamma} \bar{v}_{\delta}) \tilde{Q}_{\alpha\beta}\right).\nonumber
\end{align}

Hereafter, we shall consider only dimensionless fields and derivatives and omit the over-bar notation. Then, in terms of the stream function $\psi$, we obtain for $Q_{11}$ and $Q_{12}$,
\begin{align}
    \partial_t Q_{11}&+(\partial_y\psi,-\partial_x\psi)\cdot\nabla Q_{11}-Q_{12}\nabla^2\psi = \label{eq dt Q11} \\
     &\frac{2}{A}H_{11} -\nu  \left((1-4Q_{11}^2)d_2\psi-4Q_{11}Q_{12}d_1\psi \right),\nonumber  \\
     \partial_t Q_{12}&+(\partial_y\psi,-\partial_x\psi)\cdot\nabla Q_{12}+Q_{11}\nabla^2\psi = \label{eq dt Q12}\\
     &\frac{2}{A} H_{12} -\nu  \left((1-4Q_{12}^2)d_1\psi-4Q_{11}Q_{12}d_2\psi \right), \nonumber 
\end{align}
where we re-introduced the notation $d_1:=\frac{1}{2}(\partial_y^2-\partial_x^2)$, $d_2:=\partial_y\partial_x$.

To derive an equation for the stream function $\psi$, we evaluate the curl of \cref{eq force-balance 2}, which eliminates the pressure from the problem. The term coming from the antisymmetric stress, defined in \cref{eq antisymmetric adim}, reads
\begin{align}
&curl({\nabla}\cdot \sigma^{\text{a}})=\epsilon_{\gamma\alpha}\partial_\gamma\left(\partial_\beta\sigma_{\alpha\beta}^{\text{a}}\right) \label{eq curl div sigma a} \\
&\,=\epsilon_{\gamma\alpha}\partial_\gamma\left(\partial_\beta \epsilon_{\alpha\beta}\sigma_{12}^{\text{a}}\right) =-\delta_{\gamma\beta}\partial_\gamma\partial_\beta\sigma^{\text{a}}_{12} = -\nabla^2\sigma^{\text{a}}_{12} \nonumber \\
&\,=-\frac{2R}{A}\nabla^2\left(Q_{11}{H}_{12}-Q_{12}{H}_{11}\right), \nonumber
\end{align}
where here $\epsilon_{\alpha\beta}$ denotes the Levi-Civita tensor.

The term coming from the Ericksen stress (symmetric in our case), defined in \cref{eq ericksen adim}, reads
\begin{align}
&curl(\nabla\cdot \sigma^{\text{E}})=\epsilon_{\gamma\alpha}\partial_\gamma\left(\partial_\beta\sigma_{\alpha\beta}^{\text{E}}\right) \label{eq curl div sigma E}\\
&\,=\partial_x(\partial_x\sigma_{12}^{\text{E}} +\partial_y\sigma_{22}^{\text{E}})-\partial_y(\partial_x\sigma_{11}^{\text{E}} +\partial_y\sigma_{12}^{\text{E}} ) \nonumber\\
&\,= d_2(\sigma_{22}^\text{E}-\sigma_{11}^\text{E})-2 d_1 \sigma_{12}^\text{E} \nonumber\\
&\,=\frac{R}{A}\Bigg(d_2 \Big( (\partial_x Q_{11})^2+(\partial_x Q_{12})^2 -(\partial_y Q_{11})^2-(\partial_y Q_{12})^2\Big) \nonumber\\
&\quad\qquad+2 d_1\Big((\partial_x Q_{11})(\partial_y Q_{11}) +(\partial_x Q_{12})(\partial_y Q_{12})\Big) \Bigg), \nonumber
\end{align}

Similarly, the term coming from of the last (symmetric and trace-free) stress reads
\begin{align}
curl(\nabla\cdot \sigma)&=\epsilon_{\gamma\alpha}\partial_\gamma\left(\partial_\beta\sigma_{\alpha\beta}\right) \label{eq curl sigma} \\
&= -2d_2 \sigma_{11}-2 d_1\sigma_{12}. \nonumber
\end{align}

It is convenient at this stage to decompose $\sigma=\sigma^{\text{visc}}+\sigma^{\text{act}}+\sigma^{\text{align}}$ (respectively, the viscous, active, and flow alignment stresses). Through \cref{eq curl sigma} and \cref{{eq sigma adim}}, we obtain
\begin{align}
    &curl (\nabla\cdot \sigma^{\text{visc}})=-4\partial_y\partial_x {v}_{11}-2(\partial_y^2-\partial_x^2){v}_{12} \label{eq curl div sigma visc} \\
    &\,=-4\partial_y\partial_x^2  v_x-(\partial_y^2-\partial_x^2)(\partial_x  v_y+ \partial_y  v_x)\nonumber \\
    &\,=-4\partial_y^2\partial_x^2 \psi-(\partial_y^2-\partial_x^2)(-\partial_x^2 \psi+\partial_y^2 \psi)\nonumber
    \\
    &\,=-(\partial_x^4+2\partial_y^2\partial_x^2+\partial_y^4)\psi=-\nabla^4\psi, \nonumber \\
    &curl (\nabla\cdot \sigma^{\text{act}})=2S\left( d_2 Q_{11}+d_1 Q_{12}\right), \label{eq curl div sigma act}\\
    &curl (\nabla\cdot \sigma^{\text{align}})=
    -2d_2 \sigma_{11}^{\text{align}}
    -2d_1 \sigma_{12}^{\text{align}} \label{eq curl div sigma align} \\
    &\,= -2\frac{R\nu}{A}\Bigg(
    d_2 \Big({H}_{11} -4 Q_{11}({H}_{11}Q_{11}+{H}_{12}Q_{12})\Big) \nonumber\\
    &\qquad\qquad +d_1\Big({H}_{12} -4Q_{12}({H}_{11}Q_{11}+{H}_{12}Q_{12})\Big)\Bigg).\nonumber
\end{align}

Our problem for $\psi$ is then given by
\begin{equation}
    \nabla^4\psi= \text{\eqref{eq curl div sigma a}}+\text{\eqref{eq curl div sigma E}}+\text{\eqref{eq curl div sigma act}}+\text{\eqref{eq curl div sigma align}}, \label{eq psi adim}
\end{equation}
with the right-hand side fully explicit (i.e. independent on $\psi$).

With ${H}_{11}$ and ${H}_{12}$ defined in \crefrange{eq H_11 adim}{eq H_12 adim}, the reduced closed-form problem is given by \crefrange{eq dt Q11}{eq dt Q12} and \cref{eq psi adim}.

\subsection{The $\theta$ model ($s=1$ constraint)}
\label{Q theta model}

To verify that our Q-tensor model is formulated consistently with respect to our constrained director-based model, we derive here the equations for $\partial_t\theta$ and $\psi$ when the bulk term in $\mathbf{H}$ is replaced by a Lagrange multiplier that enforces $s=1$.

When $s=1$, we have
\begin{equation}
    Q_{11}=\frac{1}{2}\cos2\theta,\, Q_{12}=\frac{1}{2}\sin2\theta,
\end{equation}
where, as before, $\theta$ is the angle of the director $\bm{n}$ with respect to the $x$-axis. 

Now, the orientational field components are given by
\begin{align}
    H_{11}&=\frac{1}{2}\nabla^2Q_{11}+H_{b}^0 Q_{11}\\
    &=-\frac{1}{2}\sin2\theta \nabla^2\theta+\frac{1}{2}\cos2\theta\left(H_b^0-2
(\nabla\theta)^2\right), \nonumber \\
H_{12}&=\frac{1}{2}\nabla^2Q_{12}+H_{b}^0 Q_{12}\\
    &=\frac{1}{2}\cos2\theta \nabla^2\theta+\frac{1}{2}\sin2\theta\left(H_b^0-2
(\nabla\theta)^2\right), \nonumber
\end{align}
where $H_b^0$ is a Lagrange multiplier acting as a bulk forcing term.

\Cref{eq dt Q11} now reads
\begin{align}
    -\sin&2\theta\left(\partial_t\theta +(\partial_y\psi,-\partial_x\psi)\cdot\nabla\theta+\frac{1}{2}\nabla^2\psi \right)= \label{eq dt Q11 theta}\\
    &\frac{2}{A} \left(-\frac{1}{2}\sin2\theta\nabla^2\theta+\frac{1}{2}\cos 2\theta\left(H_b^0-2(\nabla\theta)^2\right)\right) \nonumber \\
    &-\nu\left((1-\cos^2 2\theta)d_2\psi-\cos2\theta\sin2\theta d_1\psi \right) \nonumber \\
    =\,& -\sin2\theta\left(\frac{1}{A}\nabla^2\theta +\nu\left(\sin2\theta\, d_2\psi - \cos2\theta\,d_1\psi\right)\right) \nonumber \\
    &+\cos 2\theta \frac{1}{A}\left(H_b^0-2(\nabla\theta)^2\right). \nonumber
\end{align}

Similirarly, \cref{eq dt Q12} now reads
\begin{align}
    \cos&2\theta\left(\partial_t\theta +(\partial_y\psi,-\partial_x\psi)\cdot\nabla\theta+\frac{1}{2}\nabla^2\psi \right)= \label{eq dt Q12 theta} \\
    &\frac{2}{A} \left(\frac{1}{2}\cos2\theta\nabla^2\theta+\frac{1}{2}\sin 2\theta\left(H_b^0-2(\nabla\theta)^2\right)\right) \nonumber \\
    &-\nu\left((1-\sin^2 2\theta)d_1\psi-\cos2\theta\sin2\theta d_2\psi \right) \nonumber \\
    =\,& \cos2\theta\left(\frac{1}{A}\nabla^2\theta-\nu(\cos2\theta\, d_1\psi -\sin2\theta\, d_2\psi)\right) \nonumber \\
    &+\sin 2\theta \frac{1}{A}\left(H_b^0-2(\nabla\theta)^2\right). \nonumber
\end{align}

Applying a perpendicular projection: $$-\sin2\theta\,(\ref{eq dt Q11 theta})+\cos2\theta\,(\ref{eq dt Q12 theta}),$$ we obtain
\begin{align}
    \partial_t\theta +&(\partial_y\psi,-\partial_x\psi)\cdot\nabla\theta+\frac{1}{2}\nabla^2\psi = \label{eq dt theta from Q} \\
    &\frac{1}{A}\nabla^2\theta-\nu\left(\cos2\theta\, d_1\psi-\sin2\theta\,d_2\psi\right). \nonumber
\end{align}

From the parallel projection: $$\cos2\theta\,(\ref{eq dt Q11 theta})+\sin 2\theta\,(\ref{eq dt Q12 theta}),$$ we obtain
\begin{equation}
0= H_b^0-2(\nabla\theta)^2. \label{eq H_b}
\end{equation}

Thus, we recover in \cref{eq dt theta from Q} an equation for $\theta$ that is \emph{identical} to \cref{eq theta}. Unlike before, however, we find in \cref{eq H_b} that the Lagrange multiplier does not depend on flow, and the orientational field components simply reduce to
\[
H_{11}=-\frac{1}{2}\sin2\theta\,\nabla^2\theta,\, 
H_{12}=\frac{1}{2}\cos 2\theta\,\nabla^2\theta.
\]

Next, we turn to the curl of momentum balance, \cref{eq psi adim}, where we aim to express each contribution on the RHS in terms of $\theta$. From \cref{eq curl div sigma a}, we obtain
\begin{align}
&curl(\nabla\cdot \sigma^{\text{a}})=-\frac{2R}{A}\nabla^2\left(Q_{11}H_{12}-Q_{12}H_{11}\right) \\
&\,=-\frac{R}{2A}\nabla^2\left(\cos^2 2\theta \nabla^2\theta+\sin^2 2\theta \nabla^2\theta\right) =-\frac{R}{2A}\nabla^4\theta. \nonumber
\end{align}

From \cref{eq curl div sigma E}, we obtain
\begin{align}
&curl(\nabla\cdot \sigma^{\text{E}}) \\
&\,=\frac{R}{A}\Bigg(d_2 \Big( (\partial_x Q_{11})^2+(\partial_x Q_{12})^2 -(\partial_y Q_{11})^2-(\partial_y Q_{12})^2\Big) \nonumber\\
&\quad\qquad+2 d_1\Big((\partial_x Q_{11})(\partial_y Q_{11}) +(\partial_x Q_{12})(\partial_y Q_{12})\Big) \Bigg) \nonumber
\\
&\,=\frac{R}{A}\left(d_2 \left( (\partial_x\theta)^2-(\partial_y\theta)^2\right) + 2 d_1\left((\partial_x\theta)(\partial_y\theta) \right) \right) \nonumber\\
&\,=\frac{R}{A}\left(\left(\partial_y\nabla^2 \theta\right)\partial_x\theta- \left(\partial_x\nabla^2 \theta\right)\partial_y\theta\right). \nonumber
\end{align}

From \cref{eq curl div sigma act}, we obtain
\begin{align}
    curl (\nabla\cdot \sigma^{\text{act}})&=2S\left( d_2 Q_{11}+d_1 Q_{12}\right) \\
    &=S\left(d_2 \cos 2\theta +d_1 \sin2\theta\right). \nonumber
\end{align}

Lastly, from \cref{eq curl div sigma align}, we obtain
\begin{align}
    &curl (\nabla\cdot \sigma^{\text{align}})  \\
    &\,= -2\frac{R\nu}{A}\Bigg(
    d_2 \Big(H_{11} -4 Q_{11}(H_{11}Q_{11}+H_{12}Q_{12})\Big) \nonumber\\
    &\qquad\qquad +d_1\Big(H_{12} -4Q_{12}(H_{11}Q_{11}+H_{12}Q_{12})\Big)\Bigg)\nonumber \\
    &\,= -\frac{R\nu}{A}\left(
    -d_2 \left(\sin2\theta\,\nabla^2\theta \right) +d_1\left(\cos2\theta\,\nabla^2\theta \right)\right).\nonumber
\end{align}

We then have, via \cref{eq psi adim},
\begin{align}
    \nabla^4\psi=&\,S(d_1 \sin2\theta+d_2\cos2\theta) \\
    &\,-\frac{R}{A}\Big(\frac{1}{2}\nabla^4\theta +\left(\partial_x\nabla^2 \theta\right)\partial_y\theta-\left(\partial_y\nabla^2 \theta\right)\partial_x\theta \nonumber \\
    &\quad\qquad+\nu\left(d_1(\cos2\theta\,\nabla^2\theta)-d_2(\sin2\theta\,\nabla^2\theta)\right). \nonumber
\end{align}
This equation recovers the same form as \cref{eq vorticity}, but with $h_{\parallel}=0$---a consequence of the orthogonal form of flow alignment in the formulation of dynamics of $\mathbf{Q}$.

\subsection{Numerical scheme}
\label{Q numerical}

Here we describe our pseudo-spectral method for numerically integrating \cref{eq psi adim} and \crefrange{eq dt Q11}{eq dt Q12}. In what follows, the index $n$ is the time iteration,  $dt$ is the timestep, and $h=N^{-1}$ is the vertex spacing on our $N\times N$ square grid. We note that for defects to be adequately resolved, one must ensure that $h<\varepsilon$.

In the computations outlined below, all derivatives including $d_1:=\frac{1}{2}(\partial_y^2-\partial_x^2)$, $d_2:=\partial_y\partial_x$ are performed in the space of the two-dimensional discrete Fourier transform (DFT). Before each real-space multiplication, we apply the 2/3 de-aliasing rule, truncating the Fourier modes with $|q_x|$ or $|q_y|$ greater than $\frac{2}{3}\pi N$. 

\subsubsection{Computing current $\mathbf{H}$ components}
We begin each iteration by computing $H_{11}^n$ and $H_{12}^n$ based on $Q_{11}^n$ and $Q_{12}^n$, the arrays representing the current Q-tensor on the vertices. Following \crefrange{eq H_11 adim}{eq H_12 adim},
\begin{align}
H_{11}^n&=\frac{1}{2}\left(\nabla^2 Q_{11}^n+\varepsilon^{-2}(1-{s^{n}}^2)Q_{11}^n\right),\\
H_{12}^n&=\frac{1}{2}\left(\nabla^2 Q_{12}^n+\varepsilon^{-2}(1-{s^{n}}^2)Q_{12}^n\right),
\end{align}
with
\[{s^n}^2=4\left({Q_{11}^n}^2+{Q_{12}^n}^2\right).\]

\subsubsection{Solving momentum balance}
We turn to solving for the current streamfunction based on $Q_{11}^n,\,Q_{12}^n$ and $H_{11}^n,\,H_{12}^n$. After rearranging terms in \cref{eq psi adim}, one has
\begin{align}
    \nabla^4&\psi^n =\, 2S \big(d_2Q_{11}^n+d_1 Q_{12}^n\big) \\
    +&\frac{R}{A    }\Bigg(2\nabla^2\left(Q_{12}^nH_{11}^n-Q_{11}^nH_{12}^n\right)\nonumber\\
    &\quad+\,d_2\left((\partial_xQ_{11}^n)^2+(\partial_xQ_{12}^n)^2- (\partial_yQ_{11}^n)^2-(\partial_yQ_{12}^n)^2\right)\nonumber \\
    &\quad+2 d_1\left((\partial_x Q_{11}^n)(\partial_y Q_{11}^n) +(\partial_x Q_{12}^n)(\partial_y Q_{12}^n)\right) \nonumber \\
    &\quad-2\nu \Big( d_2 \left({H}_{11}^n -4 Q_{11}^n({H}_{11}^nQ_{11}^n+{H}_{12}^nQ_{12}^n)\right) \nonumber\\
    & \quad \qquad +d_1\left({H}_{12}^n -4Q_{12}^n({H}_{11}^nQ_{11}^n+{H}_{12}^nQ_{12}^n)\right)\Big)\Bigg). \nonumber
\end{align}

Upon carefully computing all RHS contributions, this equation is trivially inverted in Fourier space, 
\begin{align}   &\mathcal{F}\left[{\psi}^n\right]_\mathbf{q}=\frac{1}{q^4}\mathcal{F}\left[\text{RHS}^n\right]_\mathbf{q}\quad \forall \bq\neq \mathbf{0}, \nonumber \\
&\mathcal{F}\left[{\psi}^n\right]_{\mathbf{q}=\mathbf{0}}=0. \nonumber
\end{align}

\subsubsection{Computing the flow, vorticity and additional flow gradients}
Before attending to \crefrange{eq dt Q11}{eq dt Q12}, we truncate the highest $1/3$ modes in $\psi$ and compute the following fields pseudo-spectrally:
\begin{align}
   &v_x^n = \partial_y \psi^n,\quad v_y^n=-\partial_x\psi^n,\quad \omega^n = -\nabla^2 \psi^n, \label{eq v omega E1 E2} \\
   &E_1^n=d_1\psi^n,\quad E_2^n=d_2\psi^n. \nonumber
\end{align}

\subsubsection{Evolving $\mathbf{Q}$ components (integrating factor + Adams-Bashforth)}
Let us rewrite \crefrange{eq dt Q11}{eq dt Q12} with the linear Laplacian term stemming from $H_{\alpha\beta}$ moved to the LHS, along with the time derivative, and all other terms brought to the RHS.
\begin{align}
    \partial_t Q_{11}-\frac{1}{A}&\nabla^2 Q_{11}= \frac{1-s^2}{A\varepsilon^2}Q_{11}-(v_x,v_y)\cdot\nabla Q_{11}-\omega Q_{12} \nonumber\\
     & -\nu  \left((1-4Q_{11}^2)E_2-4Q_{11}Q_{12} E_1 \right),\label{eq dt Q11 shift}  \\
     \partial_t Q_{12}-\frac{1}{A}&\nabla^2 Q_{12}= 
     \frac{1-s^2}{A\varepsilon^2}Q_{12}-(v_x,v_y)\cdot\nabla Q_{12}+\omega Q_{11}   \nonumber \\
     & -\nu  \left((1-4Q_{12}^2)E_1-4Q_{11}Q_{12}E_2 \right), \label{eq dt Q12 shift} 
\end{align}
where, as in \cref{eq v omega E1 E2}, $\omega=-\nabla^2\psi$, $E_1=d_1\psi$ and $E_2=d_2\psi$. \Cref{eq dt Q11 shift,eq dt Q12 shift} are still expressed in continuous time. Upon time discretization, the terms on the RHS will be evaluated explicitly using a multi-step method. 

We now introduce the exponential integrating factor via a change of variable in Fourier space.
\[
\hat{T}_{11}(\bm{q},t):=\hat{Q}_{11}(\bm{q},t) e^{A^{-1}q^2t}.
\]
It follows that
\begin{align}
\partial_t \hat{T}_{11}&=e^{A^{-1}q^2t}\left(\partial_t\hat{Q}_{11}+\frac{q^2}{A}\hat{Q}_{11}\right) \nonumber \\
&= e^{A^{-1}q^2 t} \mathcal{F}\left[\text{LHS}\eqref{eq dt Q11 shift}\right] = e^{A^{-1}q^2 t} \mathcal{F}\left[\text{RHS}\eqref{eq dt Q11 shift}\right]. \nonumber
\end{align}

We evolve $\hat{T}_{11}$ via the second-order Adams-Bashforth method (AB2), 
\begin{align}
\hat{T}_{11}^{n+1}=&\,\hat{T}_{11}^{n}  \label{eq Tn+1} \\
&+\frac{dt}{2}\Big(3\, e^{A^{-1}q^2 t_{n}}\mathcal{F}\left[\text{RHS}\eqref{eq dt Q11 shift}^n\right] \nonumber \\
&\qquad-e^{A^{-1}q^2 t_{n-1}}\mathcal{F}\left[\text{RHS}\eqref{eq dt Q11 shift}^{n-1}\right] \Big). \nonumber
\end{align}
where $\text{RHS}\eqref{eq dt Q11 shift}^n$ denotes the right-hand side of \cref{eq dt Q11 shift} evaluated at time $t_n$ (with $Q_{11}^n$, $Q_{12}^n$, ${s^n}$, and $v_x^n$, $v_y^n$, $\omega^n$, $E_1^n$, $E_2^n$), while $\text{RHS}\eqref{eq dt Q11 shift}^{n-1}$ is evaluated at time time $t_{n-1}$ (with $Q_{11}^{n-1}$, $Q_{12}^{n-1}$, and so on).

Substituting the definition of $\hat{T}_{11}$ back in \cref{eq Tn+1}, and dividing by $e^{A^{-1}q^2 t_{n+1}}$, we obtain
\begin{align}
\hat{Q}_{11}^{n+1}=&\,\hat{Q}_{11}^{n}e^{-A^{-1}q^2 dt} \\
&+\frac{dt}{2}\Big(3\, e^{-A^{-1}q^2 dt} \mathcal{F}\left[\text{RHS}\eqref{eq dt Q11 shift}^n\right] \nonumber \\
&\qquad-e^{-2A^{-1}q^2 dt}\mathcal{F}\left[\text{RHS}\eqref{eq dt Q11 shift}^{n-1}\right] \Big). \nonumber
\end{align}

Applying the same procedure to \cref{eq dt Q12 shift}, we obtain
\begin{align}
\hat{Q}_{12}^{n+1}=&\,\hat{Q}_{12}^{n}e^{-A^{-1}q^2 dt} \\
&+\frac{dt}{2}\Big(3\, e^{-A^{-1}q^2 dt}\mathcal{F}\left[\text{RHS}\eqref{eq dt Q12 shift}^n\right] \nonumber \\
&\qquad-e^{-2A^{-1}q^2 dt}\mathcal{F}\left[\text{RHS}\eqref{eq dt Q12 shift}^{n-1}\right] \Big). \nonumber
\end{align}

This concludes a single iteration of our explicit time-stepping scheme. The linear Laplacian term is treated exactly via an integrating factor, while all other terms are advanced explicitly using a second-order Adams–Bashforth method (AB2). All spatial derivatives are computed pseudo-spectrally and 2/3 de-aliasing is applied prior to real-space multiplications. At the end of each iteration, we update not only $Q_{11}$ and $Q_{12}$, but also the previous right-hand sides of \cref{eq dt Q11 shift,eq dt Q12 shift}, as required for the AB2 scheme. 

\begin{figure}[tbp!]
\begin{center}
\includegraphics[scale=0.48]{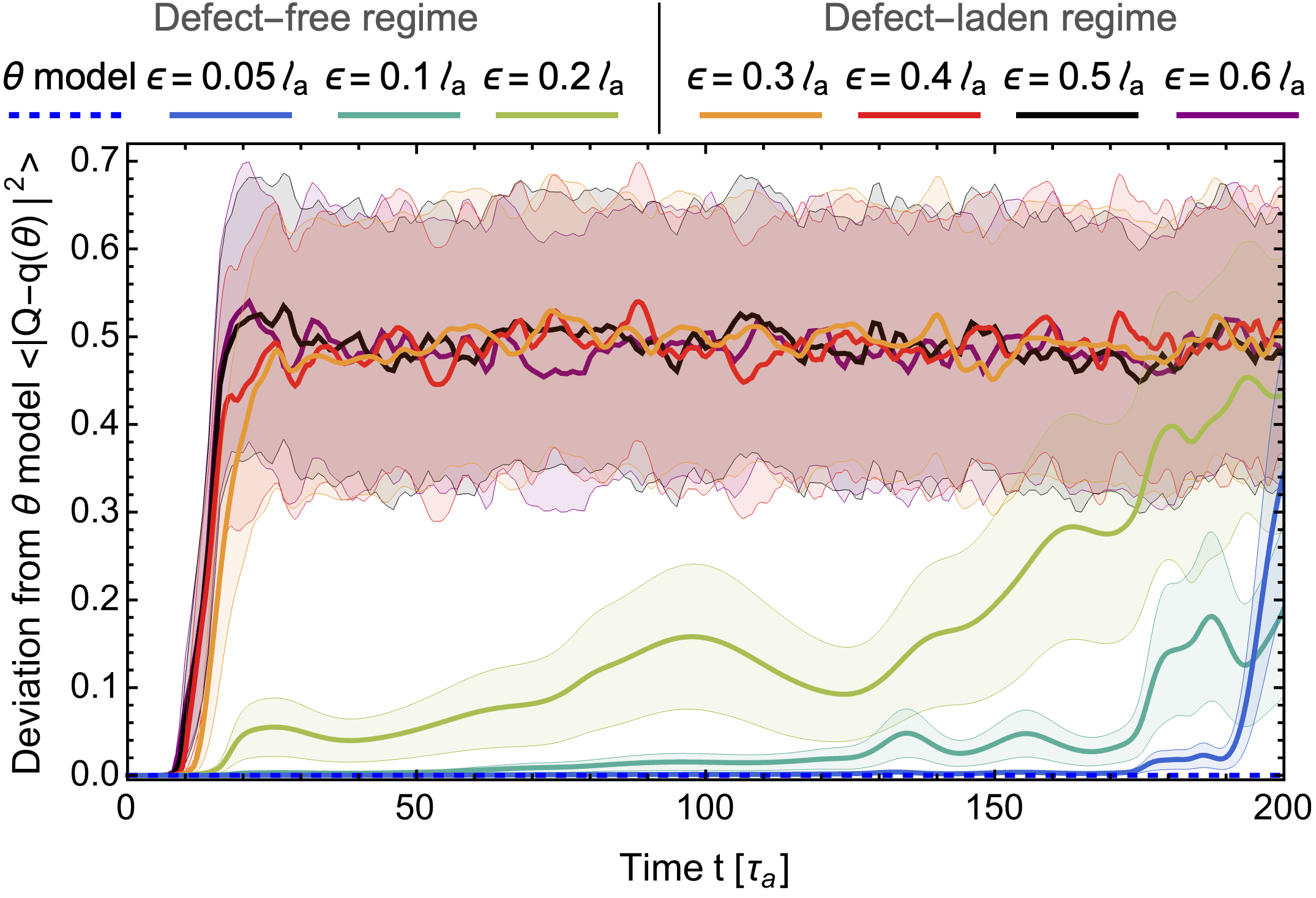}
\end{center}
\bfcaption{Agreement with the constrained model at small defect core size}{For all simulations shown in \cref{Fig Q-tensor}, which share the same initial conditions and parameters except for $\epsilon$, we plot the space-averaged deviation of the nematic tensor $\mathbf{Q}$ from that corresponding to the reference $\theta$ model simulation. Shaded regions indicating the standard deviation. As $\epsilon$ decreases, quantitative agreement (low deviation) persists systematically for longer integration times. A deviation of $\sim 0.5$ corresponds to complete statistical de-correlation. See also \Movie{6}. }
\label{Fig agreement}
\end{figure}

\begin{figure*}[tbhp!]
\begin{center}
\includegraphics[scale=0.48]{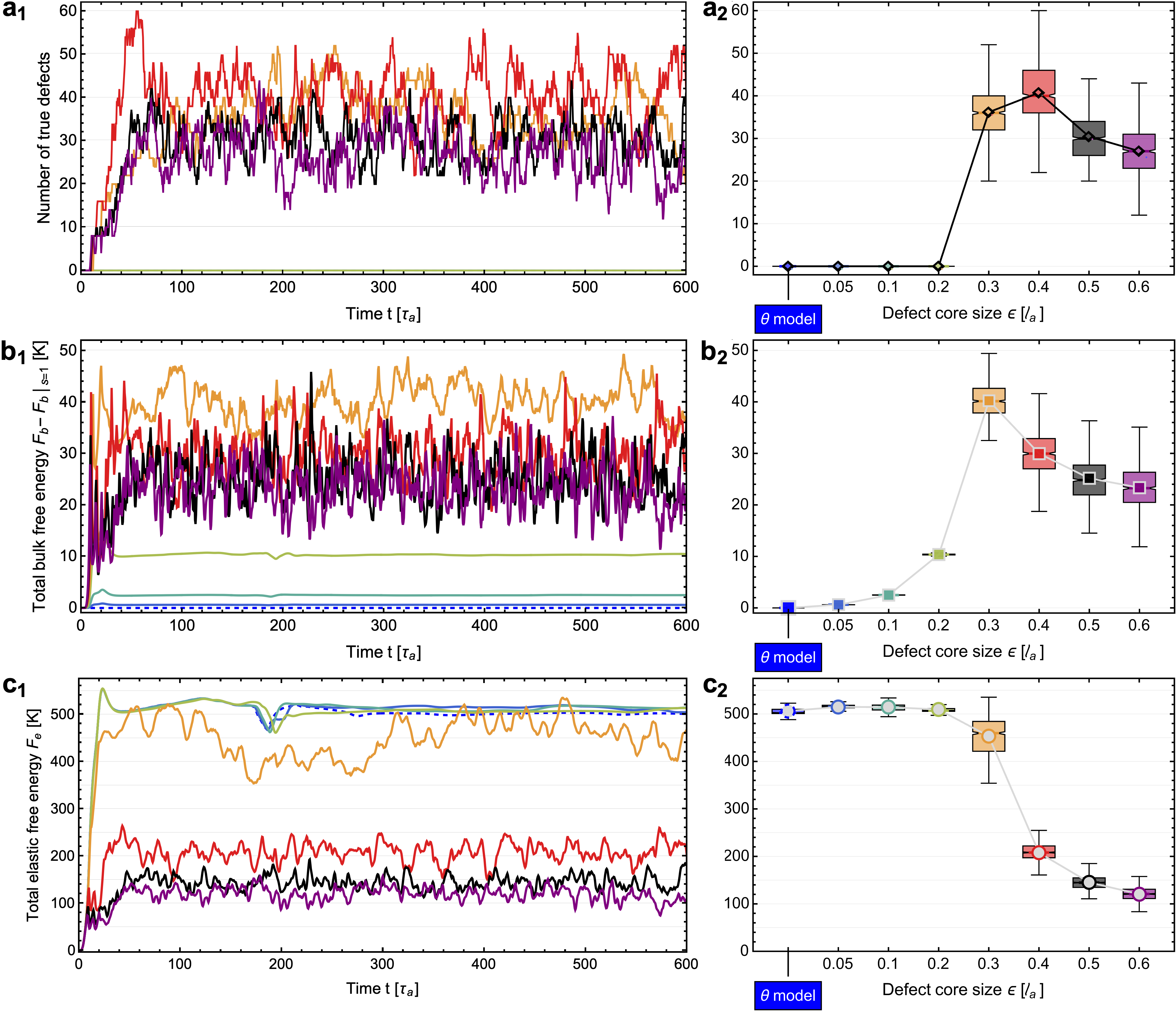}
\end{center}
  {\phantomsubcaption\label{Fig defects}}
  {\phantomsubcaption\label{Fig bulk energy}}
  {\phantomsubcaption\label{Fig elastic energy}}
\bfcaption{Timeseries and statistical breakdown of the total defect number and free energy contributions}{\subref*{Fig defects}$_1$, Number of true topological defects as a function of time in simulations where $\epsilon$ is the varied control parameter. Colors correspond to the $\epsilon$ values in \cref{Fig Q-tensor,Fig agreement}. For $\epsilon \leq 0.2\,\ell_\text{a}$, the system does not spontaneously form defects. \subref*{Fig defects}$_2$, Box–whisker plot summarizing the data in (\subref*{Fig defects}$1$) for $t > 100\,\tau\text{a}$ (excluding initial transients), as a function of $\epsilon$. The boxes span the interquartile range (25\%–75\%), with notches indicating the median and whiskers showing the upper and lower fences. Diamond markers connected by a black line denote the mean values, also shown in \cref{Fig no defects}. \subref*{Fig bulk energy}$_1$, Total bulk free energy, computed relative to its minimum at $s=1$, shown as a function of time for each simulation. Colors indicate the value of $\epsilon$ and the dashed blue curve (here identically zero) represents the $\theta$ model. \subref*{Fig bulk energy}$_2$, Same as (\subref*{Fig defects}$_2$) for the data in (\subref*{Fig bulk energy}$_1$). Square markers connected by a gray line denote the mean values, also shown in \cref{Fig no defects}. \subref*{Fig elastic energy}$_1$, Total elastic free energy shown as a function of time for each simulation. Again, colors indicate the value of $\epsilon$ and the dashed blue curve represents the $\theta$ model. \subref*{Fig elastic energy}$_2$, Same as (\subref*{Fig defects}$_2$) for the data in (\subref*{Fig elastic energy}$_1$). Circle markers connected by a gray line denote the mean values, also shown in \cref{Fig no defects}. Note the difference in scales between variations in (\subref*{Fig elastic energy}) versus (\subref*{Fig bulk energy}). }
\label{Fig timeseries}
\end{figure*}

\subsection{Agreement with the $\theta$ model at small $\epsilon/\ell_\text{a}$}
\label{Q agreement}

In \cref{Fig agreement}, we study the convergence of the full Q-tensor model to the $s=1$ constrained, $\theta$ model (\cref{Q theta model}) as the defect core size decreases. Specifically, we plot the spatial average of the squared deviation $|\mathbf{Q}-\mathbf{q}(\theta)|^2$ over time, with $\mathbf{q}(\theta)$ denoting the tensorial order parameter computed from the $\theta$ model simulation,   
\[
\mathbf{q}(\theta) = \frac{1}{2}\begin{pmatrix}
    \cos 2\theta & \sin 2\theta \\
    \sin 2\theta & -\cos 2\theta
\end{pmatrix}.
\]  
All simulations use the same parameters (except for $\epsilon$) and the same initial condition: a nematic state with $s=1$ and a small orientational perturbation. When true defects form ($\epsilon \geq 0.3\ell_\text{a}$), decorrelation is reached rapidly as defects begin to nucleate. Below the nucleation threshold ($\epsilon \leq 0.2\ell_\text{a}$), the full Q-tensor simulations converge toward the constrained model for increasingly longer integration times as $\epsilon$ decreases. See \Movie{6} for side by side evolutions.

\subsection{Defect number, bulk energy and elastic energy}
\label{Q timeseries}

In \cref{Fig timeseries}, we present timeseries and statistical breakdowns of the total defect number, bulk free energy, and elastic free energy from our Q-tensor simulations, as well as the constrained $\theta$ model for reference. This figure complements \cref{Fig no defects}, which, for compactness, shows only the time-averaged values of these quantities.

The total defect count (\cref{Fig defects}) sums over both $+1/2$ and $-1/2$ defects, identified using a thresholding method detailed in \cref{defect detection}. Near the transition from arrested wall networks to defect-laden turbulence, we find that the total number of defects (at long times) increases sharply with the ratio $\epsilon / \ell_\text{s}$. Interestingly, this number—proportional to the defect density since the system size is fixed—shows a nonmonotonic dependence on $\epsilon$. Understanding the origin of the observed decrease in defect density with increasing core size is deferred to future work.

We note that the total bulk free energy (\cref{Fig bulk energy}) is computed relative to the minimum at $s=1$, which differs for each value of $\epsilon$. The variations in this energy across the simulations are, in absolute terms, much smaller than the variations observed in the elastic free energy (\cref{Fig elastic energy}).

\section{IMAGE PROCESSING} 

\label{image processing}

\subsection{Skeleton detection and pseudo-defect classification}

\begin{figure*}[tbp!]
\begin{center}
\includegraphics[scale=0.48]{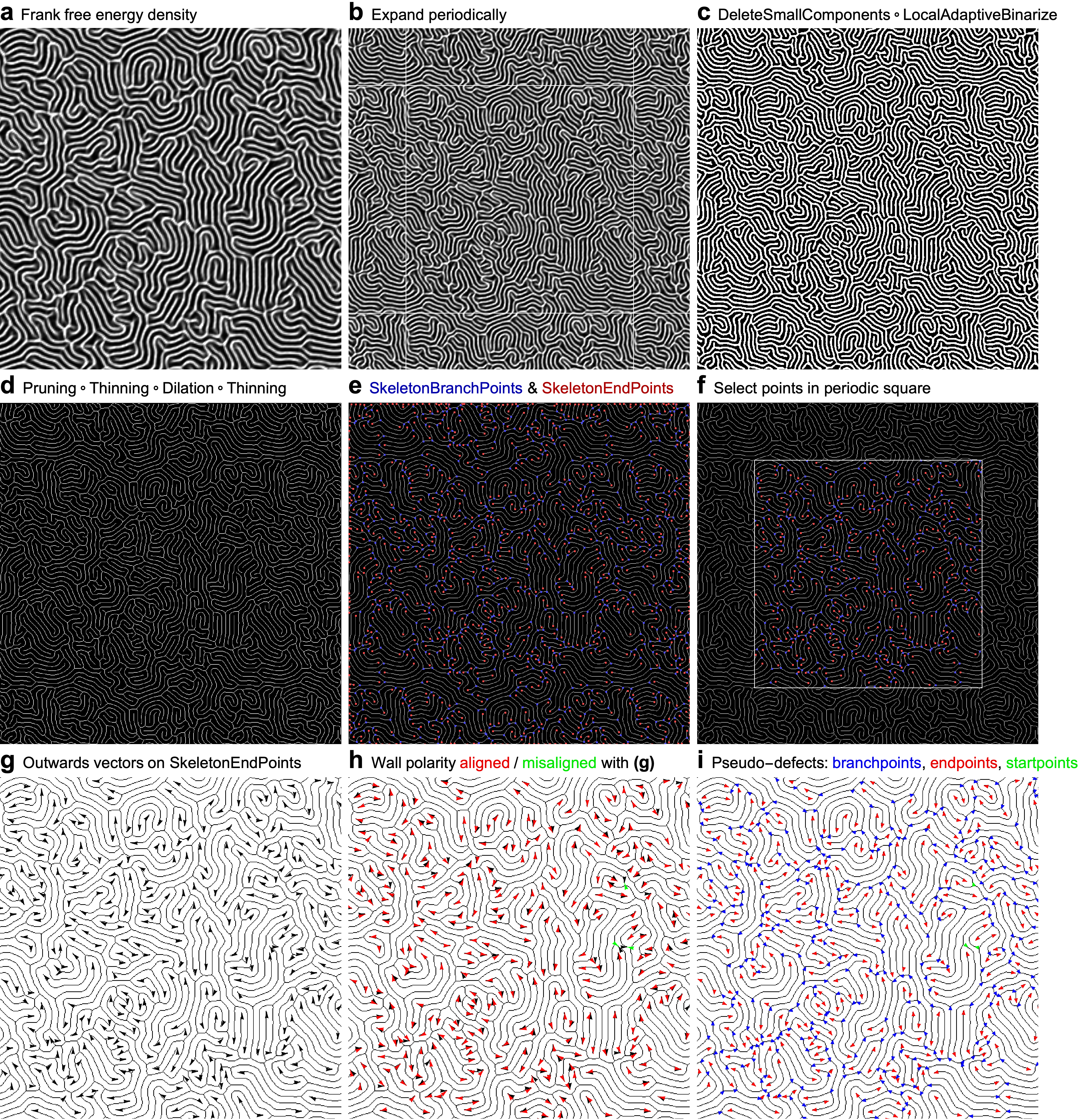}
\end{center}
  {\phantomsubcaption\label{Fig Original Frank}}
  {\phantomsubcaption\label{Fig Expanded Frank}}
  {\phantomsubcaption\label{Fig Binarized}}
  {\phantomsubcaption\label{Fig Skeleton}}
  {\phantomsubcaption\label{Fig Skeleton nodes}}
  {\phantomsubcaption\label{Fig Cropped nodes}}
  {\phantomsubcaption\label{Fig Outwards}}
  {\phantomsubcaption\label{Fig Endpoint polarity}}
  {\phantomsubcaption\label{Fig Classified}}
\bfcaption{Skeleton detection and pseudo-defect classification}{ \subref*{Fig Original Frank}, Grayscale plot of the Frank free energy density, with black corresponding to $\left|\nabla\theta\right|^2=0$ and white corresponding to $\left|\nabla \theta\right|^2 \geq 0.5\, \text{max}\left|\nabla \theta\right|^2$ to facilitate binarization. \subref*{Fig Expanded Frank}, Periodic expansion of (\subref*{Fig Original Frank}) to avoid false detection of nodes along the original square boundaries (white outline). \subref*{Fig Binarized}, Local adaptive binarization of (\subref*{Fig Expanded Frank}), followed by removal of small fragmented components not counted as walls. \subref*{Fig Skeleton}, Recursive thinning of (\subref*{Fig Binarized}), then a step of dilation to fill up tiny gaps within some walls, an additional thinning step to recover the skeleton, and a pruning step to discard open branches shorter than a small threshold. \subref*{Fig Skeleton nodes}, Using the MorphologicalTransform tool we detect and highlight the skeleton nodes. \subref*{Fig Cropped nodes}, We select only those nodes confined to the original periodic square. \subref*{Fig Outwards}, For the SkeletonEndPoints, we compute the vectors pointing outwards of their associated branch (black arrows) by finding the nearest neighbouring point on the wall skeleton. \subref*{Fig Endpoint polarity}, The wall polarity associated with each SkeletonEndPoint is shown in red (green) if its scalar product with (\subref*{Fig Outwards}) is positive (negative). \subref*{Fig Classified}, Branchpoints (blue nodes), endpoints (red nodes) and startpoints (green nodes) are plotted with their orientation given by the local wall polarity. }
\label{Fig node detection}
\end{figure*}

The process of skeleton detection and pseudo-defect classification is depicted in \cref{Fig node detection}. We apply the following pipeline using the image processing tools in Mathematica (TM). 

Initially, a plot of the Frank free energy density, $\left|\nabla\theta\right|^2$, is rendered (\cref{Fig Original Frank}). The grayscale color scheme and the plot range are chosen to aid the binarization of the walls. To avoid false node detection along the square boundaries, this plot is periodically expanded using the \textbf{ImageAssemble} function (\cref{Fig Expanded Frank}). Subsequently, we binarize the image by applying the \textbf{LocalAdaptiveBinarize} function, with neighbourhood pixel range set to 6. Then, we apply the \textbf{DeleteSmallComponents} function to remove isolated white domains, specified as totalling 20 or fewer pixels, that we do not consider part of the wall network (\cref{Fig Binarized}). 

To obtain the network skeleton, we apply the \textbf{Thinning} function with default setting on the binarized image. At times, this alone can leave tiny black gaps within some walls that would later be falsely interpreted as branching points. To avoid this, we further apply the \textbf{Dilation} function, with range set to 1 pixel, and then repeat \textbf{Thinning}. To prune short branches from the skeleton we apply the \textbf{Pruning} function, specifying the range to 2 pixels. The final skeleton is shown in \cref{Fig Skeleton}. 

The nodes of the skeleton are then detected using the \textbf{MorphologicalTransform} function, specifying "SkeletonBranchPoints" and then "SkeletonEndPoints" (highlighted in blue and red respectively in \cref{Fig Skeleton nodes}). The nodes are filtered to retain only those within the original square (\cref{Fig Cropped nodes}).
At this stage, the "SkeletonBranchPoints" correspond to our pseudo-defect branchpoints (\cref{Fig branch point}). However, the "SkeletonEndPoints" can be either pseudo-defect endpoints (\cref{Fig endpoint}) or startpoints (\cref{Fig startpoint}). 

To distinguish between endpoints and startpoints, we test whether the wall polarity faces outwards or inwards with respect to the branch associated with each "SkeletonEndPoint". To this end, we compute two vectors on each such node. The vector pointing outwards from the associated branch is calculated by locating the nearest neighboring point on the wall skeleton (\cref{Fig Outwards}). The wall polarity at each node is determined by $(\partial_y\theta, -\partial_x\theta)$. The scalar product of the polarity with the outwards vector classifies the type of pseudo-defect (\cref{Fig Endpoint polarity}). 

Finally, \cref{Fig Classified} displays the classified nodes with branchpoints (blue), endpoints (red), and startpoints (green)---all oriented according to the local wall polarity.

\begin{figure*}[tbp!]
\begin{center}
\includegraphics[scale=0.48]{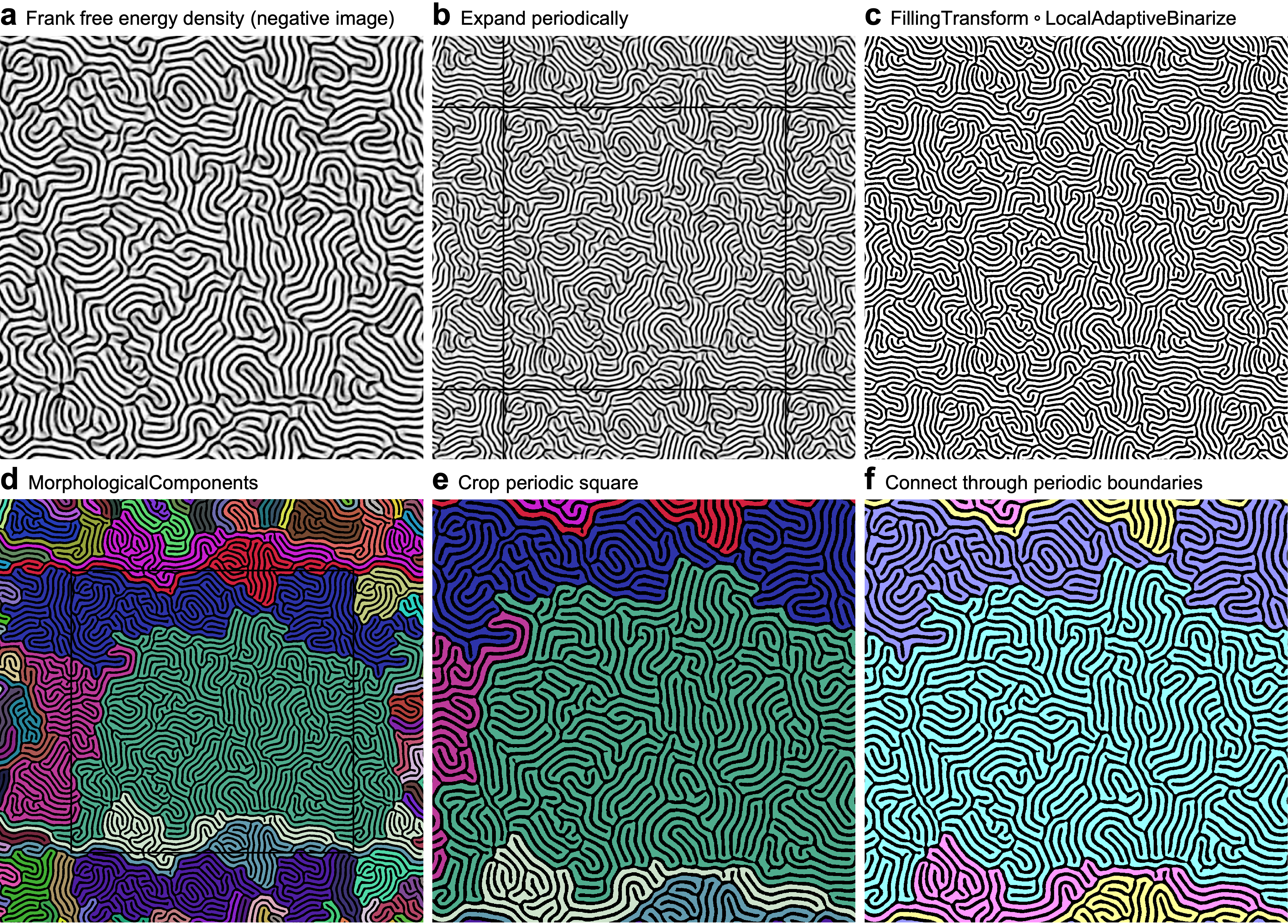}
\end{center}
  {\phantomsubcaption\label{Fig Negative Frank}}
  {\phantomsubcaption\label{Fig Expanded Negative Frank}}
  {\phantomsubcaption\label{Fig Negative Binarized}}
  {\phantomsubcaption\label{Fig MorphologicalComponents}}
  {\phantomsubcaption\label{Fig Cropped MorphologicalComponents}}
  {\phantomsubcaption\label{Fig Connected boundaries}}
\bfcaption{Labyrinths detection in arrested wall networks}{ \subref*{Fig Negative Frank}, Grayscale plot of the Frank free energy density, with white corresponding to $\left|\nabla\theta\right|^2=0$ and black corresponding to $\left|\nabla \theta\right|^2 \geq 0.5\, \text{max}\left|\nabla \theta\right|^2$ to facilitate binarization. \subref*{Fig Expanded Negative Frank}, Periodic expansion of (\subref*{Fig Negative Frank}) to capture connected domains that meander through the periodic square boundaries. \subref*{Fig Negative Binarized}, Local adaptive binarization of (\subref*{Fig Expanded Negative Frank}), followed by a filling operator to discard isolated black fragments. \subref*{Fig MorphologicalComponents}, Finding the connected morphological domains in (\subref*{Fig Negative Binarized}). \subref*{Fig Cropped MorphologicalComponents}, Cropping out the periodic square domain from (\subref*{Fig MorphologicalComponents}). \subref*{Fig Connected boundaries}, Applying our algorithm for relabeling domains that are connected through the left-right and top-bottom boundaries. }
\label{Fig labyrinth detection}
\end{figure*}

\subsection{Labyrinths detection}

The process of labyrinth detection in arrested wall networks is depicted in \cref{Fig labyrinth detection}. Here, the frames of interest are those in which no startpoints have been detected. Similarly to our skeleton detection scheme, we apply a pipeline that uses the image processing tools in Mathematica (TM).

Initially, an inverted plot of $\left|\nabla\theta\right|^2$ is rendered (\cref{Fig Negative Frank}). The color scheme and plot range are selected to aid the binarization of the gaps between the walls. To assist in identifying connected domains that frequently pass through the square boundaries, this plot is periodically expanded using the \textbf{ImageAssemble} function (\cref{Fig Expanded Negative Frank}). The \textbf{LocalAdaptiveBinarize} function is then applied with a neighborhood pixel range set to 10, followed by the \textbf{FillingTransform} function to eliminate isolated black fragments not considered as walls (\cref{Fig Negative Binarized}).

Next, we find the connected morphological domains in the binarized image using the \textbf{MorphologicalComponents} function (\cref{Fig MorphologicalComponents}). These distictnly labeled domains are then cropped to retain only those within the original periodic square (\cref{Fig Cropped MorphologicalComponents}).

To relabel domains that connect through the left-right and top-bottom boundaries, we employ a custom algorithm that ensures continuity across the periodic square (\cref{Fig Connected boundaries}). This step completes the process of detecting the unicursal labyrinths which span the system size.

\subsection{Topological defects detection}
\label{defect detection}

The detection of true defects is more straightforward than that of pseudo-defects, since the tensorial order parameter $\mathbf{Q}$ provides clear local signatures of their appearance. Specifically, we rely on the scalar order parameter $s=2\sqrt{Q_{11}^2+Q_{12}^2}$ and the so-called topological defect density:
\[
\rho_\text{top}=\frac{1}{\pi}\left(\partial_xQ_{11}\partial_y Q_{12}-\partial_yQ_{11}\partial_xQ_{12}\right).
\]

At a defect core, $s$ falls to low values, approaching zero in the continuum limit, while $\rho_\text{top}$ develops local extrema whose sign indicates the topological charge. We identified core regions using a thresholding approach: a vertex was marked as part of a $+1/2$ defect core if $s < s_0$ and $\rho_\text{top} > r_0 \max(\rho_\text{top})$, and as part of a $-1/2$ defect core if $s < s_0$ and $\rho_\text{top} < r_0 \min(\rho_\text{top})$, with $0 < s_0 < 1$ and $0 < r_0 < 1$ as threshold parameters.

For each frame, we generated two binarized images based on these criteria and tiled them across periodic boundaries to obtain an added margin. Using the \textbf{ComponentMeasurements} function, we extracted centroids of connected regions, corresponding to the locations of positive and negative defect cores. We then filtered these to retain only those centroids that lie within the original simulation domain. We explored several values of $s_0$ and $r_0$ to achieve the closest agreement between the counts of $+1/2$ and $-1/2$ defects, as required by topological charge conservation. The detected defect positions were overlaid on plots of the director field (computed as the eigenvector of $\mathbf{Q}$) to visually confirm the results.

\section{SUPPLEMENTAL MOVIE CAPTIONS} \label{movies}

\newcommand\mybar{\kern2.5pt\rule[-\dp\strutbox]{1pt}{\baselineskip}\kern2.5pt}
\noindent\textbf{Movie~S1\mybar Strong large-scale turbulence in the contractile aligning regime.} The left panel shows the evolution of the flow field $\bm{v}$, with color indicating speed and black arrows representing streamlines. The right panel depicts the evolution of the Frank free energy density, $|\nabla\theta|^2$. The parameters are the same as in \cref{Fig velocity-non-arrested,Fig Frank-non-arrested}, which are snapshots from the same simulation. The interval between movie frames is 500 computational time iterations, corresponding to a time interval of $5\tau_\text{a}$. The total duration of the movie is $2500\tau_\text{a}$.

\vspace{10pt}

\noindent\textbf{Movie~S2\mybar Arrested turbulence in the extensile aligning regime.} The left panel shows the evolution of the flow field $\bm{v}$, with color indicating speed and black arrows representing streamlines. The right panel depicts the evolution of the Frank free energy density, $|\nabla\theta|^2$. The parameters are the same as in \cref{Fig velocity-arrested,Fig Frank-arrested}, which are snapshots from the same simulation. The interval between movie frames is 500 computational time iterations, corresponding to a time interval of $5\tau_\text{a}$. The total duration of the movie is $2500\tau_\text{a}$.

\vspace{10pt}

\noindent\textbf{Movie~S3\mybar Active turbulence in the absence of flow alignment (tumbling regime).} The left panel shows the evolution of the flow field $\bm{v}$, with color indicating speed and black arrows representing streamlines. The right panel depicts the evolution of the Frank free energy density, $|\nabla\theta|^2$. The parameters are the same as in Fig.\ S1\cite{SM}, showing snapshots from the same simulation. The interval between movie frames is 500 computational time iterations, corresponding to a time interval of $5\tau_\text{a}$. The total duration of the movie is $2500\tau_\text{a}$.

\vspace{10pt}

\noindent\textbf{Movie~S4\mybar Evolution from stripes to turbulence (contractile aligning vs extensile aligning).} The evolution of the Frank free energy density, $|\nabla\theta|^2$, in the contractile aligning case ($S\nu=+1.1$, left) and the extensile aligning case ($S\nu=-1.1$, right). The wavelength of the initial stripes---with splay walls on the left and bend walls on the right---was chosen to match the selected wavelength in the fully-developed turbulent state. In both cases, the stripes coarsen and then experience a `zigzag' instability that folds the walls. In the contractile case (left), folded walls tend to dissolve and reappear perpetually. In the extensile case (right, \cref{Fig walls}), walls are more robust and tend to branch. As growing branches avoid other walls, the system reaches a grid-locked state that fluctuates only slightly. Parameters were set to $R=1$, $\nu=-1.1$, and $A=19692$ (chosen so that the system size roughly equals 6 times the selected wavelength). The interval between movie frames is 50 computational time iterations, corresponding to a time interval of $1\tau_\text{a}$. The total duration of the movie is $540\tau_\text{a}$.

\vspace{10pt}

\noindent\textbf{Movie~S5\mybar Evolution of the wall-network skeleton and its nodes.} Skeleton of the domain walls (black) with startpoints, branchpoints and endpoints shown as green, blue and red triangular nodes, respectively (see \cref{Fig topology}). The detection of these structures is described in \cref{image processing} and \cref{Fig node detection}. This movie demonstrates the pseudo-topological transitions mediating the ageing of an arrested wall network. Parameter values are as in \cref{Fig aging}. The interval between movie frames is 500 computational time iterations, corresponding to a time interval of $5\tau_\text{a}$. The total duration of the movie is $1715\tau_\text{a}$.

\vspace{10pt}

\noindent\textbf{Movie~S6\mybar Agreement between the constrained $\theta$ model and the unconstrained Q-tensor model at low defect core size.} Evolution of the flow (top panels) and the elastic energy density (bottom panels) in the $\theta$ model, compared with the $Q$-tensor model at low values of $\epsilon$. The behavior of the latter converges to the constrained case as $\epsilon$ decreases relative to $\ell_\text{a}$. See also \cref{Fig Q-tensor,Fig agreement}. Other parameters were set to $R=1$, $\nu=-1$, $S=1$ and $A=10^4$. The interval between movie frames is 500 computational time iterations, corresponding to a time interval of $0.5\tau_\text{a}$. The total duration of the movie is $220\tau_\text{a}$.

\vspace{10pt}

\noindent\textbf{Movie~S7\mybar Transition from dynamical arrest to defect-laden turbulence.} Evolution of the flow (top panels) and the elastic energy density (bottom panels) in the $Q$-tensor model, with $\epsilon$ varied across the defect proliferation threshold. See also \cref{Fig Q-tensor}. Other parameters were set to $R=1$, $\nu=-1$, $S=1$ and $A=10^4$. The interval between movie frames is 200 computational time iterations, corresponding to a time interval of $1\tau_\text{a}$. The total duration of the movie is $600\tau_\text{a}$.

\vspace{10pt}

\noindent\textbf{Movie~S8\mybar Unraveling dynamical arrest via a core-size quench.} The left panel shows the elastic energy density, and the right panel shows a line-integral-convolution plot of the director, overlaid with the scalar order parameter (black indicating $s\to 0$). We first present the evolution of an arrested state generated at a low defect core size ($\epsilon=0.2\ell_\text{a}$). We then pause to highlight the localized energy peaks associated with branchpoint pseudo-defects. Next, the simulation is continued with a larger defect core size, beyond the threshold for defect nucleation ($\epsilon=0.35\ell_\text{a}$), showing that defect pairs nucleate at suspected hotspots. Finally, we replay the last segment with the true defects explicitly traced: yellow points are $+1/2$ defects, while cyan points $-1/2$ defects. Other parameters were set to $R=1$, $\nu=-1$, $S=1$ and $A=10^4$. 

\vspace{10pt}

\noindent\textbf{Movie~S9\mybar Enlarged defect-free extensile system.} Evolution of the elastic energy density from simulations of the constrained $\theta$ model (\cref{Q theta model}) in the extensile, rod-aligning regime ($S\nu=-1.1$). For reference, the movie begins with activity $A=3.2*10^5$ on a $256\times256$ grid, as in \cref{Fig regimes}, where the system rapidly develops an arrested state with persistent labyrinths spanning the entire domain. It then shows a larger system with $A=5.12*10^6$ on a $1024\times1024$ grid, where more extensive labyrinths form but are frequently fractured, preventing persistent system-spanning connectivity.

\vspace{10pt}

\noindent\textbf{Movie~S10\mybar Enlarged defect-free contractile system.} Evolution of the elastic energy density from simulations of the constrained $\theta$ model (\cref{Q theta model}) in the contractile, rod-aligning regime ($S\nu=+1.1$). The movie begins with activity $A=3.2*10^5$ on a $256\times256$ grid, as in \cref{Fig regimes}. It then shows a larger system with $A=5.12*10^6$ on a $1024\times1024$ grid.

\section{SUPPLEMENTAL ANIMATION CAPTIONS} \label{animations}

\noindent\textbf{Animation S1\mybar Startpoint-endpoint pair birth and annihilation via wall inception and wall dissolution.} This animation illustrates both the forward and inverse processes related to \cref{Fig inception}.

\vspace{10pt}

\noindent\textbf{Animation S2\mybar Endpoint-startpoint pair birth and annihilation via local wall deletion and wall completion.} This animation illustrates both the forward and inverse processes related to \cref{Fig deletion}.

\vspace{10pt}

\noindent\textbf{Animation S3\mybar Branchpoint-endpoint pair birth and annihilation via wall branching and wall retraction.} This animation illustrates both the forward and inverse processes related to \cref{Fig branching}.

\vspace{10pt}

\noindent\textbf{Animation S4\mybar Branchpoint to startpoint transition by wall disjoining and startpoint to branchpoint transition by wall joining.} This animation illustrates both the forward and inverse processes related to \cref{Fig dislocation}.

\vspace{10pt}

\noindent\textbf{Animation S5\mybar T1 transitions of connected branchpoints.} This animation illustrates both the forward and inverse processes related to \cref{Fig T1}.

%

\end{document}